\documentclass[12pt,onecolumn]{IEEEtran}
\usepackage{color,graphicx,psfrag,epsfig,epsf,latexsym,theorem,hhline,amsmath,amssymb,array}
\renewcommand {\marginpar}[1]{}
\newcommand {\mymarginpar}[1]{\marginpar{#1}}

\def\_{\rule{.3em}{.15ex}}      

\newtheorem {definition}{Definition}
\newtheorem {axiom}[definition]{Axiom}
\newtheorem {lemma}[definition]{Lemma}
\newtheorem {theorem}[definition]{Theorem}
\newtheorem {corollary}[definition]{Corollary}
\newtheorem {property}[definition]{Property}
\newtheorem {proposition}[definition]{Proposition}
\newtheorem {example}[definition]{Example}
\newtheorem {remark}[definition]{Remark}
\newtheorem {algorithm}[definition]{Algorithm}
\newtheorem {conjecture}[definition]{Conjecture}
\newcommand {\bdefinition}[1]{\begin{definition}
                              \mymarginpar{definition:#1}
                              \label{definition:#1}}
\newcommand {\edefinition}   {\end{definition}}

\newcommand {\baxiom}[1]{\begin{axiom}
                         \mymarginpar{axiom:#1}
                         \label{axiom:#1} }
\newcommand {\eaxiom}   {\end{axiom}}

\newcommand {\blemma}[1]{\begin{lemma}
                         \mymarginpar{lemma:#1}
                         \label{lemma:#1} }
\newcommand {\elemma}   {\end{lemma}}
\newcommand {\rlemma}[1]{Lemma~\ref{lemma:#1}}

\newcommand {\btheorem}[1]{\begin{theorem}
                           \mymarginpar{theorem:#1}
                           \label{theorem:#1}}
\newcommand {\etheorem}   {\end{theorem}}
\newcommand {\rtheorem}[1]{Theorem~\ref{theorem:#1}}

\newcommand {\bcorollary}[1]{\begin{corollary}
                             \mymarginpar{corollary:#1}
                             \label{corollary:#1} }
\newcommand {\ecorollary}   {\end{corollary}}

\newcommand {\bproperty}[1]{\begin{property}
                            \mymarginpar{property:#1}
                            \label{property:#1}}
\newcommand {\eproperty}   {\end{property}}

\newcommand {\bproposition}[1]{\begin{proposition}
                               \mymarginpar{proposition:#1}
                               \label{proposition:#1}}
\newcommand {\eproposition}   {\end{proposition}}

\newcommand {\bproof}{\noindent {\bf Proof.} \ }
\newcommand {\eproof} {\hspace*{\fill}~\mbox{\rule[0pt]{1.3ex}{1.3ex}}}

\newcommand {\bexample}[1]{\begin{example}
                           \mymarginpar{example:#1}
                           \label{example:#1}}
\newcommand {\eexample}   {\end{example}}

\newcommand {\bremark}[1]{\begin{remark}
                          \mymarginpar{remark:#1}
                          \label{remark:#1}}
\newcommand {\eremark}   {\end{remark}}

\newcommand {\balgorithm}[1]{\begin{algorithm}
                         \mymarginpar{algorithm:#1}
                         \label{algorithm:#1} }
\newcommand {\ealgorithm}   {\end{algorithm}}

\newcommand {\bconjecture}[1]{\begin{conjecture}
                         \mymarginpar{conjecture:#1}
                         \label{conjecture:#1} }
\newcommand {\econjecture}   {\end{conjecture}}

\newcommand {\bexercise}[1]{\begin{exercise}
                         \mymarginpar{exercise:#1}
                         \label{exercise:#1} }
\newcommand {\eexercise}   {\end{exercise}}

\newcommand {\barray}{\begin{array}}
\newcommand {\earray}{\end{array}}

\newcommand {\beqn}[1]{\begin{equation}
                       \mymarginpar{eqn:#1}
                       \label{eqn:#1}}
\newcommand {\eeqn}   {\end{equation}}

\newcommand {\beqnarray}[1]{\begin{eqnarray}
                            \mymarginpar{eqn:#1}
                            \label{eqn:#1} }
\newcommand {\eeqnarray}   {\end{eqnarray}}

\newcommand {\beqnarraynn}[1]{\begin{eqnarray*}
                              \mymarginpar{eqn:#1}}
\newcommand {\eeqnarraynn}   {\end{eqnarray*}}

\newcommand {\reqnarray}[1]{(\ref{eqn:#1})}

\newcommand {\bcases}{\begin{cases}}
\newcommand {\ecases}{\end{cases}}

\newcommand {\bselection}{\left\{ \begin{array}{ll}}
\newcommand {\eselection}{\end{array} \right.}

\newcommand {\bsection}[2]{\section{#1}
                           \mymarginpar{section:#2}
                           \label{section:#2} }
\newcommand {\rsection}[1]{Section~\ref{section:#1}}

\newcommand {\bappendix}[2]{\section{#1}
                           \mymarginpar{appendix:#2}
                           \label{appendix:#2} }
\newcommand {\rappendix}[1]{Appendix~\ref{appendix:#1}}

\newcommand {\bfigure}[2]{\begin{figure}[htbp]
                          \centerline {
                          \epsfig{figure={#1},clip=,width={#2}}}}

\newcommand {\efigure}[2]{\caption{#2}
                          \label{figure:#1}
                          \end{figure}
                          \mymarginpar{figure:#1}}

\newcommand {\brotatefigure}[2]{\begin{figure}[htbp]
                                \centerline {
                                \epsfig{figure={#1},clip=,angle=-90,width={#2}}}}

\newcommand {\erotatefigure}[2]{\caption{#2}
                                \label{figure:#1}
                                \end{figure}
                                \mymarginpar{figure:#1}}

\newcommand {\bfigurefirst}[2] {\begin{figure}[h]
                                \centerline {
                                \setlength{\epsfxsize}{#2}
                                \epsffile{#1}}}

\newcommand {\rfigure}[1]{Figure~\ref{figure:#1}}

\newcommand {\btable}[2]{\begin{table}[#1]
                         \begin{center}
                         \begin{tabular}{#2}}
\newcommand {\etable}[2]{\end{tabular}
                         \end{center}
                         \caption{#2}
                         \label{table:#1}
                         \end{table}
                         \mymarginpar{table:#1}
                         \vspace{.1in}}
\newcommand {\rtable}[1]{Table~\ref{table:#1}}

\newcommand {\btabular}[1]{\begin{center}
                           \begin{tabular}{#1}}
\newcommand {\etabular}{\end{tabular}
                        \end{center}}

\newcommand {\btinytable}[2]{\begin{table}[#1]
                         \tiny
                         \begin{center}
                         \begin{tabular}{#2}}
\newcommand {\etinytable}[2]{\end{tabular}
                         \end{center}
                         \caption{#2}
                         \label{table:#1}
                         \end{table}
                         \mymarginpar{table:#1}
                         \vspace{.1in}}

\newcommand {\bscriptsizetable}[2]{\begin{table}[#1]
                         \scriptsize
                         \begin{center}
                         \begin{tabular}{#2}}
\newcommand {\escriptsizetable}[2]{\end{tabular}
                         \end{center}
                         \caption{#2}
                         \label{table:#1}
                         \end{table}
                         \mymarginpar{table:#1}
                         \vspace{.1in}}

\newcommand {\blargetable}[2]{\begin{table}[#1]
                         \LARGE
                         \begin{center}
                         \begin{tabular}{#2}}
\newcommand {\elargetable}[2]{\end{tabular}
                         \end{center}
                         \caption{#2}
                         \label{table:#1}
                         \end{table}
                         \mymarginpar{table:#1}
                         \vspace{.1in}}

\newcommand{\bitemize}{\begin{itemize}}
\newcommand{\eitemize}{\end{itemize}}

\newcommand{\benumerate}{\begin{enumerate}}
\newcommand{\eenumerate}{\end{enumerate}}

\newcommand {\bdescription} {\begin{description}}
\newcommand {\edescription} {\end{description}}

\newcommand{\indep}{\mathop{\lower .12em\hbox{\underbar{\raise .3ex\hbox{$\|$}}}\kern .5pt}}
\newcommand{\nn}{\nonumber}
\newcommand{\aligneq}{\hspace*{-0.1in}&=&\hspace*{-0.1in}}
\newcommand{\alignleq}{\hspace*{-0.1in}&\leq&\hspace*{-0.1in}}
\newcommand{\alignless}{\hspace*{-0.1in}&<&\hspace*{-0.1in}}
\newcommand{\aligngeq}{\hspace*{-0.1in}&\geq&\hspace*{-0.1in}}
\newcommand{\aligngreater}{\hspace*{-0.1in}&>&\hspace*{-0.1in}}
\newcommand{\alignspace}{\hspace*{-0.1in}& &\hspace*{-0.1in}}

\def\0bf{{\bf 0}}
\def\1bf{{\bf 1}}
\def\2bf{{\bf 2}}
\def\3bf{{\bf 3}}
\def\4bf{{\bf 4}}
\def\5bf{{\bf 5}}
\def\6bf{{\bf 6}}
\def\7bf{{\bf 7}}
\def\8bf{{\bf 8}}
\def\9bf{{\bf 9}}

\def\dbf{{\bf d}}

\def\nbf{{\bf n}}

\def\Zbf{{\bf Z}}

\def\Acal{{\cal A}}
\def\Bcal{{\cal B}}
\def\Ccal{{\cal C}}

\def\Gcal{{\cal G}}

\def\Ncal{{\cal N}}

\begin{document}

\baselineskip 16pt

\title{Constructions of Optical Queues With a Limited Number of
Recirculations--Part~I: Greedy Constructions}

\author{Jay Cheng, Cheng-Shang Chang, Sheng-Hua Yang, \\
Tsz-Hsuan Chao, Duan-Shin Lee, and Ching-Min Lien
\thanks{
        This paper was presented in part at the IEEE International Conference on Computer Communications
        (INFOCOM'08), Phoenix, AZ, USA, April~13--18, 2008.}
\thanks{The authors are with the Department of Electrical Engineering and
        the Institute of Communications Engineering,
        National Tsing Hua University, Hsinchu 30013, Taiwan, R.O.C.
        (e-mails: jcheng@ee.nthu.edu.tw; cschang@ee.nthu.edu.tw; u941809@oz.nthu.edu.tw;
        thchao@gibbs.ee.nthu.edu.tw; lds@cs.nthu.edu.tw; keiichi@gibbs.ee.nthu.edu.tw).}
}

\maketitle
\thispagestyle{empty}
\begin{abstract}

\baselineskip12pt

One of the main problems in all-optical packet-switched networks
is the lack of optical buffers,
and one feasible technology for the constructions of optical buffers
is to use optical crossbar Switches and fiber Delay Lines (SDL).
In this two-part paper, we consider SDL constructions of optical queues
with a limited number of recirculations through
the optical switches and the fiber delay lines.
Such a problem arises from practical feasibility considerations.
We show that the constructions of certain types of optical queues,
including linear compressors, linear decompressors, and 2-to-1 FIFO multiplexers,
under a simple packet routing scheme
and under the constraint of a limited number of recirculations
can be transformed into equivalent integer representation problems
under a corresponding constraint.
Specifically, we show that the \emph{effective} maximum delay of a linear compressor/decompressor
and the effective buffer size of a 2-to-1 FIFO multiplexer in our constructions are equal to
the \emph{maximum representable integer} $B(\dbf_1^M;k)$ with respect to $\dbf_1^M$ and $k$
(defined in \reqnarray{maximum representable integer} in \rsection{introduction}),
where $\dbf_1^M=(d_1,d_2,\ldots,d_M)$ is the sequence of the delays of the $M$ fibers
used in our constructions and $k$ is the maximum number of times
that a packet can be routed through the $M$ fibers.

Given $M$ and $k$, therefore, the problem of finding an \emph{optimal} construction,
in the sense of maximizing the maximum delay (resp., buffer size),
among our constructions of linear compressors/decompressors (resp., 2-to-1 FIFO multiplexers)
is equivalent to the problem of finding an optimal sequence
${\dbf^*}_1^M$ in $\Acal_M$ (resp., $\Bcal_M$)
such that $B({\dbf^*}_1^M;k)=\max_{\dbf_1^M\in \Acal_M}B(\dbf_1^M;k)$
(resp., $B({\dbf^*}_1^M;k)=\max_{\dbf_1^M\in \Bcal_M}B(\dbf_1^M;k)$),
where $\Acal_M$ (resp., $\Bcal_M$) is the set of all sequences of fiber delays
allowed in our constructions of linear compressors/decompressors (resp., 2-to-1 FIFO multiplexers).
In Part I, we propose a class of \emph{greedy} constructions
of linear compressors/decompressors and 2-to-1 FIFO multiplexers
by specifying a class $\Gcal_{M,k}$ of sequences
such that $\Gcal_{M,k}\subseteq \Bcal_M\subseteq \Acal_M$ and
each sequence in $\Gcal_{M,k}$ is obtained recursively in a greedy manner.
For $\dbf_1^M\in \Gcal_{M,k}$, we obtain an explicit recursive expression
for $d_i$ in terms of $d_1,d_2,\ldots,d_{i-1}$ for $i=1,2,\ldots,M$,
and obtain an explicit expression for the maximum representable integer
$B(\dbf_1^M;k)$ in terms of $d_1,d_2,\ldots,d_M$.
We then use these expressions to show that
every optimal construction must be a greedy construction.
In Part~II, we further show that there are at most two optimal constructions
and give a simple algorithm to obtain the optimal construction(s).
\end{abstract}

\begin{keywords}
FIFO multiplexers, integer representation, linear compressors, linear decompressors,
maximum representable integer, optical buffers, optical queues, packet switching, routing.
\end{keywords}

\pagestyle{empty}
\pagestyle{headings}
\pagenumbering{arabic}

\newpage
\bsection{Introduction}{introduction}

Current high-speed packet-switched networks suffer from the serious overheads
incurred by the O-E-O (optical-electrical-optical) conversion
and the accompanied signal processing requirements that prevent them from
fully exploiting the tremendous bandwidth offered by the optical fiber links
so as to achieve even higher data rates.
The O-E-O bottleneck arises from the lack of optical buffers to resolve conflicts
among packets competing for the same resources in the optical domain.
As the demand for data rates is ever increasing,
the design of optical buffers has become one of the most critically sought after
optical technologies in all-optical packet-switched networks.

Currently, the only known way to ``store'' optical packets without converting them into other media
is to direct them through a set of (bufferless) optical crossbar Switches and fiber Delay Lines (SDL).
The key idea of the SDL constructions of optical buffers is to use the fiber delay lines as storage devices
and use the optical crossbar switches to distribute optical packets over the fiber delay lines
in an appropriate manner so that the optical packets can be routed to the right place at the right time.
Such an SDL approach by using optical crossbar switches and fiber delay lines has been well recognized
as one of the promising optical technologies for the constructions of optical buffers.

As an optical packet can only enter a fiber delay line from one end of that fiber
and can only be accessed when it appears at the other end of that fiber
(before the optical packet reaches the other end of that fiber,
it is constantly moving forward inside that fiber and cannot be accessed),
the optical buffers constructed by the SDL approach do not have the random access capability;
instead, they can only be used as sequential buffers with fixed storage times.
Fortunately, results in the SDL literature (see \cite{PSS86}--\cite{Miklos08} and the references therein)
show that they can still be used to construct many types of optical queues commonly encountered in practice.
Early works on the SDL constructions of optical queues \cite{PSS86}--\cite{CFKMNPCCFHLMSSW96}
mainly focused on the feasibility of such an approach through numerical simulations
rather than through rigorous analytical studies.
Recent works on the theoretical SDL constructions of optical queues
have shown that there exist systematic methods for the constructions
of various types of optical queues,
including output-buffered switches in \cite{CT96}--\cite{CCCL07a},
FIFO multiplexers in \cite{CT96} and \cite{CCCL07a}--\cite{LT06},
FIFO queues in \cite{LT06}--\cite{WJH08d},
LIFO queues in \cite{SSB07}--\cite{HCCL07},
priority queues in \cite{SA06}--\cite{RGG09},
time slot interchanges in \cite{LT06} and \cite{JLLR94},
and linear compressors, linear decompressors, non-overtaking delay lines,
and flexible delay lines in \cite{LT06} and \cite{CCL10}--\cite{CCCL09}.
Furthermore, results on the fundamental complexity of SDL constructions of optical queues can be found in \cite{KK07b}
and performance analysis for optical queues has been addressed in \cite{LLJH09}. 
For review articles on SDL constructions of optical queues,
we refer to \cite{HCA98}--\cite{Miklos08} and the references therein.

In this two-part paper, we address an important practical feasibility issue
that is of great concern in the SDL constructions of optical queues:
the constructions of optical queues with a limited number of
recirculations through the optical switches and the fiber delay lines.
As pointed out in \cite{BB06}--\cite{GWCMKDLKY05},
crosstalk due to power leakage from other optical links,
power loss experienced during recirculations through the optical switches and the fiber delay lines,
amplified spontaneous emission (ASE) from the Erbium doped fiber amplifiers (EDFA)
that are used for boosting the signal power,
and the pattern effect of the optical switches, among others,
lead to a limitation on the number of times that an optical packet can be
recirculated through the optical switches and the fiber delay lines.
If such an issue is not taken into consideration during the design of optical queues,
then for an optical packet recirculated through the optical switches and the fiber delay lines
for a number of times exceeding a predetermined threshold,
there is a good chance that it cannot be reliably recognized
at the destined output port due to severe power loss and/or serious noise accumulation
even if it appears at the right place and at the right time.

For certain types of optical queues,
including linear compressors (see Definition~1 of \cite{CCCL09}),
linear decompressors (see Definition~10 of \cite{CCCL09}),
and 2-to-1 FIFO multiplexers (see Definition~4 of \cite{CCLC06}),
the delay of a packet is known upon its arrival and the packet is routed
according to the $\Ccal$-transform \cite{CCLC06} of its delay.
Formally, the $\Ccal$-transform of a nonnegative integer $x$
with respect to a sequence of positive integers $\dbf_1^M=(d_1,d_2,\ldots,d_M)$
is defined as the sequence $\Ccal(x;\dbf_1^M)=(I_1(x;\dbf_1^M),I_2(x;\dbf_1^M),\ldots,I_M(x;\dbf_1^M))$,
where $I_M(x;\dbf_1^M),I_{M-1}(x;\dbf_1^M),\ldots,I_1(x;\dbf_1^M)$, in that order,
are given recursively by
\beqnarray{C-transform}
I_i(x;\dbf_1^M)=
\bcases
1, \textrm{ if } x-\sum_{j=i+1}^{M}I_j(x;\dbf_1^M)d_j\geq d_i,\\
0, \textrm{ otherwise},
\ecases
\eeqnarray
with the convention that the sum in \reqnarray{C-transform} is 0
if the upper index is smaller than its lower index.
In other words, if $x\geq d_M$, then $I_M(x;\dbf_1^M)=1$, and otherwise $I_M(x;\dbf_1^M)=0$;
if the remaining value $x-I_M(x;\dbf_1^M)d_M\geq d_{M-1}$, then $I_{M-1}(x;\dbf_1^M)=1$,
and otherwise $I_{M-1}(x;\dbf_1^M)=0$; and so forth.
It is clear that if $d_i=2^{i-1}$ for $i=1,2,\ldots,M$,
then the $\Ccal$-transform becomes the well-known binary representation
for the unique representation of the nonnegative integers $0,1,\ldots,2^M-1$.
As such, the $\Ccal$-transform is a generalization of the binary representation.
Furthermore, it was also shown in Corollary~6 of \cite{CCLC06} that
the $\Ccal$-transform has the \emph{unique representation} property
such that $x=\sum_{i=1}^{M}I_i(x;\dbf_1^M)d_i$ for $x=0,1,\ldots,\sum_{i=1}^{M}d_i$
if and only if $\dbf_1^M\in \Acal_M$,
where $\Acal_M$ is given by
\beqnarray{A-M}
\Acal_M=\left\{\dbf_1^M\in (\Zbf^+)^M: d_1=1 \textrm{ and }
1 \leq d_{i+1} \leq \sum_{j=1}^{i}d_j+1 \textrm{ for } i=1,2,\ldots,M-1\right\}.
\eeqnarray

\bfigure{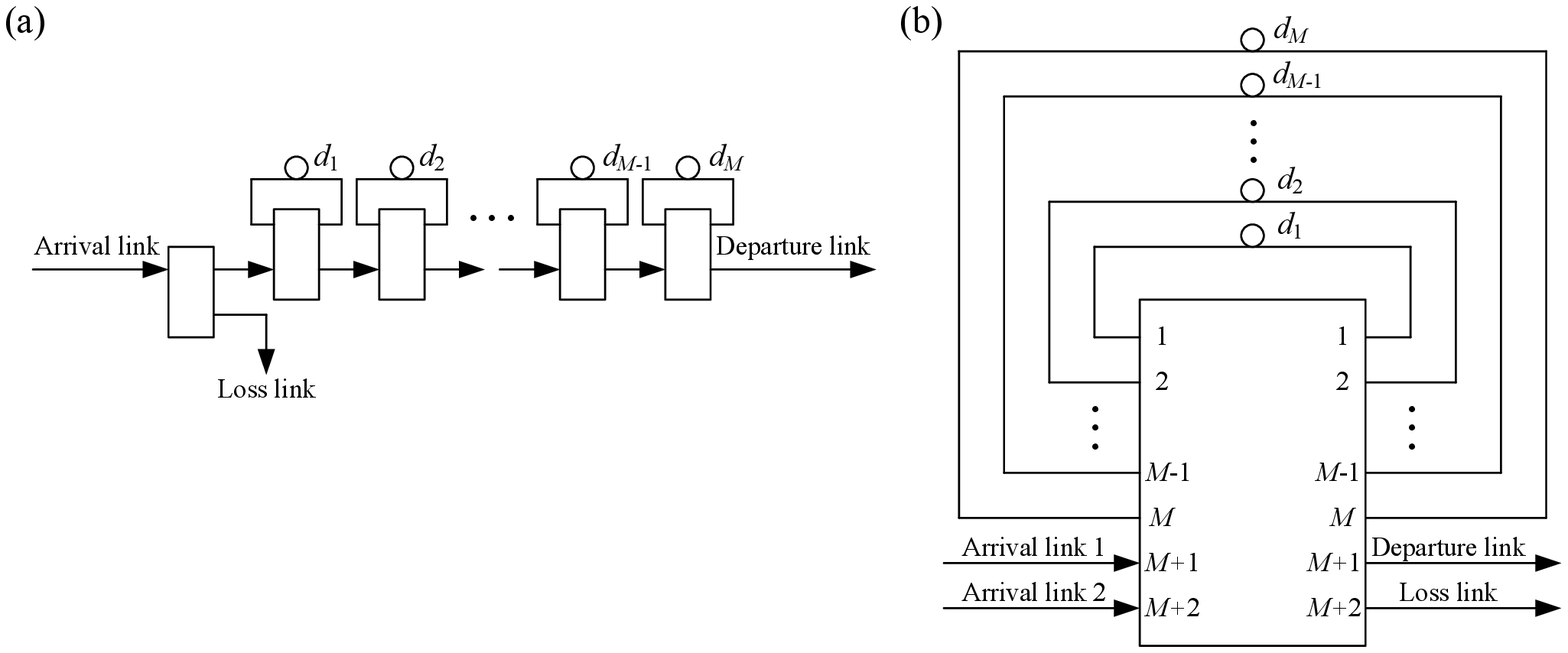}{6.0in}
\efigure{LC-2-to-1-FIFO-MUX}
{(a) A construction of a linear compressor with maximum delay $\sum_{i=1}^{M}d_i$,
where $\dbf_1^M\in \Acal_M$.
(b) A construction of a 2-to-1 FIFO multiplexer with buffer size $\sum_{i=1}^{M}d_i$,
where $\dbf_1^M\in \Bcal_M$.}

In Theorem~4 of \cite{CCCL09}, it was shown that the construction in \rfigure{LC-2-to-1-FIFO-MUX}(a)
consisting of a $1\times 2$ optical crossbar switch, $M$ $2\times 2$ optical crossbar switches,
and $M$ fiber delay lines with delays $d_1,d_2,\ldots,d_M$
can be operated as a linear compressor with maximum delay $\sum_{i=1}^{M}d_i$
under a simple packet routing scheme if and only if $\dbf_1^M\in \Acal_M$.
We note that a linear decompressor with maximum delay $\sum_{i=1}^{M}d_i$ can be similarly constructed
since it is the mirror image of a linear compressor with maximum delay $\sum_{i=1}^{M}d_i$ \cite{CCCL09}.
Furthermore, in Theorem~15 of \cite{CCLC06},
it was shown that the construction in \rfigure{LC-2-to-1-FIFO-MUX}(b)
consisting of an $(M+2)\times (M+2)$ optical crossbar switch and $M$ fiber delay lines with delays
$d_1,d_2,\ldots,d_M$ can be operated as a 2-to-1 FIFO multiplexer with buffer size $\sum_{i=1}^{M}d_i$
under a simple packet routing scheme if and only if $\dbf_1^M\in \Bcal_M$,
where $\Bcal_M$ is given by
\beqnarray{B-M}
\Bcal_M=\left\{\dbf_1^M\in (\Zbf^+)^M: d_1=1 \textrm{ and }
d_i \leq d_{i+1} \leq 2d_i \textrm{ for } i=1,2,\ldots,M-1\right\}
\eeqnarray
(note that it was shown in \cite{CCLC06} that $\Bcal_M\subseteq \Acal_M$).
Specifically, the simple packet routing scheme in \cite{CCCL09} and \cite{CCLC06}
is a \emph{self-routing} scheme and is described as follows.
Suppose that the delay of a packet arriving at time $t$ is $x$.
If $x>\sum_{i=1}^{M}d_i$, then the packet is lost and is routed to the loss link immediately.
On the other hand, if $0\leq x\leq \sum_{i=1}^{M}d_i$,
then the packet is routed to the fiber with delay $d_1$ at time $t$ if $I_1(x;\dbf_1^M)=1$,
to the fiber with delay $d_2$ at time $t+I_1(x;\dbf_1^M)d_1$ if $I_2(x;\dbf_1^M)=1,\ldots$,
to the fiber with delay $d_M$ at time $t+\sum_{i=1}^{M-1}I_i(x;\dbf_1^M)d_i$ if $I_M(x;\dbf_1^M)=1$,
and finally to the departure link at time $t+\sum_{i=1}^{M}I_i(x;\dbf_1^M)d_i=t+x$.
Therefore, the delay of the packet is indeed equal to $x$
and the packet is routed to the departure link at the right time.

The problem arises if there is a limitation on the number, say $k$,
of recirculations through the $M$ fibers in \rfigure{LC-2-to-1-FIFO-MUX}
due to the practical feasibility considerations mentioned above.
If $k\geq M$, then the limitation on the number of recirculations is redundant
as a packet can be routed to each of the $M$ fibers at most once so that it can only be
recirculated through the $M$ fibers at most $M$ times under our self-routing scheme.
It follows that the maximum delay of the linear compressor/decompressor
in \rfigure{LC-2-to-1-FIFO-MUX}(a)/mirror image of \rfigure{LC-2-to-1-FIFO-MUX}(a)
(with $\dbf_1^M\in \Acal_M$)
and the buffer size of the 2-to-1 FIFO multiplexer in \rfigure{LC-2-to-1-FIFO-MUX}(b)
(with $\dbf_1^M\in \Bcal_M$) are equal to $\sum_{i=1}^{M}d_i$.
On the other hand, if $1\leq k\leq M-1$,
then a packet routed through more than $k$ of the $M$ fibers
cannot be reliably recognized at the departure link
(e.g., a packet with delay equal to $\sum_{i=1}^{M}d_i$
will be routed to each of the $M$ fibers exactly once under our self-routing scheme
and it cannot be reliably recognized at the departure link).
It follows that the maximum delay of the linear compressor/decompressor
in \rfigure{LC-2-to-1-FIFO-MUX}(a)/mirror image of \rfigure{LC-2-to-1-FIFO-MUX}(a)
(with $\dbf_1^M\in \Acal_M$)
and the buffer size of the 2-to-1 FIFO multiplexer in \rfigure{LC-2-to-1-FIFO-MUX}(b)
(with $\dbf_1^M\in \Bcal_M$)
that can be effectively realized in this case are less than $\sum_{i=1}^{M}d_i$.
For $\dbf_1^M\in \Acal_M$ and $k\geq 1$,
we define the \emph{maximum representable integer} $B(\dbf_1^M;k)$
with respect to $\dbf_1^M$ and $k$
as the largest positive integer in $\{1,2,\ldots,\sum_{i=1}^{M}d_i\}$
such that every nonnegative integer $x$ not exceeding it satisfies the property that
the number of 1-entries in the $\Ccal$-transform
$\Ccal(x;\dbf_1^M)=(I_1(x;\dbf_1^M),I_2(x;\dbf_1^M),\ldots,I_M(x;\dbf_1^M))$
of $x$ with respect to $\dbf_1^M$ is less than or equal to $k$, i.e.,
\beqnarray{maximum representable integer}
B(\dbf_1^M;k)=\max\left\{0\leq x'\leq \sum_{i=1}^{M}d_i:
\sum_{i=1}^{M}I_i(x;\dbf_1^M)\leq k \textrm{ for } x=0,1,\ldots,x'\right\}.
\eeqnarray
For obvious reasons, we also define $B(\dbf_1^M;k)=0$ if $M=0$ or $k=0$.
Note that if $k\geq M\geq 1$, then it is easy to see from
$\sum_{i=1}^{M}I_i(x;\dbf_1^M)\leq \sum_{i=1}^{M}1=M\leq k$
for $x=0,1,\ldots,\sum_{i=1}^{M}d_i$
and the definition of $B(\dbf_1^M;k)$
in \reqnarray{maximum representable integer}
that $B(\dbf_1^M;k)=\sum_{i=1}^{M}d_i$.
If $1\leq k\leq M-1$, then it is easy to see from
$\sum_{i=1}^{M}I_i(x;\dbf_1^M)=\sum_{i=1}^{M}1=M>k$
for $x=\sum_{i=1}^{M}d_i$
and the definition of $B(\dbf_1^M;k)$
in \reqnarray{maximum representable integer}
that $B(\dbf_1^M;k)<\sum_{i=1}^{M}d_i$.
Clearly, the \emph{effective} maximum delay of the linear compressor/decompressor
in \rfigure{LC-2-to-1-FIFO-MUX}(a)/mirror image of \rfigure{LC-2-to-1-FIFO-MUX}(a)
(with $\dbf_1^M\in \Acal_M$)
and the effective buffer size of the 2-to-1 FIFO multiplexer
in \rfigure{LC-2-to-1-FIFO-MUX}(b) (with $\dbf_1^M\in \Bcal_M$)
that can be realized under our self-routing scheme and
under the limitation of at most $k$ times of recirculations through the $M$ fibers
are equal to the maximum representable integer $B(\dbf_1^M;k)$ with respect to $\dbf_1^M$ and $k$.

We call a construction of a linear compressor/decompressor
in \rfigure{LC-2-to-1-FIFO-MUX}(a)/mirror image of \rfigure{LC-2-to-1-FIFO-MUX}(a)
(resp., 2-to-1 FIFO multiplexer in \rfigure{LC-2-to-1-FIFO-MUX}(b))
with the sequence of delays ${\dbf^*}_1^M=(d_1^*,d_2^*,\ldots,d_M^*)$
in $\Acal_M$ (resp., $\Bcal_M$) an \emph{optimal} construction
if the sequence ${\dbf^*}_1^M$ gives rise to the largest effective maximum delay
(resp., the largest effective buffer size) among all of the sequences in $\Acal_M$ (resp., $\Bcal_M$)
under our self-routing scheme and under the limitation of
at most $k$ times of recirculations through the $M$ fibers,
i.e., $B({\dbf^*}_1^M;k)=\max_{\dbf_1^M\in \Acal_M}B(\dbf_1^M;k)$
(resp., $B({\dbf^*}_1^M;k)=\max_{\dbf_1^M\in \Bcal_M}B(\dbf_1^M;k)$).
Furthermore, we call such a sequence ${\dbf^*}_1^M$ an optimal $k$-constrained $M$-sequence
(or simply an optimal sequence when no confusion will arise).
Therefore, the problem of finding an optimal construction among our constructions
of linear compressors/decompressors (resp., 2-to-1 FIFO multiplexers) is equivalent to
the problem of finding an optimal sequence ${\dbf^*}_1^M\in \Acal_M$ (resp., ${\dbf^*}_1^M\in \Bcal_M$)
such that $B({\dbf^*}_1^M;k)=\max_{\dbf_1^M\in \Acal_M}B(\dbf_1^M;k)$
(resp., $B({\dbf^*}_1^M;k)=\max_{\dbf_1^M\in \Bcal_M}B(\dbf_1^M;k)$).
If $k\geq M\geq 1$, then we know that $B(\dbf_1^M;k)=\sum_{i=1}^{M}d_i$
and it is clear that there is a unique optimal construction
as there is only one optimal sequence which is given by
${\dbf^*}_1^M=\arg\max_{\dbf_1^M\in \Acal_M}\sum_{i=1}^{M}d_i=(1,2,2^2,\ldots,2^{M-1})$
(resp., ${\dbf^*}_1^M=\arg\max_{\dbf_1^M\in \Bcal_M}\sum_{i=1}^{M}d_i=(1,2,2^2,\ldots,2^{M-1})$),
and hence the largest effective maximum delay of a linear compressor/decompressor
(resp., the largest effective buffer size of a 2-to-1 FIFO multiplexer)
that can be realized in our constructions is given by $B({\dbf^*}_1^M;k)=\sum_{i=1}^{M}d_i^*=2^M-1$.
On the other hand, if $1\leq k\leq M-1$,
then the problem of finding an optimal construction/optimal sequence
turns out to be very difficult.

In \cite{Cheng08}, a dynamic programming formulation obtained through a
divide-and-conquer approach was proposed for the constructions of 2-to-1 FIFO multiplexers
with a limited number of recirculations through the $M$ fibers in \rfigure{LC-2-to-1-FIFO-MUX}(b).
However, the constructions in \cite{Cheng08} are not optimal as the fiber delays
are limited to be powers of 2.
In this two-part paper, we consider the problem of finding optimal constructions of
linear compressors/decompressors and 2-to-1 FIFO multiplexers
with at most $k$ times of recirculations through the $M$ fibers in \rfigure{LC-2-to-1-FIFO-MUX}
under our self-routing scheme.
From the discussion in the previous paragraph,
we only need to consider the nontrivial case that $M\geq 2$ and $1\leq k\leq M-1$
in the rest of the paper.

In Part~I of this paper, we will give a class of \emph{greedy} constructions
of linear compressors/decompressors and 2-to-1 FIFO multiplexers,
and will show that every optimal construction must be a greedy construction.
In Part~II of this paper, we will further show that there are at most two optimal constructions
and will give a simple algorithm to obtain the optimal construction(s).
Part~I is organized as follows.
In \rsection{OQ-LR-greedy}, we propose a class of greedy constructions
of linear compressors/decompressors and 2-to-1 FIFO multiplexers
by specifying a class $\Gcal_{M,k}$ of sequences
such that $\Gcal_{M,k}\subseteq \Bcal_M\subseteq \Acal_M$
and each sequence in $\Gcal_{M,k}$ is obtained recursively in a greedy manner.
For $\dbf_1^M\in \Gcal_{M,k}$, we obtain an explicit recursive expression
for $d_i$ in terms of $d_1,d_2,\ldots,d_{i-1}$ for $i=1,2,\ldots,M$,
and obtain an explicit expression for the maximum representable integer
$B(\dbf_1^M;k)$ in terms of $d_1,d_2,\ldots,d_M$.
In \rsection{OQ-LR-optimal is greedy},
we use the explicit expressions obtained in \rsection{OQ-LR-greedy}
to show that every optimal construction of a linear compressor/decompressor
must be a greedy construction,
i.e., if ${\dbf^*}_1^M\in \Acal_M$ and $B({\dbf^*}_1^M;k)=\max_{\dbf_1^M\in \Acal_M}B(\dbf_1^M;k)$,
then ${\dbf^*}_1^M\in \Gcal_{M,k}$,
and every optimal construction of a 2-to-1 FIFO multiplexer must also be a greedy construction,
i.e., if ${\dbf^*}_1^M\in \Bcal_M$ and $B({\dbf^*}_1^M;k)=\max_{\dbf_1^M\in \Bcal_M}B(\dbf_1^M;k)$,
then ${\dbf^*}_1^M\in \Gcal_{M,k}$.
Finally, a brief conclusion of Part~I is given in \rsection{conclusion}.

\bsection{A Class of Greedy Constructions}{OQ-LR-greedy}

From \rsection{introduction}, we know that each sequence $\dbf_1^M$ in $\Acal_M$
corresponds to a construction of a linear compressor/decompressor
in \rfigure{LC-2-to-1-FIFO-MUX}(a)/mirror image of \rfigure{LC-2-to-1-FIFO-MUX}(a)
with effective maximum delay $B(\dbf_1^M;k)$ under our self-routing scheme
and under the limitation of at most $k$ times of recirculations through the $M$ fibers,
where $B(\dbf_1^M;k)$ is the maximum representable integer with respect to $\dbf_1^M$ and $k$
as defined in \reqnarray{maximum representable integer}.
To find an optimal construction of a linear compressor/decompressor,
it suffices to find a sequence ${\dbf^*}_1^M\in \Acal_M$
such that $B({\dbf^*}_1^M;k)=\max_{\dbf_1^M\in \Acal_M}B(\dbf_1^M;k)$.
Similarly, to find an optimal construction of a 2-to-1 FIFO multiplexer,
it suffices to find a sequence ${\dbf^*}_1^M\in \Bcal_M$
such that $B({\dbf^*}_1^M;k)=\max_{\dbf_1^M\in \Bcal_M}B(\dbf_1^M;k)$.
For this, in this section we propose a class of greedy constructions
of linear compressors/decompressors and 2-to-1 FIFO multiplexers
by specifying to a class $\Gcal_{M,k}$ of sequences
such that $\Gcal_{M,k}\subseteq \Bcal_M\subseteq \Acal_M$ and
each sequence $\dbf_1^M=(d_1,d_2,\ldots,d_M)$ in $\Gcal_{M,k}$ is obtained recursively
so that $d_i$ is obtained from $d_1,d_2,\ldots,d_{i-1}$ in a greedy manner for $i=1,2,\ldots,M$.
Then in \rsection{OQ-LR-optimal is greedy},
we show that every optimal construction must be a greedy construction.

Consider the case that $M=6$ and $k=2$.
Suppose that ${\dbf'}_1^6=(1,2,4,8,16,32)\in \Acal_6$
(note that ${\dbf'}_1^6=\arg\max_{\dbf_1^6\in \Acal_6}\sum_{i=1}^{6}d_i$).
Although the nonnegative integers $0,1,\ldots,\sum_{i=1}^{6}d'_i=63$
can be uniquely represented by their $\Ccal$-transforms with respect to ${\dbf'}_1^6$
according to the unique representation property of the $\Ccal$-transform,
it is clear from \reqnarray{maximum representable integer}
that the maximum representable integer with respect to ${\dbf'}_1^6$ and $2$
is given by $B({\dbf'}_1^6;2)=6$
(as $\sum_{i=1}^{6}I_i(x;{\dbf'}_1^6)\leq 2$ for $x=0,1,\ldots,6$
and $\sum_{i=1}^{6}I_i(x;{\dbf'}_1^6)=3>2$ for $x=7$).
As another example, suppose that ${\dbf''}_1^6=(1,2,3,5,6,8)\in \Acal_6$.
Then the nonnegative integers $0,1,\ldots,\sum_{i=1}^{6}d''_i=25$
can be uniquely represented by their $\Ccal$-transforms with respect to ${\dbf''}_1^6$
according to the unique representation property of the $\Ccal$-transform
(this can also be easily verified from
the $\Ccal$-transform of $x$ with respect to ${\dbf''}_1^6$
for $x=0,1,\ldots,\sum_{i=1}^{6}d''_i=25$ in \rtable{C-transform-1}).
Although $\sum_{i=1}^{6}d''_i=25$ is smaller than $\sum_{i=1}^{6}d'_i=63$,
it is clear from \rtable{C-transform-1} that the maximum representable integer
with respect to ${\dbf''}_1^6$ and $2$ is given by $B({\dbf''}_1^6;2)=11$
(as $\sum_{i=1}^{6}I_i(x;{\dbf''}_1^6)\leq 2$ for $x=0,1,\ldots,11$
and $\sum_{i=1}^{6}I_i(x;{\dbf''}_1^6)=3>2$ for $x=12$),
which is larger than $B({\dbf'}_1^6;2)=6$.
It follows that ${\dbf''}_1^6$ is a better choice than ${\dbf'}_1^6$ for our purpose
as it gives rise to a larger maximum representable integer.

\btable{htbp}{|c||c|c|c|c|c|c|} \hline
$x$ & $I_1(x;{\dbf''}_1^6)$ & $I_2(x;{\dbf''}_1^6)$  & $I_3(x;{\dbf''}_1^6)$
& $I_4(x;{\dbf''}_1^6)$ & $I_5(x;{\dbf''}_1^6)$  & $I_6(x;{\dbf''}_1^6)$ \\ \hline
0  & 0 & 0 & 0 & 0 & 0 & 0\\ \hline
1  & 1 & 0 & 0 & 0 & 0 & 0\\ \hline
2  & 0 & 1 & 0 & 0 & 0 & 0\\ \hline
3  & 0 & 0 & 1 & 0 & 0 & 0\\ \hline
4  & 1 & 0 & 1 & 0 & 0 & 0\\ \hline
5  & 0 & 0 & 0 & 1 & 0 & 0\\ \hline
6  & 0 & 0 & 0 & 0 & 1 & 0\\ \hline
7  & 1 & 0 & 0 & 0 & 1 & 0\\ \hline
8  & 0 & 0 & 0 & 0 & 0 & 1\\ \hline
9  & 1 & 0 & 0 & 0 & 0 & 1\\ \hline
10 & 0 & 1 & 0 & 0 & 0 & 1\\ \hline
11 & 0 & 0 & 1 & 0 & 0 & 1\\ \hline
\textbf{12} & \textbf{1} & \textbf{0} & \textbf{1}
& \textbf{0} & \textbf{0} & \textbf{1}\\ \hline
13 & 0 & 0 & 0 & 1 & 0 & 1\\ \hline
14 & 0 & 0 & 0 & 0 & 1 & 1\\ \hline
15 & 1 & 0 & 0 & 0 & 1 & 1\\ \hline
16 & 0 & 1 & 0 & 0 & 1 & 1\\ \hline
17 & 0 & 0 & 1 & 0 & 1 & 1\\ \hline
18 & 1 & 0 & 1 & 0 & 1 & 1\\ \hline
19 & 0 & 0 & 0 & 1 & 1 & 1\\ \hline
20 & 1 & 0 & 0 & 1 & 1 & 1\\ \hline
21 & 0 & 1 & 0 & 1 & 1 & 1\\ \hline
22 & 0 & 0 & 1 & 1 & 1 & 1\\ \hline
23 & 1 & 0 & 1 & 1 & 1 & 1\\ \hline
24 & 0 & 1 & 1 & 1 & 1 & 1\\ \hline
25 & 1 & 1 & 1 & 1 & 1 & 1\\ \hline
\etable{C-transform-1}{The $\Ccal$-transform of $x$ with respect to
${\dbf''}_1^6=(1,2,3,5,6,8)$ for $x=0,1,\ldots,\sum_{i=1}^{6}d''_i=25$.}

A natural question we would like to ask is then: can we do better and how to do that?
In other words, are there any methods for choosing a sequence $\dbf_1^6$ in $\Acal_6$
such that $B(\dbf_1^6;2)>B({\dbf''}_1^6;2)$.
The answer is affirmative as we now show it.
A direct approach to choose a sequence $\dbf_1^6$ in $\Acal_6$
is to divide the choice into two parts,
say the choice of $d_1,d_2,d_3$ and the choice of $d_4,d_5,d_6$,
so that there is at most one $1$-entry in $(I_1(x;\dbf_1^6),I_2(x;\dbf_1^6),I_3(x;\dbf_1^6))$
and there is at most one $1$-entry in $(I_4(x;\dbf_1^6),I_5(x;\dbf_1^6),I_6(x;\dbf_1^6))$
(and hence there are at most two $1$-entries in $\Ccal(x;\dbf_1^6)$)
for as many consecutive nonnegative integers (starting from zero) as possible.
For instance, we can first choose
\beqnarray{}
d_1=1,\ d_2=2, \textrm{ and } d_3=3. \nn
\eeqnarray
Clearly, we have $B(\dbf_1^3;1)=3$.
Then we can choose
\beqnarray{}
\alignspace d_4=B(\dbf_1^3;1)+1=4, \nn\\
\alignspace d_5=(d_4+B(\dbf_1^3;1))+1=2(B(\dbf_1^3;1)+1)=8, \nn\\
\alignspace d_6=(d_5+B(\dbf_1^3;1))+1=3(B(\dbf_1^3;1)+1)=12. \nn
\eeqnarray
For $4\leq i\leq 6$, it is easy to see from \rtable{C-transform-2}
that for $x=d_i,d_i+1,\ldots,d_i+B(\dbf_1^3;1)$,
there is at most one $1$-entry in $(I_1(x;\dbf_1^6),I_2(x;\dbf_1^6),I_3(x;\dbf_1^6))$
and there is exactly one $1$-entry in $(I_4(x;\dbf_1^6),I_5(x;\dbf_1^6),I_6(x;\dbf_1^6))$
(as $I_i(x;\dbf_1^6)=1$ and $I_j(x;\dbf_1^6)=0$ for $j\in\{4,5,6\}\verb+\+ \{i\}$)
so that there are at most two $1$-entries in $\Ccal(x;\dbf_1^6)$.
It follows that such a direct approach guarantees that $B(\dbf_1^6;2)\geq d_6+B(\dbf_1^3;1)=15$.
Indeed, from \rtable{C-transform-2} we see that $B(\dbf_1^6;2)=16$
(as $\sum_{i=1}^{6}I_i(x;\dbf_1^6)\leq 2$ for $x=0,1,\ldots,16$
and $\sum_{i=1}^{6}I_i(x;\dbf_1^6)=3>2$ for $x=17$),
which is greater than $B({\dbf''}_1^6;2)=11$ in the above paragraph.

\btable{htbp}{|c||c|c|c||c|c|c|} \hline
$x$ & $I_1(x;\dbf_1^6)$ & $I_2(x;\dbf_1^6)$  & $I_3(x;\dbf_1^6)$
& $I_4(x;\dbf_1^6)$ & $I_5(x;\dbf_1^6)$  & $I_6(x;\dbf_1^6)$ \\ \hline
0  & 0 & 0 & 0 & 0 & 0 & 0\\ \hline
1  & 1 & 0 & 0 & 0 & 0 & 0\\ \hline
2  & 0 & 1 & 0 & 0 & 0 & 0\\ \hline
3  & 0 & 0 & 1 & 0 & 0 & 0\\ \hline
4  & 0 & 0 & 0 & 1 & 0 & 0\\ \hline
5  & 1 & 0 & 0 & 1 & 0 & 0\\ \hline
6  & 0 & 1 & 0 & 1 & 0 & 0\\ \hline
7  & 0 & 0 & 1 & 1 & 0 & 0\\ \hline
8  & 0 & 0 & 0 & 0 & 1 & 0\\ \hline
9  & 1 & 0 & 0 & 0 & 1 & 0\\ \hline
10 & 0 & 1 & 0 & 0 & 1 & 0\\ \hline
11 & 0 & 0 & 1 & 0 & 1 & 0\\ \hline
12 & 0 & 0 & 0 & 0 & 0 & 1\\ \hline
13 & 1 & 0 & 0 & 0 & 0 & 1\\ \hline
14 & 0 & 1 & 0 & 0 & 0 & 1\\ \hline
15 & 0 & 0 & 1 & 0 & 0 & 1\\ \hline
16 & 0 & 0 & 0 & 1 & 0 & 1\\ \hline
\textbf{17} & \textbf{1} & \textbf{0} & \textbf{0}
& \textbf{1} & \textbf{0} & \textbf{1}\\ \hline
\etable{C-transform-2}{The $\Ccal$-transform of $x$ with respect to
$\dbf_1^6=(1,2,3,4,8,12)$ for $x=0,1,\ldots,17$.}

An even better approach, called a \emph{greedy} approach in this paper,
is described as follows.
We still divide the choice of a sequence $\dbf_1^6$ in $\Acal_6$ into two parts,
say the choice of $d_1,d_2,d_3$ and the choice of $d_4,d_5,d_6$,
as in the direct approach above.
First we choose $d_1,d_2,d_3$ recursively so that $d_i$ is obtained from
$d_1,d_2,\ldots,d_{i-1}$ in such a way that $B(\dbf_1^i;1)$
is at least one more than $B(\dbf_1^{i-1};1)$,
and this is possible by simply choosing $d_i=B(\dbf_1^{i-1};1)+1$ for $i=1,2,3$.
It follows that we can choose
\beqnarray{}
\alignspace d_1=B(\dbf_1^0;1)+1=0+1=1, \nn\\
\alignspace d_2=B(\dbf_1^1;1)+1=1+1=2, \nn\\
\alignspace d_3=B(\dbf_1^2;1)+1=2+1=3. \nn
\eeqnarray
Then we choose $d_4,d_5,d_6$ recursively so that $d_i$ is obtained from
$d_1,d_2,\ldots,d_{i-1}$ in such a way that $B(\dbf_1^i;2)$
is at least one more than $B(\dbf_1^{i-1};2)$,
and this is possible by simply choosing $d_i=B(\dbf_1^{i-1};2)+1$ for $i=4,5,6$.
It follows that we can choose
\beqnarray{}
\alignspace d_4=B(\dbf_1^3;2)+1=5+1=6, \nn\\
\alignspace d_5=B(\dbf_1^4;2)+1=9+1=10, \nn\\
\alignspace d_6=B(\dbf_1^5;2)+1=13+1=14. \nn
\eeqnarray
From \rtable{C-transform-3}, we see that $B(\dbf_1^6;2)=17$
(as $\sum_{i=1}^{6}I_i(x;\dbf_1^6)\leq 2$ for $x=0,1,\ldots,17$
and $\sum_{i=1}^{6}I_i(x;\dbf_1^6)=3>2$ for $x=18$),
which is larger than that in the direct approach above.

\btable{htbp}{|c||c|c|c||c|c|c|} \hline
$x$ & $I_1(x;\dbf_1^6)$ & $I_2(x;\dbf_1^6)$  & $I_3(x;\dbf_1^6)$
& $I_4(x;\dbf_1^6)$ & $I_5(x;\dbf_1^6)$  & $I_6(x;\dbf_1^6)$ \\ \hline
0  & 0 & 0 & 0 & 0 & 0 & 0\\ \hline
1  & 1 & 0 & 0 & 0 & 0 & 0\\ \hline
2  & 0 & 1 & 0 & 0 & 0 & 0\\ \hline
3  & 0 & 0 & 1 & 0 & 0 & 0\\ \hline
4  & 1 & 0 & 1 & 0 & 0 & 0\\ \hline
5  & 0 & 1 & 1 & 0 & 0 & 0\\ \hline
6  & 0 & 0 & 0 & 1 & 0 & 0\\ \hline
7  & 1 & 0 & 0 & 1 & 0 & 0\\ \hline
8  & 0 & 1 & 0 & 1 & 0 & 0\\ \hline
9  & 0 & 0 & 1 & 1 & 0 & 0\\ \hline
10 & 0 & 0 & 0 & 0 & 1 & 0\\ \hline
11 & 1 & 0 & 0 & 0 & 1 & 0\\ \hline
12 & 0 & 1 & 0 & 0 & 1 & 0\\ \hline
13 & 0 & 0 & 1 & 0 & 1 & 0\\ \hline
14 & 0 & 0 & 0 & 0 & 0 & 1\\ \hline
15 & 1 & 0 & 0 & 0 & 0 & 1\\ \hline
16 & 0 & 1 & 0 & 0 & 0 & 1\\ \hline
17 & 0 & 0 & 1 & 0 & 0 & 1\\ \hline
\textbf{18} & \textbf{1} & \textbf{0} & \textbf{1}
& \textbf{0} & \textbf{0} & \textbf{1}\\ \hline
\etable{C-transform-3}{The $\Ccal$-transform of $x$ with respect to
$\dbf_1^6=(1,2,3,6,10,14)$ for $x=0,1,\ldots,18$.}

We are now in a position to formally describe our greedy approach in a general setting.
Suppose that $M\geq 2$ and $1\leq k\leq M-1$.
Let $\nbf_1^k=(n_1,n_2,\ldots,n_k)$ be a sequence of positive integers such that $\sum_{i=1}^{k}n_i=M$,
and let $s_0=0$ and $s_i=\sum_{\ell=1}^{i}n_{\ell}$ for $i=1,2,\ldots,k$
(note that $s_k=\sum_{\ell=1}^{k}n_{\ell}=M$).
In our greedy approach, we divide the choice of a sequence $\dbf_1^M$ in $\Acal_M$ into $k$ parts,
first the choice of $d_1,d_2,\ldots,d_{n_1}=d_{s_1}$,
then the choice of $d_{s_1+1},d_{s_1+2},\ldots,d_{s_1+n_2}=d_{s_2},\ldots$,
and finally the choice of $d_{s_{k-1}+1},d_{s_{k-1}+2},\ldots,d_{s_{k-1}+n_k}=d_{s_k}=d_M$.
In the $(i+1)^{\textrm{th}}$ part, where $0\leq i\leq k-1$,
we choose $d_{s_i+1},d_{s_i+2},\ldots,d_{s_{i+n_{i+1}}}=d_{s_{i+1}}$ recursively
so that $d_{s_i+j}$ is obtained from $d_1,d_2,\ldots,d_{s_i+j-1}$
in such a way that $B(\dbf_1^{s_i+j};i+1)$ is at least one more than
$B(\dbf_1^{s_i+j-1};i+1)$ for $j=1,2,\ldots,n_{i+1}$.
We do so by choosing
\beqnarray{OQ-LR-delays-greedy-1}
d_{s_i+j}=B(\dbf_1^{s_i+j-1};i+1)+1,
\emph{ for } i=0,1,\ldots,k-1 \textrm{ and } j=1,2,\ldots,n_{i+1}.
\eeqnarray
For example, in \rtable{OQ-LR-delays-greedy-1} we show the sequence $\dbf_1^M$
given by \reqnarray{OQ-LR-delays-greedy-1} for the case that
$M=18$, $k=6$, and $\nbf_1^k=(3,4,2,5,1,3)$.

{\tiny
\btable{htbp}{||c||ccc|cccc|cc|ccccc|c|ccc||}
\hline
$i$   &  1 &  2 &  3 &  4 &  5 &  6 &  7 &  8 &  9
      & 10 & 11 & 12 & 13 & 14 & 15 & 16 & 17 & 18 \\
\hline
$d_i$ &  1  &   2 &   3 &   6 &  10 &  14 &   18 &   36 &   58
      & 116 & 196 & 276 & 356 & 436 & 872 & 1744 & 3132 & 4520 \\
\hline
\etable{OQ-LR-delays-greedy-1}{The sequence $\dbf_1^M$ given by \reqnarray{OQ-LR-delays-greedy-1}
for the case that $M=18$, $k=6$, and $\nbf_1^k=(3,4,2,5,1,3)$.}}

The reason why we choose $d_{s_i+j}=B(\dbf_1^{s_i+j-1};i+1)+1$ as in \reqnarray{OQ-LR-delays-greedy-1}
for $i=0,1,\ldots,k-1$ and $j=1,2,\ldots,n_{i+1}$ can be explained as follows.
Initially, we choose $d_1=d_{s_0+1}=B(\dbf_1^{s_0};1)+1=B(\dbf_1^0;1)+1=0+1=1$
so that $d_1\in \Acal_1$.
After $\dbf_1^{s_i+j-1}\in \Acal_{s_i+j-1}$ has been chosen
for some $0\leq i\leq k-1$ and $1\leq j\leq n_{i+1}$, where $1\leq s_i+j-1\leq M-1$,
the maximum representable integer with respect to $\dbf_1^{s_i+j-1}$ and $i+1$ is $B(\dbf_1^{s_i+j-1};i+1)$.
The key idea in our greedy approach is to choose $d_{s_i+j}$ such that
$\dbf_1^{s_i+j}\in \Acal_{s_i+j}$
and the maximum representable integer $B(\dbf_1^{s_i+j};i+1)$
with respect to $\dbf_1^{s_i+j}$ and $i+1$
is greater than $B(\dbf_1^{s_i+j-1};i+1)$ and is as large as possible
(that is why we call such an approach a ``greedy'' approach in this paper).
As $\dbf_1^{s_i+j-1}\in \Acal_{s_i+j-1}$,
we need to choose $d_{s_i+j}$ such that
$1\leq d_{s_i+j}\leq \sum_{\ell=1}^{s_i+j-1}d_{\ell}+1$
in order to have $\dbf_1^{s_i+j}\in \Acal_{s_i+j}$.
If we choose $d_{s_i+j}$ such that
$B(\dbf_1^{s_i+j-1};i+1)+1<d_{s_i+j}\leq \sum_{\ell=1}^{s_i+j-1}d_{\ell}+1$
(note that this is possible only in the case that
$B(\dbf_1^{s_i+j-1};i+1)<\sum_{\ell=1}^{s_i+j-1}d_{\ell}$),
then for $0\leq x\leq B(\dbf_1^{s_i+j-1};i+1)+1$,
we have $x<d_{s_i+j}$ and it is clear from \reqnarray{C-transform} that
\beqnarray{}
\alignspace
I_{s_i+j}(x;\dbf_1^{s_i+j})=0,
\label{eqn:proof-OQ-LR-delays-greedy-1-111} \\
\alignspace
I_{\ell}(x;\dbf_1^{s_i+j})=I_{\ell}(x;\dbf_1^{s_i+j-1}),
\textrm{ for } \ell=s_i+j-1,s_i+j-2,\ldots,1.
\label{eqn:proof-OQ-LR-delays-greedy-1-222}
\eeqnarray
From \reqnarray{proof-OQ-LR-delays-greedy-1-111},
\reqnarray{proof-OQ-LR-delays-greedy-1-222},
$B(\dbf_1^{s_i+j-1};i+1)<\sum_{\ell=1}^{s_i+j-1}d_{\ell}$,
and the definition of $B(\dbf_1^{s_i+j-1};i+1)$
in \reqnarray{maximum representable integer}, we have
\beqnarray{}
\alignspace
\sum_{\ell=1}^{s_i+j}I_{\ell}(x;\dbf_1^{s_i+j})
=\sum_{\ell=1}^{s_i+j-1}I_{\ell}(x;\dbf_1^{s_i+j-1})\leq i+1,
\textrm{ for } 0\leq x\leq B(\dbf_1^{s_i+j-1};i+1),
\label{eqn:proof-OQ-LR-delays-greedy-1-333} \\
\alignspace
\sum_{\ell=1}^{s_i+j}I_{\ell}(x;\dbf_1^{s_i+j})
=\sum_{\ell=1}^{s_i+j-1}I_{\ell}(x;\dbf_1^{s_i+j-1})>i+1,
\textrm{ for } x=B(\dbf_1^{s_i+j-1};i+1)+1.
\label{eqn:proof-OQ-LR-delays-greedy-1-444}
\eeqnarray
It follows from \reqnarray{proof-OQ-LR-delays-greedy-1-333},
\reqnarray{proof-OQ-LR-delays-greedy-1-444},
$B(\dbf_1^{s_i+j-1};i+1)<\sum_{\ell=1}^{s_i+j-1}d_{\ell}<\sum_{\ell=1}^{s_i+j}d_{\ell}$,
and the definition of $B(\dbf_1^{s_i+j};i+1)$
in \reqnarray{maximum representable integer} that
\beqnarray{proof-OQ-LR-delays-greedy-1-555}
B(\dbf_1^{s_i+j};i+1)=B(\dbf_1^{s_i+j-1};i+1).
\eeqnarray
This shows that there is no gain in the maximum representable integer
if we choose $d_{s_i+j}$ such that
$B(\dbf_1^{s_i+j-1};i+1)+1<d_{s_i+j}\leq \sum_{\ell=1}^{s_i+j-1}d_{\ell}+1$.

On the other hand, if we choose $d_{s_i+j}$ such that
$1\leq d_{s_i+j}\leq B(\dbf_1^{s_i+j-1};i+1)+1$,
then for $0\leq x\leq d_{s_i+j}-1$,
it is also clear from \reqnarray{C-transform}
that \reqnarray{proof-OQ-LR-delays-greedy-1-111}
and \reqnarray{proof-OQ-LR-delays-greedy-1-222} still hold.
From \reqnarray{proof-OQ-LR-delays-greedy-1-111},
\reqnarray{proof-OQ-LR-delays-greedy-1-222},
$d_{s_i+j}-1\leq B(\dbf_1^{s_i+j-1};i+1)$,
and the definition of $B(\dbf_1^{s_i+j-1};i+1)$
in \reqnarray{maximum representable integer},
we have
\beqnarray{proof-OQ-LR-delays-greedy-1-666}
\alignspace
\sum_{\ell=1}^{s_i+j}I_{\ell}(x;\dbf_1^{s_i+j})
=\sum_{\ell=1}^{s_i+j-1}I_{\ell}(x;\dbf_1^{s_i+j-1})\leq i+1,
\textrm{ for } 0\leq x\leq d_{s_i+j}-1.
\eeqnarray
For $d_{s_i+j}\leq x\leq d_{s_i+j}+B(\dbf_1^{s_i+j-1};i)+1$,
we can see from \reqnarray{C-transform} that
\beqnarray{}
\alignspace
I_{s_i+j}(x;\dbf_1^{s_i+j})=1,
\label{eqn:proof-OQ-LR-delays-greedy-1-777} \\
\alignspace
I_{\ell}(x;\dbf_1^{s_i+j})=I_{\ell}(x-d_{s_i+j};\dbf_1^{s_i+j-1}),
\textrm{ for } \ell=s_i+j-1,s_i+j-2,\ldots,1.
\label{eqn:proof-OQ-LR-delays-greedy-1-888}
\eeqnarray
In the case that $B(\dbf_1^{s_i+j-1};i)<\sum_{\ell=1}^{s_i+j-1}d_{\ell}$,
we have from \reqnarray{proof-OQ-LR-delays-greedy-1-777},
\reqnarray{proof-OQ-LR-delays-greedy-1-888},
and the definition of $B(\dbf_1^{s_i+j-1};i)$
in \reqnarray{maximum representable integer} that
\beqnarray{}
\alignspace
\sum_{\ell=1}^{s_i+j}I_{\ell}(x;\dbf_1^{s_i+j})
=\sum_{\ell=1}^{s_i+j-1}I_{\ell}(x-d_{s_i+j};\dbf_1^{s_i+j-1})+1\leq i+1, \nn\\
\alignspace \hspace*{2.0in}
\textrm{ for } d_{s_i+j}\leq x\leq d_{s_i+j}+B(\dbf_1^{s_i+j-1};i),
\label{eqn:proof-OQ-LR-delays-greedy-1-999} \\
\alignspace
\sum_{\ell=1}^{s_i+j}I_{\ell}(x;\dbf_1^{s_i+j})
=\sum_{\ell=1}^{s_i+j-1}I_{\ell}(x-d_{s_i+j};\dbf_1^{s_i+j-1})+1>i+1, \nn\\
\alignspace \hspace*{2.0in}
\textrm{ for } x=d_{s_i+j}+B(\dbf_1^{s_i+j-1};i)+1.
\label{eqn:proof-OQ-LR-delays-greedy-1-aaa}
\eeqnarray
It follows from \reqnarray{proof-OQ-LR-delays-greedy-1-666},
\reqnarray{proof-OQ-LR-delays-greedy-1-999},
\reqnarray{proof-OQ-LR-delays-greedy-1-aaa},
$d_{s_i+j}+B(\dbf_1^{s_i+j-1};i)<d_{s_i+j}+\sum_{\ell=1}^{s_i+j-1}d_{\ell}
=\sum_{\ell=1}^{s_i+j}d_{\ell}$,
and the definition of $B(\dbf_1^{s_i+j};i+1)$
in \reqnarray{maximum representable integer} that
\beqnarray{proof-OQ-LR-delays-greedy-1-bbb}
B(\dbf_1^{s_i+j};i+1)=d_{s_i+j}+B(\dbf_1^{s_i+j-1};i).
\eeqnarray
Similarly, in the case that $B(\dbf_1^{s_i+j-1};i)=\sum_{\ell=1}^{s_i+j-1}d_{\ell}$,
we have from \reqnarray{proof-OQ-LR-delays-greedy-1-777},
\reqnarray{proof-OQ-LR-delays-greedy-1-888},
and the definition of $B(\dbf_1^{s_i+j-1};i)$
in \reqnarray{maximum representable integer} that
\reqnarray{proof-OQ-LR-delays-greedy-1-999} holds,
and it follows from \reqnarray{proof-OQ-LR-delays-greedy-1-666},
\reqnarray{proof-OQ-LR-delays-greedy-1-999},
$d_{s_i+j}+B(\dbf_1^{s_i+j-1};i)=d_{s_i+j}+\sum_{\ell=1}^{s_i+j-1}d_{\ell}
=\sum_{\ell=1}^{s_i+j}d_{\ell}$,
and the definition of $B(\dbf_1^{s_i+j};i+1)$
in \reqnarray{maximum representable integer} that
\beqnarray{proof-OQ-LR-delays-greedy-1-ccc}
B(\dbf_1^{s_i+j};i+1)=\sum_{\ell=1}^{s_i+j}d_{\ell}=d_{s_i+j}+B(\dbf_1^{s_i+j-1};i).
\eeqnarray
As our purpose is to choose $d_{s_i+j}$ such that $B(\dbf_1^{s_i+j};i+1)$
is greater than $B(\dbf_1^{s_i+j-1};i+1)$ and is as large as possible,
it is immediate from \reqnarray{proof-OQ-LR-delays-greedy-1-bbb}
and \reqnarray{proof-OQ-LR-delays-greedy-1-ccc}
(note that \reqnarray{proof-OQ-LR-delays-greedy-1-bbb}
and \reqnarray{proof-OQ-LR-delays-greedy-1-ccc}
hold for $1\leq d_{s_i+j}\leq B(\dbf_1^{s_i+j-1};i+1)+1$)
that the best choice is $d_{s_i+j}=B(\dbf_1^{s_i+j-1};i+1)+1$
as given in \reqnarray{OQ-LR-delays-greedy-1}.

Note that in the rest of the paper we assume that $M\geq 2$ and $1\leq k\leq M-1$,
and in our greedy approach as described above
$\nbf_1^k$ is a sequence of positive integers such that $\sum_{i=1}^{k}n_i=M$.
As such, we must have $n_i\geq 2$ for some $1\leq i\leq k$
(otherwise we have $n_i=1$ for $i=1,2,\ldots,k$ and hence $\sum_{i=1}^{k}n_i=k\leq M-1$,
contradicting to $\sum_{i=1}^{k}n_i=M$).
In the following theorem, we show that it suffices to consider only the cases with $n_1\geq 2$.

\btheorem{OQ-LR-delays-greedy-redundant}
Suppose that $M\geq 2$ and $1\leq k\leq M-1$,
and suppose that $\nbf_1^k=(n_1,n_2,\ldots,n_k)$ is a sequence of positive integers
such that $n_1=1$ and $\sum_{i=1}^{k}n_i=M$.
Let $a=\min\{2\leq i\leq k:\ n_i\geq 2\}$
(note that $a$ is well defined as $n_1=1$
and hence we must have $n_i\geq 2$ for some $2\leq i\leq k$),
and let $n'_1=n_1+1=2$, $n'_i=n_i=1$ for $i=2,3,\ldots,a-1$,
$n'_a=n_a-1\geq 1$, and $n'_i=n_i$ for $i=a+1,a+2,\ldots,k$
(note that ${\nbf'}_1^k=(n'_1,n'_2,\ldots,n'_k)$
is a sequence of positive integers such that $\sum_{i=1}^{k}n'_i=\sum_{i=1}^{k}n_i=M$).
Furthermore, let $s_0=0$ and $s_i=\sum_{\ell=1}^{i}n_{\ell}$ for $i=1,2,\ldots,k$,
let $s'_0=0$ and $s'_i=\sum_{\ell=1}^{i}n'_{\ell}$ for $i=1,2,\ldots,k$,
and let
\beqnarray{}
\alignspace
d_{s_i+j}=B(\dbf_1^{s_i+j-1};i+1)+1,
\textrm{ for } i=0,1,\ldots,k-1 \textrm{ and } j=1,2,\ldots,n_{i+1},
\label{eqn:OQ-LR-delays-greedy-redundant-1} \\
\alignspace
d'_{s'_i+j}=B({\dbf'}_1^{s'_i+j-1};i+1)+1,
\textrm{ for } i=0,1,\ldots,k-1 \textrm{ and } j=1,2,\ldots,n'_{i+1}.
\label{eqn:OQ-LR-delays-greedy-redundant-2}
\eeqnarray
Then $d_{\ell}=d'_{\ell}$ for $\ell=1,2,\ldots,M$.
\etheorem

\bproof
See \rappendix{OQ-LR-delays-greedy-redundant}.
\eproof

We give an illustration of \rtheorem{OQ-LR-delays-greedy-redundant}
in the following two tables.
In \rtable{OQ-LR-delays-greedy-redundant-1}, we show the sequence $\dbf_1^M$
given by \reqnarray{OQ-LR-delays-greedy-redundant-1} for the case that
$M=18$, $k=7$, and $\nbf_1^k=(1,1,1,4,2,6,3)$.
In \rtable{OQ-LR-delays-greedy-redundant-2}, we show the sequence ${\dbf'}_1^M$
given by \reqnarray{OQ-LR-delays-greedy-redundant-2} for the case that
$M=18$, $k=7$, and ${\nbf'}_1^k=(2,1,1,3,2,6,3)$.
As $n_1=1$, $\min\{2\leq i\leq 7:\ n_i\geq 2\}=4$,
$n'_1=2=n_1+1$, $n'_i=1=n_i$ for $i=2,3$, $n'_4=3=n_4-1$, and $n'_i=n_i$ for $i=5,6,7$,
it follows from \rtheorem{OQ-LR-delays-greedy-redundant}
that $d_{\ell}=d'_{\ell}$ for $\ell=1,2,\ldots,18$
(this can also be easily verified from
\rtable{OQ-LR-delays-greedy-redundant-1} and \rtable{OQ-LR-delays-greedy-redundant-2}).

{\tiny
\btable{htbp}{||c||c|c|c|cccc|cc|cccccc|ccc||}
\hline
$i$   &  1 &  2 &  3 &  4 &  5 &  6 &  7 &  8 &  9
      & 10 & 11 & 12 & 13 & 14 & 15 & 16 & 17 & 18 \\
\hline
$d_i$ &  1  &   2 &   4 &   8 &   16 &   31 &   46 &   92 &  153
      & 306 & 520 & 734 & 948 & 1162 & 1376 & 2752 & 4342 & 5932 \\
\hline
\etable{OQ-LR-delays-greedy-redundant-1}
{The sequence $\dbf_1^M$ given by \reqnarray{OQ-LR-delays-greedy-redundant-1}
for the case that $M=18$, $k=7$, and $\nbf_1^k=(1,1,1,4,2,6,3)$.}}

{\tiny
\btable{htbp}{||c||cc|c|c|ccc|cc|cccccc|ccc||}
\hline
$i$   &  1 &  2 &  3 &  4 &  5 &  6 &  7 &  8 &  9
      & 10 & 11 & 12 & 13 & 14 & 15 & 16 & 17 & 18 \\
\hline
$d'_i$ &  1  &   2 &   4 &   8 &   16 &   31 &   46 &   92 &  153
       & 306 & 520 & 734 & 948 & 1162 & 1376 & 2752 & 4342 & 5932 \\
\hline
\etable{OQ-LR-delays-greedy-redundant-2}
{The sequence ${\dbf'}_1^M$ given by \reqnarray{OQ-LR-delays-greedy-redundant-2}
for the case that $M=18$, $k=7$, and ${\nbf'}_1^k=(2,1,1,3,2,6,3)$.}}

As a result of \rtheorem{OQ-LR-delays-greedy-redundant},
we let $\Ncal_{M,k}$ be the set of sequences of positive integers
$\nbf_1^k$ such that $n_1\geq 2$ and $\sum_{i=1}^{k}n_i=M$,
i.e.,
\beqnarray{N-M-k}
\Ncal_{M,k}=
\left\{\nbf_1^k\in (\Zbf^+)^k: n_1\geq 2 \textrm{ and } \sum_{i=1}^{k}n_i=M\right\}.
\eeqnarray
Furthermore, we let $\Gcal_{M,k}$ be the set of sequences of positive integers
$\dbf_1^M$ given by \reqnarray{OQ-LR-delays-greedy-1} for some $\nbf_1^k\in \Ncal_{M,k}$,
i.e.,
\beqnarray{G-M-k}
\Gcal_{M,k}=
\left\{\dbf_1^M\in (\Zbf^+)^M: \dbf_1^M \textrm{ is given by } \reqnarray{OQ-LR-delays-greedy-1}
\textrm{ for some } \nbf_1^k\in \Ncal_{M,k}\right\}.
\eeqnarray
In this paper, we call a construction of
a linear compressor/decompressor
in \rfigure{LC-2-to-1-FIFO-MUX}(a)/mirror image of \rfigure{LC-2-to-1-FIFO-MUX}(a)
or a 2-to-1 FIFO multiplexer in \rfigure{LC-2-to-1-FIFO-MUX}(b)
with the sequence of fiber delays $\dbf_1^M\in \Gcal_{M,k}$ a greedy construction.

In the following theorem, we show that if $\dbf_1^M\in \Gcal_{M,k}$,
then $\dbf_1^M\in \Bcal_M$ in \reqnarray{OQ-LR-delays-greedy-4},
i.e., $\Gcal_{M,k}\subseteq \Bcal_M$.
As we have from \cite{CCLC06} that $\Bcal_M\subseteq \Acal_M$,
it follows that
\beqnarray{N-M-k-B-M-A-M}
\Gcal_{M,k}\subseteq \Bcal_M\subseteq \Acal_M.
\eeqnarray
For $\dbf_1^M\in \Gcal_{M,k}$, we also obtain an explicit recursive expression
for $d_i$ in terms of $d_1,d_2,\ldots,d_{i-1}$ for $i=1,2,\ldots,M$
in \reqnarray{OQ-LR-delays-greedy-2}--\reqnarray{OQ-LR-delays-greedy-3},
and obtain an explicit expression for $B(\dbf_1^M;k)$ in terms of $d_1,d_2,\ldots,d_M$
in \reqnarray{OQ-LR-delays-greedy-7}.

\btheorem{OQ-LR-delays-greedy}
Suppose that $M\geq 2$ and $1\leq k\leq M-1$.
Let $\dbf_1^M\in \Gcal_{M,k}$ so that there exists a sequence $\nbf_1^k\in \Ncal_{M,k}$
such that $d_{s_i+j}$ is given by \reqnarray{OQ-LR-delays-greedy-1},
i.e., $d_{s_i+j}=B(\dbf_1^{s_i+j-1};i+1)+1$
for $i=0,1,\ldots,k-1$ and $j=1,2,\ldots,n_{i+1}$,
where $s_0=0$ and $s_i=\sum_{\ell=1}^{i}n_{\ell}$ for $i=1,2,\ldots,k$.
Then $d_1,d_2,\ldots,d_M$ can be recursively expressed as
\beqnarray{}
\alignspace
d_j=j, \textrm{ for } j=1,2,\ldots,s_1,
\label{eqn:OQ-LR-delays-greedy-2}\\
\alignspace
d_{s_i+j}=2d_{s_i}+(j-1)(d_{s_1}+d_{s_2}+\cdots+d_{s_i}+1), \nn\\
\alignspace \hspace*{0.2in}
\textrm{ for } i=1,2,\ldots,k-1 \textrm{ and } j=1,2,\ldots,n_{i+1},
\label{eqn:OQ-LR-delays-greedy-3}
\eeqnarray
and we have
\beqnarray{OQ-LR-delays-greedy-4}
\dbf_1^{s_i+j}\in \Bcal_{s_i+j},
\textrm{ for } i=0,1,\ldots,k-1 \textrm{ and } j=1,2,\ldots,n_{i+1}.
\eeqnarray
Furthermore, we have
\beqnarray{}
\alignspace
B(\dbf_1^j;1)=j, \textrm{ for } j=1,2,\ldots,s_1,
\label{eqn:OQ-LR-delays-greedy-5}\\
\alignspace
B(\dbf_1^{s_i+j};i+1)=d_{s_i+j}+d_{s_1}+d_{s_2}+\cdots+d_{s_i}, \nn\\
\alignspace \hspace*{0.1in}
\textrm{ for } i=1,2,\ldots,k-1 \textrm{ and } j=1,2,\ldots,n_{i+1}.
\label{eqn:OQ-LR-delays-greedy-6}
\eeqnarray
In particular, we have
\beqnarray{OQ-LR-delays-greedy-7}
B(\dbf_1^{s_i};i)=d_{s_1}+d_{s_2}+\cdots+d_{s_i}, \textrm{ for } i=1,2,\ldots,k.
\eeqnarray
\etheorem

We need the following five lemmas for the proof of \rtheorem{OQ-LR-delays-greedy}.

\blemma{sum of C-transform}
Suppose that $\dbf_1^m\in \Acal_m$ and $m\geq 1$.

(i) If $0\leq x<\min\{d_{\ell'+1},d_{\ell'+2},\ldots,d_m\}$
for some $0\leq \ell'\leq m-1$,
then we have
\beqnarray{sum of C-transform-1}
\sum_{\ell=1}^{m}I_{\ell}(x;\dbf_1^m)=\sum_{\ell=1}^{m-1}I_{\ell}(x;\dbf_1^{m-1})
=\cdots=\sum_{\ell=1}^{\ell'}I_{\ell}(x;\dbf_1^{\ell'}).
\eeqnarray

(ii) If $\sum_{\ell=\ell'+1}^{m}d_{\ell}\leq x\leq \sum_{\ell=1}^{m}d_{\ell}$
for some $0\leq \ell'\leq m-1$, then we have
\beqnarray{sum of C-transform-2}
\sum_{\ell=1}^{m}I_{\ell}(x;\dbf_1^m)
=\sum_{\ell=1}^{m-1}I_{\ell}(x-d_m;\dbf_1^{m-1})+1=\cdots
=\sum_{\ell=1}^{\ell'}I_{\ell}\left(x-\sum_{\ell=\ell'+1}^{m}d_{\ell};\dbf_1^{\ell'}\right)+m-\ell'.
\eeqnarray
\elemma

\bproof
See \rappendix{sum of C-transform}.
\eproof

\blemma{OQ-LR-MRI-equivalent conditions}
Suppose that $\dbf_1^m\in \Acal_m$, $m\geq 1$, $i\geq 1$,
and $d_{\ell'+1}=\min\{d_{\ell'+1},d_{\ell'+2},\ldots,d_m\}$ for some $1\leq \ell'\leq m-1$
(note that this condition holds trivially for $\ell'=m-1$).
Then the following three conditions are equivalent:

(i) $B(\dbf_1^m;i)\geq d_{\ell'+1}-1$.

(ii) $B(\dbf_1^m;i)\geq d_{\ell'+1}$.

(iii) $B(\dbf_1^{m-1};i)\geq d_{\ell'+1}-1$.
\elemma

\bproof
See \rappendix{OQ-LR-MRI-equivalent conditions}.
\eproof

We remark that if $\dbf_1^m\in \Acal_m$, $m\geq 1$, $i\geq 1$,
and $d_{\ell'+1}=\min\{d_{\ell'+1},d_{\ell'+2},\ldots,d_m\}$ for some $1\leq \ell'\leq m-1$,
then it follows from the two equivalent conditions in
\rlemma{OQ-LR-MRI-equivalent conditions}(i) and \rlemma{OQ-LR-MRI-equivalent conditions}(ii)
that $B(\dbf_1^m;i)\neq d_{\ell'+1}-1$.

\blemma{OQ-LR-MRI-general}
Suppose that $\dbf_1^m\in \Acal_m$, $m\geq 1$, and $i\geq 1$.

(i) Suppose that $d_{\ell'+1}=\min\{d_{\ell'+1},d_{\ell'+2},\ldots,d_m\}$ for some $1\leq \ell'\leq m-1$.
If $B(\dbf_1^m;i)<d_{\ell'+1}-1$ or $B(\dbf_1^m;i)<d_{\ell'+1}$
or $B(\dbf_1^{m-1};i)<d_{\ell'+1}-1$
(note that these three conditions are equivalent by \rlemma{OQ-LR-MRI-equivalent conditions}),
then we have
\beqnarray{OQ-LR-MRI-general-1}
B(\dbf_1^m;i)=B(\dbf_1^{m-1};i)=\cdots=B(\dbf_1^{\ell'};i).
\eeqnarray

(ii) If $B(\dbf_1^m;i)\geq d_m-1$ or $B(\dbf_1^m;i)\geq d_m$ or $B(\dbf_1^{m-1};i)\geq d_m-1$
(note that these three conditions are equivalent by \rlemma{OQ-LR-MRI-equivalent conditions}),
then we have
\beqnarray{OQ-LR-MRI-general-2}
B(\dbf_1^m;i)=d_m+B(\dbf_1^{m-1};i-1).
\eeqnarray

(iii) Suppose that $d_1\leq d_2\leq \cdots\leq d_m$.
Let
\beqnarray{OQ-LR-MRI-general-3}
\ell'=\max\{1\leq \ell\leq m:\ d_{\ell}\leq B(\dbf_1^m;i)\}
\eeqnarray
(note that $\ell'$ is well defined as it is clear from $\dbf_1^m\in \Acal_m$,
$m\geq 1$, and $i\geq 1$ that $B(\dbf_1^m;i)\geq 1=d_1$).
Then we have
\beqnarray{OQ-LR-MRI-general-4}
B(\dbf_1^m;i)=B(\dbf_1^{m-1};i)=\cdots=B(\dbf_1^{\ell'};i)
=d_{\ell'}+B(\dbf_1^{\ell'-1};i-1).
\eeqnarray

(iv) Suppose that $d_{\ell'+1}=\min\{d_{\ell'+1},d_{\ell'+2},\ldots,d_m\}$ for some $1\leq \ell'\leq m-1$.
If $B(\dbf_1^{\ell'};i)<d_{\ell'+1}-1$,
then we have
\beqnarray{OQ-LR-MRI-general-5}
B(\dbf_1^{m};i)=B(\dbf_1^{m-1};i)=\cdots=B(\dbf_1^{\ell'};i).
\eeqnarray
\elemma

\bproof
See \rappendix{OQ-LR-MRI-general}.
\eproof

We remark that the definition of $\ell'$ in \reqnarray{OQ-LR-MRI-general-3}
is essential for \reqnarray{OQ-LR-MRI-general-4} to hold.
This is because $B(\dbf_1^m;i)\geq d_{\ell'}$ does not always guarantee that
$B(\dbf_1^m;i)=d_{\ell'}+B(\dbf_1^{\ell'-1};i-1)$
unless $\ell'=\max\{1\leq \ell\leq m:\ d_{\ell}\leq B(\dbf_1^m;i)\}$
as given in \reqnarray{OQ-LR-MRI-general-3}.
We illustrate this by an example.
If $\dbf_1^4=(1,2,4,8)$, then we can see that $B(\dbf_1^4;2)=6\geq d_2$ and $B(\dbf_1^1;1)=1$,
but $B(\dbf_1^4;2)\neq d_2+B(\dbf_1^1;1)$.
However, we have $d_3\leq B(\dbf_1^4;2)<d_4$ and $B(\dbf_1^2;1)=2$,
and hence $B(\dbf_1^4;2)=d_3+B(\dbf_1^2;1)$.

\blemma{OQ-LR-MRI-greedy-1}
Suppose that $M\geq 2$ and $1\leq k\leq M-1$.
Let $\dbf_1^M\in \Gcal_{M,k}$ so that there exists a sequence $\nbf_1^k\in \Ncal_{M,k}$
such that $d_{s_i+j}$ is given by \reqnarray{OQ-LR-delays-greedy-1},
i.e., $d_{s_i+j}=B(\dbf_1^{s_i+j-1};i+1)+1$
for $i=0,1,\ldots,k-1$ and $j=1,2,\ldots,n_{i+1}$,
where $s_0=0$ and $s_i=\sum_{\ell=1}^{i}n_{\ell}$ for $i=1,2,\ldots,k$.
Suppose that $\dbf_1^{s_i+j}\in \Acal_{s_i+j}$
for some $1\leq i\leq k-1$ and $0\leq j\leq n_{i+1}$.
Then we have
\beqnarray{OQ-LR-MRI-greedy-1}
B(\dbf_1^{s_i+j};i+1)=d_{s_i+j}+B(\dbf_1^{s_i+j-1};i).
\eeqnarray
\elemma

\bproof
See \rappendix{OQ-LR-MRI-greedy-1}.
\eproof

\blemma{OQ-LR-MRI-greedy-2}
Suppose that $M\geq 2$ and $1\leq k\leq M-1$.
Let $\nbf_1^k\in \Ncal_{M,k}$ and let $s_0=0$
and $s_i=\sum_{\ell=1}^{i}n_{\ell}$ for $i=1,2,\ldots,k$.

(i) Suppose that the sequence $\dbf_1^{s_i+j}=(d_1,d_2,\ldots,d_{s_i+j})$
is given by \reqnarray{OQ-LR-delays-greedy-2} and \reqnarray{OQ-LR-delays-greedy-3}
for some $1\leq i\leq k-1$ and $1\leq j\leq n_{i+1}$.
Then we have $\dbf_1^{s_i+j}\in \Bcal_{s_i+j}$.

(ii) Suppose that the sequence $\dbf_1^{s_i+j}=(d_1,d_2,\ldots,d_{s_i+j})$
is given by \reqnarray{OQ-LR-delays-greedy-2} and \reqnarray{OQ-LR-delays-greedy-3}
for some $1\leq i\leq k-1$ and $1\leq j\leq n_{i+1}$,
and suppose that
\beqnarray{OQ-LR-MRI at break point-greedy}
B(\dbf_1^{s_i};i)=d_{s_1}+d_{s_2}+\cdots+d_{s_i}.
\eeqnarray
Then we have
\beqnarray{OQ-LR-MRI-greedy-2}
B(\dbf_1^{s_i+j};i)=B(\dbf_1^{s_i+j-1};i)=\cdots=B(\dbf_1^{s_i};i)
=d_{s_1}+d_{s_2}+\cdots+d_{s_i}.
\eeqnarray
\elemma

\bproof
See \rappendix{OQ-LR-MRI-greedy-2}.
\eproof

\bproof \textbf{(Proof of \rtheorem{OQ-LR-delays-greedy})}
From \reqnarray{OQ-LR-delays-greedy-1} and $B(\dbf_1^0;1)=0$, we have
\beqnarray{}
d_1 \aligneq d_{s_0+1}=B(\dbf_1^0;1)+1=1,\ B(\dbf_1^1;1)=B((1);1)=1, \nn\\
d_2 \aligneq d_{s_0+2}=B(\dbf_1^1;1)+1=2,\ B(\dbf_1^2;1)=B((1,2);1)=2, \nn\\
d_3 \aligneq d_{s_0+3}=B(\dbf_1^2;1)+1=3,\ B(\dbf_1^3;1)=B((1,2,3);1)=3, \nn\\
&\vdots& \nn\\
d_{s_1} \aligneq d_{s_0+n_1}=B(\dbf_1^{s_1-1};1)+1=s_1,\
B(\dbf_1^{s_1};1)=B((1,2,\ldots,s_1);1)=s_1. \nn
\eeqnarray
Therefore, \reqnarray{OQ-LR-delays-greedy-2} and \reqnarray{OQ-LR-delays-greedy-5} are proved.
As it is clear from \reqnarray{OQ-LR-delays-greedy-2} that
$\dbf_1^j\in \Bcal_j$ for $j=1,2,\ldots,s_1$,
we see that \reqnarray{OQ-LR-delays-greedy-4} holds for $i=0$ and $j=1,2,\ldots,s_1=n_1$.

In the following, we show by induction that \reqnarray{OQ-LR-delays-greedy-3},
\reqnarray{OQ-LR-delays-greedy-4}, and \reqnarray{OQ-LR-delays-greedy-6}
hold for $1\leq i\leq k-1$ and $1\leq j\leq n_{i+1}$.
From $\dbf_1^{s_1}\in \Bcal_{s_1}\subseteq \Acal_{s_1}$,
\rlemma{OQ-LR-MRI-greedy-1} (with $i=1$ and $j=0$ in \rlemma{OQ-LR-MRI-greedy-1}),
$B(\dbf_1^{s_1-1};1)=s_1-1$ in \reqnarray{OQ-LR-delays-greedy-5},
and $d_{s_1}=s_1$ in \reqnarray{OQ-LR-delays-greedy-2}, we have
\beqnarray{proof-OQ-LR-delays-greedy-111}
B(\dbf_1^{s_1};2)=d_{s_1}+B(\dbf_1^{s_1-1};1)=d_{s_1}+(s_1-1)=d_{s_1}+(d_{s_1}-1)=2d_{s_1}-1.
\eeqnarray
It follows from \reqnarray{OQ-LR-delays-greedy-1}
(with $i=1$ and $j=1$ in \reqnarray{OQ-LR-delays-greedy-1})
and \reqnarray{proof-OQ-LR-delays-greedy-111} that
\beqnarray{proof-OQ-LR-delays-greedy-222}
d_{s_1+1}=B(\dbf_1^{s_1};2)+1=2d_{s_1}.
\eeqnarray
From \reqnarray{OQ-LR-delays-greedy-2} and \reqnarray{proof-OQ-LR-delays-greedy-222},
we see that $\dbf_1^{s_1+1}$ is given by
\reqnarray{OQ-LR-delays-greedy-2} and \reqnarray{OQ-LR-delays-greedy-3},
and it follows from \rlemma{OQ-LR-MRI-greedy-2}(i)
(with $i=1$ and $j=1$ in \rlemma{OQ-LR-MRI-greedy-2}(i)) that
\beqnarray{proof-OQ-LR-delays-greedy-333}
\dbf_1^{s_1+1}\in \Bcal_{s_1+1}.
\eeqnarray
Furthermore, from $\dbf_1^{s_1+1}\in \Bcal_{s_1+1}\subseteq \Acal_{s_1+1}$,
\rlemma{OQ-LR-MRI-greedy-1} (with $i=1$ and $j=1$ in \rlemma{OQ-LR-MRI-greedy-1}),
$B(\dbf_1^{s_1};1)=s_1$ in \reqnarray{OQ-LR-delays-greedy-5},
and $d_{s_1}=s_1$ in \reqnarray{OQ-LR-delays-greedy-2}, we have
\beqnarray{proof-OQ-LR-delays-greedy-444}
B(\dbf_1^{s_1+1};2)=d_{s_1+1}+B(\dbf_1^{s_1};1)=d_{s_1+1}+s_1=d_{s_1+1}+d_{s_1}.
\eeqnarray
Thus, we see from \reqnarray{proof-OQ-LR-delays-greedy-222}--\reqnarray{proof-OQ-LR-delays-greedy-444}
that \reqnarray{OQ-LR-delays-greedy-3}, \reqnarray{OQ-LR-delays-greedy-4},
and \reqnarray{OQ-LR-delays-greedy-6} hold for $i=1$ and $j=1$.

Now assume as the induction hypothesis that \reqnarray{OQ-LR-delays-greedy-3},
\reqnarray{OQ-LR-delays-greedy-4}, and \reqnarray{OQ-LR-delays-greedy-6}
hold up to some $1\leq i\leq k-1$ and $1\leq j\leq n_{i+1}$, where $s_i+j<M$.
We then consider the following two cases.

\emph{Case 1: $1\leq j\leq n_{i+1}-1$.}
In this case, we have $2\leq j+1\leq n_{i+1}$,
and it follows from \reqnarray{OQ-LR-delays-greedy-1} and the induction hypothesis that
\beqnarray{proof-OQ-LR-delays-greedy-555}
d_{s_i+j+1}
\aligneq B(\dbf_1^{s_i+j};i+1)+1 \nn\\
\aligneq d_{s_i+j}+d_{s_1}+d_{s_2}+\cdots+d_{s_i}+1 \nn\\
\aligneq (2d_{s_i}+(j-1)(d_{s_1}+d_{s_2}+\cdots+d_{s_i}+1))
+d_{s_1}+d_{s_2}+\cdots+d_{s_i}+1 \nn\\
\aligneq 2d_{s_i}+j(d_{s_1}+d_{s_2}+\cdots+d_{s_i}+1).
\eeqnarray
From the induction hypothesis and \reqnarray{proof-OQ-LR-delays-greedy-555},
we see that $\dbf_1^{s_i+j+1}$ is given by
\reqnarray{OQ-LR-delays-greedy-2} and \reqnarray{OQ-LR-delays-greedy-3},
and it follows from $1\leq i\leq k-1$, $2\leq j+1\leq n_{i+1}$,
and \rlemma{OQ-LR-MRI-greedy-2}(i) that
\beqnarray{proof-OQ-LR-delays-greedy-666}
\dbf_1^{s_i+j+1}\in \Bcal_{s_i+j+1}.
\eeqnarray
As such, we have from $\dbf_1^{s_i+j+1}\in \Bcal_{s_i+j+1}\subseteq \Acal_{s_i+j+1}$,
$1\leq i\leq k-1$, $2\leq j+1\leq n_{i+1}$, and \rlemma{OQ-LR-MRI-greedy-1} that
\beqnarray{proof-OQ-LR-delays-greedy-777}
B(\dbf_1^{s_i+j+1};i+1)=d_{s_i+j+1}+B(\dbf_1^{s_i+j};i).
\eeqnarray
As it is easy to see from the induction hypothesis that
$\dbf_1^{s_i+j}$ is given by
\reqnarray{OQ-LR-delays-greedy-2} and \reqnarray{OQ-LR-delays-greedy-3}
and $B(\dbf_1^{s_i};i)=d_{s_1}+d_{s_2}+\cdots+d_{s_i}$,
it then follows from \reqnarray{proof-OQ-LR-delays-greedy-777},
$1\leq i\leq k-1$, $2\leq j+1\leq n_{i+1}$,
and \rlemma{OQ-LR-MRI-greedy-2}(ii) that
\beqnarray{proof-OQ-LR-delays-greedy-888}
B(\dbf_1^{s_i+j+1};i+1)=d_{s_i+j+1}+B(\dbf_1^{s_i+j};i)
=d_{s_i+j+1}+d_{s_1}+d_{s_2}+\cdots+d_{s_i}.
\eeqnarray

The induction is completed by combining \reqnarray{proof-OQ-LR-delays-greedy-555},
\reqnarray{proof-OQ-LR-delays-greedy-666}, and \reqnarray{proof-OQ-LR-delays-greedy-888}
in this case.

\emph{Case 2: $j=n_{i+1}$.}
In this case, we have $s_i+j=s_{i+1}$.
Since we assume that $s_i+j=s_{i+1}<M=s_k$, it must be the case that $i+1\leq k-1$.
From the induction hypothesis, we have $\dbf_1^{s_i+j}\in \Bcal_{s_i+j}$,
i.e., $\dbf_1^{s_{i+1}}\in \Bcal_{s_{i+1}}$.
It then follows from $2\leq i+1\leq k-1$, \reqnarray{OQ-LR-delays-greedy-1},
$\dbf_1^{s_{i+1}}\in \Bcal_{s_{i+1}}\subseteq \Acal_{s_{i+1}}$,
and \rlemma{OQ-LR-MRI-greedy-1} that
\beqnarray{proof-OQ-LR-delays-greedy-999}
d_{s_{i+1}+1}=B(\dbf_1^{s_{i+1}};i+2)+1=d_{s_{i+1}}+B(\dbf_1^{s_{i+1}-1};i+1)+1.
\eeqnarray
If $n_{i+1}=1$, then we have $s_{i+1}-1=s_{i+1}-n_{i+1}=s_i$,
and it follows from $\dbf_1^{s_i}\in \Bcal_{s_i}\subseteq \Acal_{s_i}$
in the induction hypothesis, $1\leq i\leq k-2$, and \rlemma{OQ-LR-MRI-greedy-1} that
\beqnarray{proof-OQ-LR-delays-greedy-aaa}
B(\dbf_1^{s_{i+1}-1};i+1)+1=B(\dbf_1^{s_i};i+1)+1
=d_{s_i}+B(\dbf_1^{s_i-1};i)+1.
\eeqnarray
From \reqnarray{OQ-LR-delays-greedy-1}, we have
\beqnarray{proof-OQ-LR-delays-greedy-bbb}
d_{s_i}=d_{s_{i-1}+n_i}=B(\dbf_1^{s_{i-1}+n_i-1};i)+1=B(\dbf_1^{s_i-1};i)+1.
\eeqnarray
Thus, we have from \reqnarray{proof-OQ-LR-delays-greedy-aaa},
\reqnarray{proof-OQ-LR-delays-greedy-bbb},
$d_{s_i+1}=2d_{s_i}$ in the induction hypothesis,
and $s_{i+1}=s_i+1$ that
\beqnarray{proof-OQ-LR-delays-greedy-ccc}
B(\dbf_1^{s_{i+1}-1};i+1)+1=d_{s_i}+B(\dbf_1^{s_i-1};i)+1=d_{s_i}+d_{s_i}=d_{s_i+1}=d_{s_{i+1}}.
\eeqnarray
On the other hand, if $n_{i+1}\geq 2$,
then we have $n_{i+1}-1\geq 1$ and it follows from the induction hypothesis that
\beqnarray{proof-OQ-LR-delays-greedy-ddd}
\alignspace B(\dbf_1^{s_{i+1}-1};i+1)+1 \nn\\
\alignspace =B(\dbf_1^{s_i+n_{i+1}-1};i+1)+1 \nn\\
\alignspace =d_{s_i+n_{i+1}-1}+d_{s_1}+d_{s_2}+\cdots+d_{s_i}+1 \nn\\
\alignspace =2d_{s_i}+(n_{i+1}-2)(d_{s_1}+d_{s_2}+\cdots+d_{s_i}+1)
+d_{s_1}+d_{s_2}+\cdots+d_{s_i}+1\nn\\
\alignspace =2d_{s_i}+(n_{i+1}-1)(d_{s_1}+d_{s_2}+\cdots+d_{s_i}+1) \nn\\
\alignspace =d_{s_i+n_{i+1}}=d_{s_{i+1}}.
\eeqnarray
Therefore, we have from \reqnarray{proof-OQ-LR-delays-greedy-999},
\reqnarray{proof-OQ-LR-delays-greedy-ccc}, and \reqnarray{proof-OQ-LR-delays-greedy-ddd} that
\beqnarray{proof-OQ-LR-delays-greedy-eee}
d_{s_{i+1}+1}=d_{s_{i+1}}+B(\dbf_1^{s_{i+1}-1};i+1)+1=d_{s_{i+1}}+d_{s_{i+1}}=2d_{s_{i+1}}.
\eeqnarray

From the induction hypothesis and \reqnarray{proof-OQ-LR-delays-greedy-eee},
we see that $\dbf_1^{s_{i+1}+1}$ is given by
\reqnarray{OQ-LR-delays-greedy-2} and \reqnarray{OQ-LR-delays-greedy-3},
and it follows from $2\leq i+1\leq k-1$ and \rlemma{OQ-LR-MRI-greedy-2}(i) that
\beqnarray{proof-OQ-LR-delays-greedy-fff}
\dbf_1^{s_{i+1}+1}\in \Bcal_{s_{i+1}+1}.
\eeqnarray
As it is easy to see from the induction hypothesis that
$B(\dbf_1^{s_{i+1}};i+1)=d_{s_1}+d_{s_2}+\cdots+d_{s_i}+d_{s_{i+1}}$,
we have from $\dbf_1^{s_{i+1}+1}\in \Bcal_{s_{i+1}+1}\subseteq \Acal_{s_{i+1}+1}$,
$2\leq i+1\leq k-1$, and \rlemma{OQ-LR-MRI-greedy-1} that
\beqnarray{proof-OQ-LR-delays-greedy-ggg}
B(\dbf_1^{s_{i+1}+1};i+2)=d_{s_{i+1}+1}+B(\dbf_1^{s_{i+1}};i+1)
=d_{s_{i+1}+1}+d_{s_1}+d_{s_2}+\cdots+d_{s_i}+d_{s_{i+1}}.
\eeqnarray

The induction is completed by combining
\reqnarray{proof-OQ-LR-delays-greedy-eee}--\reqnarray{proof-OQ-LR-delays-greedy-ggg}
in this case.
\eproof

\bsection{Optimal Constructions Must Be Greedy Constructions}{OQ-LR-optimal is greedy}

In this section, we show that every optimal construction of a linear compressor/decompressor
in \rfigure{LC-2-to-1-FIFO-MUX}(a)/mirror image of \rfigure{LC-2-to-1-FIFO-MUX}(a)
must be a greedy construction,
and every optimal construction of a 2-to-1 FIFO multiplexer
in \rfigure{LC-2-to-1-FIFO-MUX}(b) must also be a greedy construction.
As $|\Acal_M|=\Omega(2^M)$, $|\Bcal_M|=\Omega(2^M)$ \cite{Cheng07},
and $|\Gcal_{M,k}|=\binom{M-2}{k-1}=O(M^k)$,
the complexity of searching for an optimal construction can be greatly reduced
by searching through the set $\Gcal_{M,k}$
rather than performing an exhaustive search through the set $\Acal_M$ or $\Bcal_M$
(polynomial time vs. exponential time).
Certainly, it will be great if we can obtain an optimal construction directly
without even having to search through the set $\Gcal_{M,k}$,
and in Part~II of this paper,
we will further show that there are at most two optimal constructions
and will give a simple algorithm to obtain the optimal construction(s).

In the following theorem,
we show that every optimal construction of a linear compressor/decompressor
in \rfigure{LC-2-to-1-FIFO-MUX}(a)/mirror image of \rfigure{LC-2-to-1-FIFO-MUX}(a)
must be a greedy construction.

\btheorem{OQ-LR-optimal is greedy}
Suppose that $M\geq 2$ and $1\leq k\leq M-1$.
If ${\dbf^*}_1^M\in \Acal_M$
and $B({\dbf^*}_1^M;k)=\max_{\dbf_1^M\in \Acal_M}B(\dbf_1^M;k)$,
then ${\dbf^*}_1^M\in \Gcal_{M,k}$.
In other words, every optimal construction of a linear compressor/decompressor
in \rfigure{LC-2-to-1-FIFO-MUX}(a)/mirror image of \rfigure{LC-2-to-1-FIFO-MUX}(a)
under our self-routing scheme and under the limitation of at most $k$ times
of recirculations through the $M$ fibers must be a greedy construction.
\etheorem

In the following corollary to \rtheorem{OQ-LR-optimal is greedy},
we show that every optimal construction of a 2-to-1 FIFO multiplexer
in \rfigure{LC-2-to-1-FIFO-MUX}(b) must also be a greedy construction.

\bcorollary{FIFO-MUX-LR-optimal is greedy}
Suppose that $M\geq 2$ and $1\leq k\leq M-1$.
If ${\dbf^*}_1^M\in \Bcal_M$
and $B({\dbf^*}_1^M;k)=\max_{\dbf_1^M\in \Bcal_M}B(\dbf_1^M;k)$,
then ${\dbf^*}_1^M\in \Gcal_{M,k}$.
In other words, every optimal construction of a 2-to-1 FIFO multiplexer
in \rfigure{LC-2-to-1-FIFO-MUX}(b) under our self-routing scheme
and under the limitation of at most $k$ times of recirculations
through the $M$ fibers must be a greedy construction.
\ecorollary

\bproof
Suppose that ${\dbf^*}_1^M\in \Bcal_M$ and $B({\dbf^*}_1^M;k)=\max_{\dbf_1^M\in \Bcal_M}B(\dbf_1^M;k)$.
Note that it is easy to see from $\Gcal_{M,k}\subseteq \Acal_M$ and \rtheorem{OQ-LR-optimal is greedy}
that $\max_{\dbf_1^M\in \Acal_M}B(\dbf_1^M;k)=\max_{\dbf_1^M\in \Gcal_{M,k}}B(\dbf_1^M;k)$.
It then follows from $\Gcal_{M,k}\subseteq \Bcal_M\subseteq \Acal_M$
and $\max_{\dbf_1^M\in \Acal_M}B(\dbf_1^M;k)=\max_{\dbf_1^M\in \Gcal_{M,k}}B(\dbf_1^M;k)$ that
\beqnarray{}
\alignspace
B({\dbf^*}_1^M;k)=\max_{\dbf_1^M\in \Bcal_M}B(\dbf_1^M;k)\leq \max_{\dbf_1^M\in \Acal_M}B(\dbf_1^M;k),
\label{eqn:proof-FIFO-MUX-LR-optimal is greedy-111} \\
\alignspace
B({\dbf^*}_1^M;k)=\max_{\dbf_1^M\in \Bcal_M}B(\dbf_1^M;k)
\geq \max_{\dbf_1^M\in \Gcal_{M,k}}B(\dbf_1^M;k)=\max_{\dbf_1^M\in \Acal_M}B(\dbf_1^M;k).
\label{eqn:proof-FIFO-MUX-LR-optimal is greedy-222}
\eeqnarray
From \reqnarray{proof-FIFO-MUX-LR-optimal is greedy-111}
and \reqnarray{proof-FIFO-MUX-LR-optimal is greedy-222},
we have $B({\dbf^*}_1^M;k)=\max_{\dbf_1^M\in \Acal_M}B(\dbf_1^M;k)$.
Therefore, we see from ${\dbf^*}_1^M\in \Bcal_M\subseteq \Acal_M$,
$B({\dbf^*}_1^M;k)=\max_{\dbf_1^M\in \Acal_M}B(\dbf_1^M;k)$,
and \rtheorem{OQ-LR-optimal is greedy} that ${\dbf^*}_1^M\in \Gcal_{M,k}$.
\eproof

We need the following three lemmas for the proof of \rtheorem{OQ-LR-optimal is greedy}.

\blemma{OQ-LR-MRI-monotone}
Suppose that $\dbf_1^m\in \Acal_m$, $m\geq 1$, and $0\leq i\leq i'$.
Then we have

(i) $B(\dbf_1^m;i)\leq B(\dbf_1^m;i')$.

(ii) $B(\dbf_1^m;i)=B(\dbf_1^m;i+1)$ if and only if $i\geq m$.
\elemma

\bproof
See \rappendix{OQ-LR-MRI-monotone}.
\eproof

\blemma{OQ-LR-optimal-is-nondecreasing}
Suppose that $M\geq 2$ and $1\leq k\leq M-1$.
Assume that ${\dbf^*}_1^M\in \Acal_M$
and $B({\dbf^*}_1^M;k)=\max_{\dbf_1^M\in \Acal_M}B(\dbf_1^M;k)$.
Then we have

(i) $B({\dbf^*}_1^M;k)\geq d_M^*$.

(ii) $d_1^*\leq d_2^*\leq \cdots\leq d_M^*$.
\elemma

\bproof
See \rappendix{OQ-LR-optimal-is-nondecreasing}.
\eproof

\blemma{OQ-LR-optimal-properties}
Suppose that $M\geq 2$ and $1\leq k\leq M-1$.
Assume that ${\dbf^*}_1^M\in \Acal_M$
and $B({\dbf^*}_1^M;k)=\max_{\dbf_1^M\in \Acal_M}B(\dbf_1^M;k)$.
Let $s_k,s_{k-1},\ldots,s_1$, in that order, be recursively given by
\beqnarray{}
\alignspace
s_k=\max\{1\leq \ell\leq M: d_{\ell}^*\leq B({\dbf^*}_1^M;k)\},
\label{eqn:OQ-LR-optimal-properties-1} \\
\alignspace
s_i=\max\{1\leq \ell\leq s_{i+1}-1: d_{\ell}^*\leq B({\dbf^*}_1^{s_{i+1}-1};i)\},
\textrm{ for } i=k-1,k-2,\ldots,1.
\label{eqn:OQ-LR-optimal-properties-2}
\eeqnarray

(i) We have
\beqnarray{OQ-LR-optimal-properties-3}
s_k=M \textrm{ and } B({\dbf^*}_1^{s_k};k)=d_{s_k}^*+B({\dbf^*}_1^{s_k-1};k-1).
\eeqnarray

(ii) We have
\beqnarray{}
\alignspace
s_i\geq i+1, \textrm{ for } i=1,2,\ldots,k,
\label{eqn:OQ-LR-optimal-properties-4} \\
\alignspace
B({\dbf^*}_1^{s_{i+1}-1};i)=B({\dbf^*}_1^{s_{i+1}-2};i)=\cdots
=B({\dbf^*}_1^{s_i};i) \nn\\
\alignspace \hspace*{1.0in}
=d_{s_i}^*+B({\dbf^*}_1^{s_i-1};i-1), \textrm{ for } i=1,2,\ldots,k-1,
\label{eqn:OQ-LR-optimal-properties-5} \\
\alignspace
B({\dbf^*}_1^{s_i};i)=d_{s_1}^*+d_{s_2}^*+\cdots +d_{s_i}^*,
\textrm{ for } i=1,2,\ldots,k.
\label{eqn:OQ-LR-optimal-properties-6}
\eeqnarray

(iii) Furthermore, we have
\beqnarray{}
\alignspace
B({\dbf^*}_1^{s_i};i+1)\geq d_{s_i+1}^*-1,
\textrm{ for } i=1,2,\ldots,k-1,
\label{eqn:OQ-LR-optimal-properties-7} \\
\alignspace
B({\dbf^*}_1^{s_i-1};i)\geq d_{s_i}^*-1,
\textrm{ for } i=1,2,\ldots,k,
\label{eqn:OQ-LR-optimal-properties-8} \\
\alignspace
B({\dbf^*}_1^{s_i};i+1)=d_{s_i}^*+B({\dbf^*}_1^{s_i-1};i),
\textrm{ for } i=1,2,\ldots,k-1,
\label{eqn:OQ-LR-optimal-properties-9} \\
\alignspace
B({\dbf^*}_1^{s_i-1};i)=d_{s_i-1}^*+B({\dbf^*}_1^{s_i-2};i-1),
\textrm{ for } i=1,2,\ldots,k.
\label{eqn:OQ-LR-optimal-properties-10}
\eeqnarray
\elemma

\bproof
See \rappendix{OQ-LR-optimal-properties}.
\eproof

\bproof \textbf{(Proof of \rtheorem{OQ-LR-optimal is greedy})}
Suppose that ${\dbf^*}_1^M\in \Acal_M$
and $B({\dbf^*}_1^M;k)=\max_{\dbf_1^M\in \Acal_M}B(\dbf_1^M;k)$.
Let $n_i=s_i-s_{i-1}$ for $i=1,2,\ldots,k$,
where $s_0=0$ and $s_1,s_2,\ldots,s_k$ are given by
\reqnarray{OQ-LR-optimal-properties-1}
and \reqnarray{OQ-LR-optimal-properties-2}.
Then we have from \reqnarray{OQ-LR-optimal-properties-4}
that $n_1=s_1-s_0=s_1\geq 2$,
we have from the definition of $s_i$ in \reqnarray{OQ-LR-optimal-properties-2}
that $n_{i+1}=s_{i+1}-s_i\geq 1$ for $i=1,2,\ldots,k-1$,
and we have from $s_k=M$ in \reqnarray{OQ-LR-optimal-properties-3}
that $\sum_{i=1}^{k}n_i=s_k-s_0=M$.
Therefore, we immediately see that $\nbf_1^k\in \Ncal_{M,k}$.

Let $\dbf_1^M\in \Gcal_{M,k}$
be given by $d_{s_i+j}=B(\dbf_1^{s_i+j-1};i+1)+1$
for $i=0,1,\ldots,k-1$ and $j=1,2,\ldots,n_{i+1}$
as in \reqnarray{OQ-LR-delays-greedy-1}.
In the following, we prove that ${\dbf^*}_1^M\in \Gcal_{M,k}$
by showing that ${\dbf^*}_1^M=\dbf_1^M$.
We divide the proof of ${\dbf^*}_1^M=\dbf_1^M$ into the following two parts.
Note that from \rlemma{OQ-LR-optimal-is-nondecreasing}(ii),
we have $d_1^*\leq d_2^*\leq \cdots\leq d_M^*$.

(i) In the first part, we show by induction on $\ell$ that $d_{\ell}^*\leq d_{\ell}$
for $\ell=1,2,\ldots,M$,
and if $d_{\ell'}^*<d_{\ell'}$ for some $2\leq \ell'\leq M$ (note that $d_1^*=d_1=1$),
then $d_{\ell}^*<d_{\ell}$ for $\ell=\ell',\ell'+1,\ldots,M$.

We first show that $d_1^*\leq d_1,d_2^*\leq d_2,\ldots,d_{s_1}^*\leq d_{s_1}$,
and if $d_{\ell'}^*<d_{\ell'}$ for some $2\leq \ell'\leq s_1$,
then we have $d_{\ell'}^*<d_{\ell'},d_{\ell'+1}^*<d_{\ell'+1},\ldots,d_{s_1}^*<d_{s_1}$.
From $d_1^*=1$, $d_1^*\leq d_2^*\leq \cdots\leq d_{s_2-1}^*$,
and the definition of $B({\dbf^*}_1^{s_2-1};1)$
in \reqnarray{maximum representable integer},
it is easy to see that $B({\dbf^*}_1^{s_2-1};1)=d_{s'_1}^*$,
where
\beqnarray{proof-OQ-LR-optimal is greedy-(i)-111}
s'_1=\max\{2\leq \ell\leq s_2-1:d_2^*-d_1^*\leq 1,d_3^*-d_2^*\leq 1,
\ldots,d_{\ell}^*-d_{\ell-1}^*\leq 1\}.
\eeqnarray
Note that $s'_1$ is well defined as we have from
\reqnarray{OQ-LR-optimal-properties-4} that $s_2-1\geq 2$
and we have from ${\dbf^*}_1^M\in \Acal_M$ that $d_2^*-d_1^*\leq (d_1^*+1)-d_1^*=1$.

We claim that
\beqnarray{proof-OQ-LR-optimal is greedy-(i)-222}
s_1=s'_1.
\eeqnarray
If $2\leq s'_1\leq s_2-2$,
then we see from the definition of $s'_1$
in \reqnarray{proof-OQ-LR-optimal is greedy-(i)-111}
that $d_{s'_1+1}^*-d_{s'_1}^*\geq 2$
and hence we have from $d_{s'_1}^*=B({\dbf^*}_1^{s_2-1};1)$ that
\beqnarray{proof-OQ-LR-optimal is greedy-(i)-333}
d_{s'_1+1}^*\geq d_{s'_1}^*+2>d_{s'_1}^*=B({\dbf^*}_1^{s_2-1};1).
\eeqnarray
Thus, it follows from $d_{s'_1}^*=B({\dbf^*}_1^{s_2-1};1)$,
$d_{s'_1+1}^*>B({\dbf^*}_1^{s_2-1};1)$
in \reqnarray{proof-OQ-LR-optimal is greedy-(i)-333},
$d_{s'_1+1}^*\leq d_{s'_1+2}^*\leq \cdots\leq d_{s_2-1}^*$,
and the definition of $s_1$ in \reqnarray{OQ-LR-optimal-properties-2}
that $s_1=s'_1$.
On the other hand, if $s'_1=s_2-1$,
then we have $d_{s_2-1}^*=d_{s'_1}^*=B({\dbf^*}_1^{s_2-1};1)$,
and it follows from the definition of $s_1$ in \reqnarray{OQ-LR-optimal-properties-2}
that $s_1=s_2-1=s'_1$.

From $d_1^*=1$, $d_{\ell}^*-d_{\ell-1}^*\leq 1$
for $\ell=2,3,\ldots,s'_1$ in \reqnarray{proof-OQ-LR-optimal is greedy-(i)-111},
and $s_1=s'_1$ in \reqnarray{proof-OQ-LR-optimal is greedy-(i)-222},
we can see that $d_{\ell}^*\leq \ell$ for $\ell=1,2,\ldots,s'_1=s_1$.
Since $\dbf_1^M\in \Gcal_{M,k}$,
we have from \reqnarray{OQ-LR-delays-greedy-2}
that $d_{\ell}=\ell$ for $\ell=1,2,\ldots,s_1$.
It then follows that
\beqnarray{}
d_{\ell}^*\leq \ell=d_{\ell}, \textrm{ for } \ell=1,2,\ldots,s_1. \nn
\eeqnarray
Furthermore, if $d_{\ell'}^*<d_{\ell'}=\ell'$ for some $2\leq \ell'\leq s_1$,
then we can see from $d_{\ell}^*-d_{\ell-1}^*\leq 1$ for $\ell=2,3,\ldots,s_1$
and $d_{\ell}=\ell$ for $\ell=1,2,\ldots,s_1$ that
\beqnarray{}
d_{\ell}^*\leq d_{\ell'}^*+(\ell-\ell')<\ell'+(\ell-\ell')=\ell=d_{\ell},
\textrm{ for } \ell=\ell',\ell'+1,\ldots,s_1. \nn
\eeqnarray

Now assume as the induction hypothesis that
for some $s_1\leq \ell\leq M-1$,
we have $d_1^*\leq d_1,d_2^*\leq d_2,\ldots,d_{\ell}^*\leq d_{\ell}$,
and if $d_{\ell'}^*<d_{\ell'}$ for some $2\leq \ell'\leq \ell$,
then we have $d_{\ell'}^*<d_{\ell'},d_{\ell'+1}^*<d_{\ell'+1},\ldots,d_{\ell}^*<d_{\ell}$.
We then consider the following two cases.

\emph{Case 1: $\ell=s_i$, where $1\leq i\leq k-1$.}
Let $i'=\max\{1\leq i''\leq i: n_{i''}\geq 2\}$
(note that $i'$ is well defined as $i\geq 1$ and $n_1\geq 2$).
It is clear that $n_i=n_{i-1}=\cdots=n_{i'+1}=1$ and $n_{i'}\geq 2$.
It then follows from \reqnarray{OQ-LR-optimal-properties-7},
\reqnarray{OQ-LR-optimal-properties-9},
$s_i-1=s_i-n_i=s_{i-1}$, $s_{i-1}-1=s_{i-1}-n_{i-1}=s_{i-2},\ldots$,
and $s_{i'+1}-1=s_{i'+1}-n_{i'+1}=s_{i'}$ that
\beqnarray{proof-OQ-LR-optimal is greedy-(i)-case 1-111}
d_{s_i+1}^*
\alignleq B({\dbf^*}_1^{s_i};i+1)+1=d_{s_i}^*+B({\dbf^*}_1^{s_i-1};i)+1 \nn\\
\aligneq d_{s_i}^*+B({\dbf^*}_1^{s_{i-1}};i)+1
=d_{s_i}^*+d_{s_{i-1}}^*+B({\dbf^*}_1^{s_{i-1}-1};i-1)+1 \nn\\
\aligneq d_{s_i}^*+d_{s_{i-1}}^*+B({\dbf^*}_1^{s_{i-2}};i-1)+1 \nn\\
&\vdots& \nn\\
\aligneq d_{s_i}^*+d_{s_{i-1}}^*+\cdots+d_{s_{i'+1}}^*+B({\dbf^*}_1^{s_{i'}};i'+1)+1 \nn\\
\aligneq d_{s_i}^*+d_{s_{i-1}}^*+\cdots+d_{s_{i'+1}}^*+d_{s_{i'}}^*+B({\dbf^*}_1^{s_{i'}-1};i')+1.
\eeqnarray
Thus, we have from \reqnarray{proof-OQ-LR-optimal is greedy-(i)-case 1-111},
\reqnarray{OQ-LR-optimal-properties-10}, $n_{i'}\geq 2$,
\reqnarray{OQ-LR-optimal-properties-5},
and \reqnarray{OQ-LR-optimal-properties-6} that
\beqnarray{proof-OQ-LR-optimal is greedy-(i)-case 1-222}
d_{s_i+1}^*
\alignleq d_{s_i}^*+d_{s_{i-1}}^*+\cdots+d_{s_{i'}}^*+B({\dbf^*}_1^{s_{i'}-1};i')+1 \nn\\
\aligneq d_{s_i}^*+d_{s_{i-1}}^*+\cdots+d_{s_{i'}}^*+d_{s_{i'-1}}^*+B({\dbf^*}_1^{s_{i'}-2};i'-1)+1 \nn\\
\aligneq d_{s_i}^*+d_{s_{i-1}}^*+\cdots+d_{s_{i'}}^*+d_{s_{i'-1}}^*+B({\dbf^*}_1^{s_{i'-1}};i'-1)+1 \nn\\
\aligneq d_{s_i}^*+d_{s_{i-1}}^*+\cdots+d_{s_{i'}}^*+d_{s_{i'-1}}^*
+(d_{s_1}^*+d_{s_2}^*+\cdots +d_{s_{i'-1}}^*+1).
\eeqnarray
As $\dbf_1^M\in \Gcal_{M,k}$,
we have from \reqnarray{OQ-LR-delays-greedy-3}
and $n_i=n_{i-1}=\cdots=n_{i'+1}=1$ that
\beqnarray{proof-OQ-LR-optimal is greedy-(i)-case 1-333}
d_{s_i+1}
\aligneq 2d_{s_i}=d_{s_i}+d_{s_i} \nn\\
\aligneq d_{s_i}+2d_{s_{i-1}}+(n_i-1)(d_{s_1}+d_{s_2}+\cdots +d_{s_{i-1}}+1) \nn\\
\aligneq d_{s_i}+2d_{s_{i-1}}=d_{s_i}+d_{s_{i-1}}+d_{s_{i-1}} \nn\\
&\vdots& \nn\\
\aligneq d_{s_i}+d_{s_{i-1}}+\cdots+d_{s_{i'+1}}+2d_{s_{i'}}
=d_{s_i}+d_{s_{i-1}}+\cdots+d_{s_{i'+1}}+d_{s_{i'}}+d_{s_{i'}} \nn\\
\aligneq d_{s_i}+d_{s_{i-1}}+\cdots+d_{s_{i'}}+2d_{s_{i'-1}}
+(n_{i'}-1)(d_{s_1}+d_{s_2}+\cdots +d_{s_{i'-1}}+1).
\eeqnarray
As such, we see from \reqnarray{proof-OQ-LR-optimal is greedy-(i)-case 1-222},
\reqnarray{proof-OQ-LR-optimal is greedy-(i)-case 1-333},
the induction hypothesis, and $n_{i'}\geq 2$ that
\beqnarray{}
d_{s_i+1}^*
\alignleq d_{s_i}^*+d_{s_{i-1}}^*+\cdots+d_{s_{i'}}^*+d_{s_{i'-1}}^*
+(d_{s_1}^*+d_{s_2}^*+\cdots +d_{s_{i'-1}}^*+1) \nn\\
\alignleq d_{s_i}+d_{s_{i-1}}+\cdots+d_{s_{i'}}+2d_{s_{i'-1}}
+(n_{i'}-1)(d_{s_1}+d_{s_2}+\cdots +d_{s_{i'-1}}+1)
\label{eqn:proof-OQ-LR-optimal is greedy-(i)-case 1-444}\\
\aligneq d_{s_i+1}. \nn
\eeqnarray

Furthermore, if $d_{\ell'}^*<d_{\ell'}$ for some $2\leq \ell'\leq s_i$,
then we have $d_{\ell'}^*<d_{\ell'},d_{\ell'+1}^*<d_{\ell'+1},\ldots,d_{s_i}^*<d_{s_i}$
from the induction hypothesis.
Therefore, it is clear from $d_{s_i}^*<d_{s_i}$ that the inequality
in \reqnarray{proof-OQ-LR-optimal is greedy-(i)-case 1-444} is strict
so that we have $d_{s_i+1}^*<d_{s_i+1}$, and the induction is completed in this case.

\emph{Case 2: $\ell=s_i+j$, where $1\leq i\leq k-1$ and $1\leq j\leq n_{i+1}-1$.}
We first show that
\beqnarray{proof-OQ-LR-optimal is greedy-(i)-case 2-111}
d_{s_i+j+1}^*-1\leq d_{s_i+j}^*+d_{s_1}^*+d_{s_2}^*+\cdots+d_{s_i}^*.
\eeqnarray
If $d_{s_i+j+1}^*-1<d_{s_i+j}^*$, then there is nothing to prove.
On the other hand, if $d_{s_i+j+1}^*-1\geq d_{s_i+j}^*$,
then for $d_{s_i+j}^*\leq x\leq d_{s_i+j+1}^*-1$,
we have from ${\dbf^*}_1^{s_{i+1}}\in \Acal_{s_{i+1}}$,
$x<d_{s_i+j+1}^*=\min\{d_{s_i+j+1},d_{s_i+j+2},\ldots,d_{s_{i+1}}\}$,
$0<s_i+j\leq s_i+n_{i+1}-1=s_{i+1}-1$,
\rlemma{sum of C-transform}(i)
(with $m=s_{i+1}$ and $\ell'=s_i+j$ in \rlemma{sum of C-transform}(i)),
and \rlemma{sum of C-transform}(ii)
(with $m=s_i+j$ and $\ell'=m-1=s_i+j-1$
in \rlemma{sum of C-transform}(ii)), that
\beqnarray{proof-OQ-LR-optimal is greedy-(i)-case 2-222}
\sum_{\ell=1}^{s_{i+1}}I_{\ell}(x;{\dbf^*}_1^{s_{i+1}})
\aligneq \sum_{\ell=1}^{s_{i+1}-1}I_{\ell}(x;{\dbf^*}_1^{s_{i+1}-1})
=\cdots=\sum_{\ell=1}^{s_i+j}I_{\ell}(x;{\dbf^*}_1^{s_i+j}) \nn\\
\aligneq \sum_{\ell=1}^{s_i+j-1}I_{\ell}(x-d_{s_i+j}^*;{\dbf^*}_1^{s_i+j-1})+1,
\textrm{ for } d_{s_i+j}^*\leq x\leq d_{s_i+j+1}^*-1.
\eeqnarray
As we have from $s_i+j+1\leq s_i+n_{i+1}=s_{i+1}$,
$d_1^*\leq d_2^*\leq \cdots\leq d_M^*$,
and \reqnarray{OQ-LR-optimal-properties-6} that
$d_{s_i+j+1}^*-1<d_{s_{i+1}}^*<d_{s_1}^*+d_{s_2}^*+\cdots+d_{s_{i+1}}^*
=B({\dbf^*}_1^{s_{i+1}};i+1)$,
it then follows from \reqnarray{proof-OQ-LR-optimal is greedy-(i)-case 2-222}
and the definition of $B({\dbf^*}_1^{s_{i+1}};i+1)$
in \reqnarray{maximum representable integer} that
\beqnarray{proof-OQ-LR-optimal is greedy-(i)-case 2-333}
\sum_{\ell=1}^{s_i+j-1}I_{\ell}(x-d_{s_i+j}^*;{\dbf^*}_1^{s_i+j-1})
=\sum_{\ell=1}^{s_{i+1}}I_{\ell}(x;{\dbf^*}_1^{s_{i+1}})-1\leq (i+1)-1=i, \nn\\
\textrm{ for } d_{s_i+j}^*\leq x\leq d_{s_i+j+1}^*-1.
\eeqnarray
From \reqnarray{proof-OQ-LR-optimal is greedy-(i)-case 2-333},
we have $\sum_{\ell=1}^{s_i+j-1}I_{\ell}(x;{\dbf^*}_1^{s_i+j-1})\leq i$
for $0\leq x\leq d_{s_i+j+1}^*-d_{s_i+j}^*-1$.
As such, it follows from the definition of $B({\dbf^*}_1^{s_i+j-1};i)$
in \reqnarray{maximum representable integer},
$s_i+j-1\geq s_i$, $s_i+j-1\leq s_i+n_{i+1}-2=s_{i+1}-2$,
\reqnarray{OQ-LR-optimal-properties-5},
and \reqnarray{OQ-LR-optimal-properties-6} that
\beqnarray{}
d_{s_i+j+1}^*-d_{s_i+j}^*-1 \leq B({\dbf^*}_1^{s_i+j-1};i)
=B({\dbf^*}_1^{s_i};i)=d_{s_1}^*+d_{s_2}^*+\cdots+d_{s_i}^*, \nn
\eeqnarray
which is the desired result in \reqnarray{proof-OQ-LR-optimal is greedy-(i)-case 2-111}.

As $1\leq j\leq n_{i+1}-1$, we have $2\leq j+1\leq n_{i+1}$.
It then follows from \reqnarray{proof-OQ-LR-optimal is greedy-(i)-case 2-111},
$d_1^*\leq d_1,d_2^*\leq d_2,\ldots,d_{s_i+j}^*\leq d_{s_i+j}$
in the induction hypothesis,
and \reqnarray{OQ-LR-delays-greedy-3} that
\beqnarray{}
d_{s_i+j+1}^*
\alignleq d_{s_i+j}^*+d_{s_1}^*+d_{s_2}^*+\cdots+d_{s_i}^*+1 \nn\\
\alignleq d_{s_i+j}+d_{s_1}+d_{s_2}+\cdots+d_{s_i}+1
\label{eqn:proof-OQ-LR-optimal is greedy-(i)-case 2-444} \\
\aligneq 2d_{s_i}+(j-1)(d_{s_1}+d_{s_2}+\cdots+d_{s_i}+1)+d_{s_1}+d_{s_2}+\cdots+d_{s_i}+1 \nn\\
\aligneq 2d_{s_i}+j(d_{s_1}+d_{s_2}+\cdots+d_{s_i}+1) \nn\\
\aligneq d_{s_i+j+1}. \nn
\eeqnarray
Furthermore, if $d_{\ell'}^*<d_{\ell'}$ for some $2\leq \ell'\leq s_i+j$,
then we have $d_{\ell'}^*<d_{\ell'},d_{\ell'+1}^*<d_{\ell'+1},\ldots$, $d_{s_i+j}^*<d_{s_i+j}$
from the induction hypothesis.
Therefore, it is clear from $d_{s_i+j}^*<d_{s_i+j}$ that the inequality
in \reqnarray{proof-OQ-LR-optimal is greedy-(i)-case 2-444} is strict
so that we have $d_{s_i+j+1}^*<d_{s_i+j+1}$, and the induction is also completed in this case.

(ii) In the second part, we show that $d_{\ell}^*=d_{\ell}$ for $\ell=1,2,\ldots,M$,
and the proof is completed.
From $s_k=M$ in \reqnarray{OQ-LR-optimal-properties-3},
$B({\dbf^*}_1^{s_k};k)=d_{s_1}^*+d_{s_2}^*+\cdots+d_{s_k}^*$
in \reqnarray{OQ-LR-optimal-properties-6},
$d_{\ell}^*\leq d_{\ell}$ for $\ell=1,2,\ldots,M$ in part (i) above,
$B(\dbf_1^{s_k};k)=d_{s_1}+d_{s_2}+\cdots+d_{s_k}$
in \reqnarray{OQ-LR-delays-greedy-7}, $\dbf_1^M\in \Gcal_{M,k}\subseteq \Acal_M$,
and $B({\dbf^*}_1^M;k)=\max_{{\dbf'}_1^M\in \Acal_M}B({\dbf'}_1^M;k)$,
we have
\beqnarray{}
B({\dbf^*}_1^M;k)
\aligneq B({\dbf^*}_1^{s_k};k)=d_{s_1}^*+d_{s_2}^*+\cdots+d_{s_k}^* \nn\\
\alignleq d_{s_1}+d_{s_2}+\cdots+d_{s_k}
\label{eqn:proof-OQ-LR-optimal is greedy-(ii)-111} \\
\aligneq B(\dbf_1^{s_k};k)=B(\dbf_1^M;k) \nn\\
\alignleq \max_{{\dbf'}_1^M\in \Acal_M}B({\dbf'}_1^M;k)=B({\dbf^*}_1^M;k).
\label{eqn:proof-OQ-LR-optimal is greedy-(ii)-222}
\eeqnarray
It is clear that the two inequalities in
\reqnarray{proof-OQ-LR-optimal is greedy-(ii)-111}
and \reqnarray{proof-OQ-LR-optimal is greedy-(ii)-222}
must hold with equalities.
As such, it follows from the equality
in \reqnarray{proof-OQ-LR-optimal is greedy-(ii)-111}
and $d_{s_i}^*\leq d_{s_i}$ for $i=1,2,\ldots,k$ that
\beqnarray{proof-OQ-LR-optimal is greedy-(ii)-333}
d_{s_i}^*=d_{s_i}, \textrm{ for } i=1,2,\ldots,k.
\eeqnarray
From \reqnarray{proof-OQ-LR-optimal is greedy-(ii)-333},
we can see that $d_{\ell}^*=d_{\ell}$ for $\ell=1,2,\ldots,M$.
Otherwise, if $d_{\ell'}^*<d_{\ell'}$ for some $2\leq \ell'\leq M$
(note that $d_1^*=d_1=1$),
then we have from the result in part (i) above that
$d_{\ell}^*<d_{\ell}$ for $\ell=\ell',\ell'+1,\ldots,M$,
contradicting to $d_M^*=d_{s_k}^*=d_{s_k}=d_M$
in \reqnarray{proof-OQ-LR-optimal is greedy-(ii)-333}.
\eproof

\bsection{Conclusion}{conclusion}

In this paper, we considered an important problem arising from
practical feasibility considerations in the SDL constructions of optical queues:
the constructions of optical queues with a limited number of recirculations
through the optical switches and the fiber delay lines.
In Part~I of this paper,
we first showed that the constructions of certain types of optical queues,
including linear compressors, linear decompressors, and 2-to-1 FIFO multiplexers,
under a simple packet routing scheme
and under a limited number of recirculations constraint
can be transformed into equivalent integer representation problems under a corresponding constraint,
and showed that the problem of finding an optimal construction
(in the sense of maximizing the effective maximum delay of a linear compressor/decompressor
or the effective buffer size of a 2-to-1 FIFO multiplexer)
is equivalent to the problem of finding an optimal sequence
(in the sense of maximizing the maximum representable integer)
for the corresponding integer representation problem.
Then we proposed a class of greedy constructions
of linear compressors/decompressors and 2-to-1 FIFO multiplexers,
and showed that every optimal construction must be a greedy construction.

In the sequel to this paper, we will further show that
there are at most two optimal constructions
and will give a simple algorithm to obtain the optimal construction(s).


\appendices

\bappendix{Proof of \rtheorem{OQ-LR-delays-greedy-redundant}}{OQ-LR-delays-greedy-redundant}

We will show that $d_{\ell}=d'_{\ell}$ for $\ell=1,2,\ldots,M$ by induction on $\ell$.
From $n_1=1$ and $a=\min\{2\leq i\leq k:\ n_i\geq 2\}$,
we see that $n_i=1$ for $i=1,2,\ldots,a-1$ and $n_a\geq 2$.
It then follows from $s_i=\sum_{\ell=1}^{i}n_{\ell}$ for $i=1,2,\ldots,k$
that $s_i=i$ for $i=1,2,\ldots,a-1$ and $s_a\geq a+1$.
As such, it is easy to see from \reqnarray{OQ-LR-delays-greedy-redundant-1},
$B(\dbf_1^0;1)=0$, and $B(\dbf_1^m;i)=\sum_{\ell=1}^{m}d_{\ell}$ for $i\geq m\geq 1$ that
\beqnarray{}
d_1\aligneq d_{s_0+1}=B(\dbf_1^{s_0};1)+1=B(\dbf_1^0;1)+1=0+1=1,
\label{eqn:proof-OQ-LR-delays-greedy-redundant-111} \\
d_2\aligneq d_{s_1+1}=B(\dbf_1^{s_1};2)+1=B(\dbf_1^1;2)+1=d_1+1=2,
\label{eqn:proof-OQ-LR-delays-greedy-redundant-222} \\
d_3\aligneq d_{s_2+1}=B(\dbf_1^{s_2};3)+1=B(\dbf_1^2;3)+1=\sum_{\ell=1}^{2}d_{\ell}+1=2^2,
\label{eqn:proof-OQ-LR-delays-greedy-redundant-333} \\
&\vdots& \nn\\
d_a\aligneq d_{s_{a-1}+1}=B(\dbf_1^{s_{a-1}};a)+1=B(\dbf_1^{a-1};a)+1
=\sum_{\ell=1}^{a-1}d_{\ell}+1=2^{a-1},
\label{eqn:proof-OQ-LR-delays-greedy-redundant-444} \\
d_{a+1}\aligneq d_{s_{a-1}+2}=B(\dbf_1^{s_{a-1}+1};a)+1=B(\dbf_1^a;a)+1
=\sum_{\ell=1}^{a}d_{\ell}+1=2^a.
\label{eqn:proof-OQ-LR-delays-greedy-redundant-555}
\eeqnarray

Furthermore, from $n'_1=n_1+1=2$, $n'_i=n_i=1$ for $i=2,3,\ldots,a-1$,
$n'_a=n_a-1$, $n'_i=n_i$ for $i=a+1,a+2,\ldots,k$,
and $s'_i=\sum_{\ell=1}^{i}n'_{\ell}$ for $i=1,2,\ldots,k$,
we see that $s'_i=i+1=s_i+1$ for $i=1,2,\ldots,a-1$, and
$s'_i=s_i$ for $i=a,a+1,\ldots,k$.
It then follows from \reqnarray{OQ-LR-delays-greedy-redundant-2},
$B({\dbf'}_1^0;1)=0$, and $B({\dbf'}_1^m;i)=\sum_{\ell=1}^{m}d'_{\ell}$ for $i\geq m\geq 1$ that
\beqnarray{}
d'_1\aligneq d'_{s'_0+1}=B({\dbf'}_1^{s'_0};1)+1=B({\dbf'}_1^0;1)+1=0+1=1,
\label{eqn:proof-OQ-LR-delays-greedy-redundant-666} \\
d'_2\aligneq d'_{s'_0+2}=B({\dbf'}_1^{s'_0+1};1)+1=B({\dbf'}_1^1;1)+1=d'_1+1=2,
\label{eqn:proof-OQ-LR-delays-greedy-redundant-777} \\
d'_3\aligneq d'_{s'_1+1}=B({\dbf'}_1^{s'_1};2)+1=B({\dbf'}_1^2;2)+1=\sum_{\ell=1}^{2}d'_{\ell}+1=2^2,
\label{eqn:proof-OQ-LR-delays-greedy-redundant-888} \\
&\vdots& \nn\\
d'_a\aligneq d'_{s'_{a-2}+1}=B({\dbf'}_1^{s'_{a-2}};a-1)+1=B({\dbf'}_1^{a-1};a-1)+1
=\sum_{\ell=1}^{a-1}d'_{\ell}+1=2^{a-1},
\label{eqn:proof-OQ-LR-delays-greedy-redundant-999} \\
d'_{a+1}\aligneq d'_{s'_{a-1}+1}=B({\dbf'}_1^{s'_{a-1}};a)+1=B({\dbf'}_1^a;a)+1
=\sum_{\ell=1}^{a}d'_{\ell}+1=2^a.
\label{eqn:proof-OQ-LR-delays-greedy-redundant-aaa}
\eeqnarray
Therefore, it is clear from
\reqnarray{proof-OQ-LR-delays-greedy-redundant-111}--\reqnarray{proof-OQ-LR-delays-greedy-redundant-aaa}
and $a+1=s_{a-1}+2$ that $d_1=d'_1,d_2=d'_2,\ldots,d_{s_{a-1}+2}=d'_{s_{a-1}+2}$.

Now assume as the induction hypothesis that $d_1=d'_1,d_2=d'_2,\ldots,d_{\ell}=d'_{\ell}$
for some $s_{a-1}+2\leq \ell\leq M-1$.
In the following two possible cases,
we show that $d_{\ell+1}=d'_{\ell+1}$,
and the induction is completed.

\emph{Case 1: $\ell=s_{a-1}+j$, where $2\leq j\leq n_a-1$.}
In this case, we have from $3\leq j+1\leq n_a$
and \reqnarray{OQ-LR-delays-greedy-redundant-1} that
\beqnarray{proof-OQ-LR-delays-greedy-redundant-bbb}
d_{s_{a-1}+j+1}=B(\dbf_1^{s_{a-1}+j};a)+1.
\eeqnarray
Also, we have from $s'_{a-1}=s_{a-1}+1$, $2\leq j\leq n_a-1=n'_a$,
\reqnarray{OQ-LR-delays-greedy-redundant-2},
and $\dbf_1^{s_{a-1}+j}={\dbf'}_1^{s_{a-1}+j}$ in the induction hypothesis that
\beqnarray{proof-OQ-LR-delays-greedy-redundant-ccc}
d'_{s_{a-1}+j+1}=d'_{s'_{a-1}+j}=B({\dbf'}_1^{s'_{a-1}+j-1};a)+1
=B({\dbf'}_1^{s_{a-1}+j};a)+1=B(\dbf_1^{s_{a-1}+j};a)+1.
\eeqnarray
Thus, we see from \reqnarray{proof-OQ-LR-delays-greedy-redundant-bbb}
and \reqnarray{proof-OQ-LR-delays-greedy-redundant-ccc}
that $d_{s_{a-1}+j+1}=d'_{s_{a-1}+j+1}$,
i.e., $d_{\ell+1}=d'_{\ell+1}$.

\emph{Case 2: $\ell=s_i+j$, where $a\leq i\leq k-1$ and $0\leq j\leq n_{i+1}-1$.}
In this case, we have from $1\leq j+1\leq n_{i+1}$
and \reqnarray{OQ-LR-delays-greedy-redundant-1} that
\beqnarray{proof-OQ-LR-delays-greedy-redundant-ddd}
d_{s_i+j+1}=B(\dbf_1^{s_i+j};i+1)+1.
\eeqnarray
As $a\leq i\leq k-1$ and $0\leq j\leq n_{i+1}-1$,
we have $s'_i=s_i$ and $1\leq j+1\leq n_{i+1}=n'_{i+1}$.
It then follows from \reqnarray{OQ-LR-delays-greedy-redundant-2}
and $\dbf_1^{s_i+j}={\dbf'}_1^{s_i+j}$ in the induction hypothesis that
\beqnarray{proof-OQ-LR-delays-greedy-redundant-eee}
d'_{s_i+j+1}=d'_{s'_i+j+1}=B({\dbf'}_1^{s'_i+j};i+1)+1
=B({\dbf'}_1^{s_i+j};i+1)+1=B(\dbf_1^{s_i+j};i+1)+1.
\eeqnarray
Thus, we see from \reqnarray{proof-OQ-LR-delays-greedy-redundant-ddd}
and \reqnarray{proof-OQ-LR-delays-greedy-redundant-eee}
that $d_{s_i+j+1}=d'_{s_i+j+1}$,
i.e., $d_{\ell+1}=d'_{\ell+1}$.

\bappendix{Proof of \rlemma{sum of C-transform}}{sum of C-transform}

In this lemma, we have $\dbf_1^m\in \Acal_m$ and $m\geq 1$.

(i) Suppose that  $0\leq x<\min\{d_{\ell'+1},d_{\ell'+2},\ldots,d_m\}$ for some $1\leq \ell'\leq m-1$.
Let $\ell'\leq m'\leq m-1$.
Then it is clear from
$0\leq x<\min\{d_{\ell'+1},d_{\ell'+2},\ldots,d_m\}\leq \min\{d_{m'+1},d_{m'+2},\ldots,d_m\}$
and \reqnarray{C-transform} that
\beqnarray{}
\alignspace
I_{\ell}(x;\dbf_1^m)=0,
\textrm{ for } \ell=m,m-1,\ldots,m'+1,
\label{eqn:proof-sum of C-transform-(i)-111}\\
\alignspace
I_{\ell}(x;\dbf_1^m)=I_{\ell}(x;\dbf_1^{m'}),
\textrm{ for } \ell=m',m'-1,\ldots,1.
\label{eqn:proof-sum of C-transform-(i)-222}
\eeqnarray
It follows from \reqnarray{proof-sum of C-transform-(i)-111}
and \reqnarray{proof-sum of C-transform-(i)-222} that
\beqnarray{}
\sum_{\ell=1}^{m}I_{\ell}(x;\dbf_1^m)=\sum_{\ell=1}^{m'}I_{\ell}(x;\dbf_1^m)
=\sum_{\ell=1}^{m'}I_{\ell}(x;\dbf_1^{m'}),
\textrm{ for } \ell'\leq m'\leq m-1, \nn
\eeqnarray
which is the desired result in \reqnarray{sum of C-transform-1}.

(ii) Suppose that $\sum_{\ell=\ell'+1}^{m}d_{\ell}\leq x\leq \sum_{\ell=1}^{m}d_{\ell}$
for some $0\leq \ell'\leq m-1$.
Let $\ell'\leq m'\leq m-1$.
Then it is clear from
$\sum_{\ell=m'+1}^{m}d_{\ell}\leq\sum_{\ell=\ell'+1}^{m}d_{\ell}
\leq x\leq \sum_{\ell=1}^{m}d_{\ell}$
and \reqnarray{C-transform} that
\beqnarray{}
\alignspace
I_{\ell}(x;\dbf_1^m)=1,
\textrm{ for } \ell=m,m-1,\ldots,m'+1,
\label{eqn:proof-sum of C-transform-(ii)-111}\\
\alignspace
I_{\ell}(x;\dbf_1^m)=I_{\ell}\left(x-\sum_{\ell=m'+1}^{m}d_{\ell};\dbf_1^{m'}\right),
\textrm{ for } \ell=m',m'-1,\ldots,1.
\label{eqn:proof-sum of C-transform-(ii)-222}
\eeqnarray
It follows from \reqnarray{proof-sum of C-transform-(ii)-111}
and \reqnarray{proof-sum of C-transform-(ii)-222} that
\beqnarray{}
\sum_{\ell=1}^{m}I_{\ell}(x;\dbf_1^m)=\sum_{\ell=1}^{m'}I_{\ell}(x;\dbf_1^m)+m-m'
=\sum_{\ell=1}^{m'}I_{\ell}\left(x-\sum_{\ell=m'+1}^{m}d_{\ell};\dbf_1^{m'}\right)+m-m', \nn\\
\textrm{ for } \ell'\leq m'\leq m-1, \nn
\eeqnarray
which is the desired result in \reqnarray{sum of C-transform-2}.

\bappendix{Proof of \rlemma{OQ-LR-MRI-equivalent conditions}}{OQ-LR-MRI-equivalent conditions}

In this lemma, we have $\dbf_1^m\in \Acal_m$, $m\geq 1$, $i\geq 1$,
and $d_{\ell'+1}=\min\{d_{\ell'+1},d_{\ell'+2},\ldots,d_m\}$ for some $1\leq \ell'\leq m-1$.

(i) $\Rightarrow$ (ii):
Suppose that $B(\dbf_1^m;i)\geq d_{\ell'+1}-1$.
For $0\leq x\leq d_{\ell'+1}-1$,
it is clear from $\dbf_1^m\in \Acal_m$, $x\leq d_{\ell'+1}-1\leq B(\dbf_1^m;i)$,
and the definition of $B(\dbf_1^m;i)$
in \reqnarray{maximum representable integer} that
\beqnarray{proof-OQ-LR-MRI-equivalent conditions-111}
\sum_{\ell=1}^{m}I_{\ell}(x;\dbf_1^m)\leq i,
\textrm{ for } 0\leq x\leq d_{\ell'+1}-1.
\eeqnarray
Furthermore, for $x=d_{\ell'+1}$,
it is clear from $d_{\ell'+1}=\min\{d_{\ell'+1},d_{\ell'+2},\ldots,d_m\}$
and \reqnarray{C-transform} that
$I_{\ell''}(x;\dbf_1^m)=1$ and $I_{\ell}(x;\dbf_1^m)=0$ for all $\ell\neq \ell''$,
where $\ell''=\max\{\ell'+1\leq \ell\leq m: d_{\ell}=d_{\ell'+1}\}$.
Thus, it follows from $i\geq 1$ that
\beqnarray{proof-OQ-LR-MRI-equivalent conditions-222}
\sum_{\ell=1}^{m}I_{\ell}(x;\dbf_1^m)=1\leq i,
\textrm{ for } x=d_{\ell'+1}.
\eeqnarray
As $d_{\ell'+1}\leq \sum_{\ell=1}^{m}d_{\ell}$,
we see from \reqnarray{proof-OQ-LR-MRI-equivalent conditions-111},
\reqnarray{proof-OQ-LR-MRI-equivalent conditions-222},
and the definition of $B(\dbf_1^m;i)$
in \reqnarray{maximum representable integer}
that $B(\dbf_1^m;i)\geq d_{\ell'+1}$.

(ii) $\Rightarrow$ (iii):
Suppose that $B(\dbf_1^m;i)\geq d_{\ell'+1}$.
We will show that $B(\dbf_1^{m-1};i)\geq d_{\ell'+1}-1$ by contradiction.
So assume on the contrary that $B(\dbf_1^{m-1};i)<d_{\ell'+1}-1$.
For $x=d_{\ell'+1}-1$,
we have $x=d_{\ell'+1}-1<d_{\ell'+1}$,
$x=d_{\ell'+1}-1>B(\dbf_1^{m-1};i)$,
and $x=d_{\ell'+1}-1\leq \sum_{\ell=1}^{\ell'}d_{\ell}\leq \sum_{\ell=1}^{m-1}d_{\ell}$
(as $\dbf_1^m\in \Acal_m$ and $\ell'\leq m-1$).
It is then clear from $\dbf_1^m\in \Acal_m$,
$0\leq x<d_{\ell'+1}=\min\{d_{\ell'+1},d_{\ell'+2},\ldots,d_m\}$,
$1\leq \ell'\leq m-1$, \rlemma{sum of C-transform}(i),
$B(\dbf_1^{m-1};i)<x\leq \sum_{\ell=1}^{m-1}d_{\ell}$,
and the definition of $B(\dbf_1^{m-1};i)$
in \reqnarray{maximum representable integer} that
\beqnarray{proof-OQ-LR-MRI-equivalent conditions-333}
\sum_{\ell=1}^{m}I_{\ell}(x;\dbf_1^m)
=\sum_{\ell=1}^{m-1}I_{\ell}(x;\dbf_1^{m-1})
>i, \textrm{ for } x=d_{\ell'+1}-1.
\eeqnarray
However, we also have from $\dbf_1^m\in \Acal_m$,
$x=d_{\ell'+1}-1<d_{\ell'+1}\leq B(\dbf_1^m;i)$,
and the definition of $B(\dbf_1^m;i)$
in \reqnarray{maximum representable integer} that
\beqnarray{}
\sum_{\ell=1}^{m}I_{\ell}(x;\dbf_1^m)\leq i,
\textrm{ for } x=d_{\ell'+1}-1, \nn
\eeqnarray
and we have reached a contradiction to $\sum_{\ell=1}^{m}I_{\ell}(x;\dbf_1^m)>i$
for $x=d_{\ell'+1}-1$ in \reqnarray{proof-OQ-LR-MRI-equivalent conditions-333}.

(iii) $\Rightarrow$ (i):
Suppose that $B(\dbf_1^{m-1};i)\geq d_{\ell'+1}-1$.
For $0\leq x\leq d_{\ell'+1}-1$,
it is clear from $\dbf_1^m\in \Acal_m$,
$0\leq x<d_{\ell'+1}=\min\{d_{\ell'+1},d_{\ell'+2},\ldots,d_m\}$,
$1\leq \ell'\leq m-1$, \rlemma{sum of C-transform}(i),
$x\leq d_{\ell'+1}-1\leq B(\dbf_1^{m-1};i)$,
and the definition of $B(\dbf_1^{m-1};i)$
in \reqnarray{maximum representable integer} that
\beqnarray{proof-OQ-LR-MRI-equivalent conditions--444}
\sum_{\ell=1}^{m}I_{\ell}(x;\dbf_1^m)
=\sum_{\ell=1}^{m-1}I_{\ell}(x;\dbf_1^{m-1})
\leq i,
\textrm{ for } 0\leq x\leq d_{\ell'+1}-1.
\eeqnarray
As $d_{\ell'+1}-1<\sum_{\ell=1}^{m}d_{\ell}$,
we see from \reqnarray{proof-OQ-LR-MRI-equivalent conditions--444}
and the definition of $B(\dbf_1^m;i)$
in \reqnarray{maximum representable integer}
that $B(\dbf_1^m;i)\geq d_{\ell'+1}-1$.

\bappendix{Proof of \rlemma{OQ-LR-MRI-general}}{OQ-LR-MRI-general}

In this lemma, we have $\dbf_1^m\in \Acal_m$, $m\geq 1$, and $i\geq 1$.

(i) In \rlemma{OQ-LR-MRI-general}(i),
we have $d_{\ell'+1}=\min\{d_{\ell'+1},d_{\ell'+2},\ldots,d_m\}$ for some $1\leq \ell'\leq m-1$.
Suppose that $B(\dbf_1^m;i)<d_{\ell'+1}-1$ or $B(\dbf_1^m;i)<d_{\ell'+1}$
or $B(\dbf_1^{m-1};i)<d_{\ell'+1}-1$
(note that these three conditions are equivalent by \rlemma{OQ-LR-MRI-equivalent conditions}).
For $0\leq x\leq B(\dbf_1^{m-1};i)+1$,
it is clear from $\dbf_1^m\in \Acal_m$,
$0\leq x\leq B(\dbf_1^{m-1};i)+1<d_{\ell'+1}=\min\{d_{\ell'+1},d_{\ell'+2},\ldots,d_m\}$,
$1\leq \ell'\leq m-1$, and \rlemma{sum of C-transform}(i) that
\beqnarray{proof-OQ-LR-MRI-general-(i)-111}
\sum_{\ell=1}^{m}I_{\ell}(x;\dbf_1^m)=\sum_{\ell=1}^{m-1}I_{\ell}(x;\dbf_1^{m-1})
=\cdots=\sum_{\ell=1}^{\ell'}I_{\ell}(x;\dbf_1^{\ell'}),
\textrm{ for } 0\leq x\leq B(\dbf_1^{m-1};i)+1.
\eeqnarray

Let $\ell'\leq m'\leq m$.
Note that we have $B(\dbf_1^{m-1};i)<d_{\ell'+1}-1\leq \sum_{\ell=1}^{\ell'}d_{\ell}
\leq \sum_{\ell=1}^{m-1}d_{\ell}$ (as $\dbf_1^m\in \Acal_m$ and $\ell'\leq m-1$).
It then follows from \reqnarray{proof-OQ-LR-MRI-general-(i)-111},
$B(\dbf_1^{m-1};i)<\sum_{\ell=1}^{m-1}d_{\ell}$,
and the definition of $B(\dbf_1^{m-1};i)$
in \reqnarray{maximum representable integer} that
\beqnarray{}
\alignspace
\sum_{\ell=1}^{m'}I_{\ell}(x;\dbf_1^{m'})=
\sum_{\ell=1}^{m-1}I_{\ell}(x;\dbf_1^{m-1})\leq i,
\textrm{ for } 0\leq x\leq B(\dbf_1^{m-1};i),
\label{eqn:proof-OQ-LR-MRI-general-(i)-222} \\
\alignspace
\sum_{\ell=1}^{m'}I_{\ell}(x;\dbf_1^{m'})=
\sum_{\ell=1}^{m-1}I_{\ell}(x;\dbf_1^{m-1})>i,
\textrm{ for } x=B(\dbf_1^{m-1};i)+1.
\label{eqn:proof-OQ-LR-MRI-general-(i)-333}
\eeqnarray
Also note that we have $B(\dbf_1^{m-1};i)<d_{\ell'+1}-1\leq \sum_{\ell=1}^{\ell'}d_{\ell}
\leq \sum_{\ell=1}^{m'}d_{\ell}$ (as $\dbf_1^m\in \Acal_m$ and $\ell'\leq m'$).
Therefore, we see from
\reqnarray{proof-OQ-LR-MRI-general-(i)-222},
\reqnarray{proof-OQ-LR-MRI-general-(i)-333},
$B(\dbf_1^{m-1};i)<\sum_{\ell=1}^{m'}d_{\ell}$,
and the definition of $B(\dbf_1^{m'};i)$
in \reqnarray{maximum representable integer} that
\beqnarray{}
B(\dbf_1^{m'};i)=B(\dbf_1^{m-1};i), \textrm{ for } \ell'\leq m'\leq m, \nn
\eeqnarray
which is the desired result that $B(\dbf_1^m;i)=B(\dbf_1^{m-1};i)=\cdots=B(\dbf_1^{\ell'};i)$
in \reqnarray{OQ-LR-MRI-general-1}.

(ii) Suppose that $B(\dbf_1^m;i)\geq d_m-1$ or $B(\dbf_1^m;i)\geq d_m$ or $B(\dbf_1^{m-1};i)\geq d_m-1$
(note that these three conditions are equivalent by \rlemma{OQ-LR-MRI-equivalent conditions}).
If $i\geq m$, then we have
\beqnarray{proof-OQ-LR-MRI-general-(ii)-111}
B(\dbf_1^m;i)=\sum_{\ell=1}^{m}d_{\ell}
\textrm{ and } B(\dbf_1^{m-1};i-1)=\sum_{\ell=1}^{m-1}d_{\ell}.
\eeqnarray
It is clear from \reqnarray{proof-OQ-LR-MRI-general-(ii)-111}
that $B(\dbf_1^m;i)=d_m+B(\dbf_1^{m-1};i-1)$,
which is the desired result in \reqnarray{OQ-LR-MRI-general-2}.

On the other hand, if $1\leq i\leq m-1$,
then $i-1<m-1$ and hence we have
\beqnarray{proof-OQ-LR-MRI-general-(ii)-222}
B(\dbf_1^{m-1};i-1)<\sum_{\ell=1}^{m-1}d_{\ell}.
\eeqnarray
For $0\leq x\leq d_m-1$,
it is clear from $x\leq d_m-1<d_m\leq B(\dbf_1^m;i)$
and the definition of $B(\dbf_1^m;i)$
in \reqnarray{maximum representable integer} that
\beqnarray{proof-OQ-LR-MRI-general-(ii)-333}
\sum_{\ell=1}^{m}I_{\ell}(x;\dbf_1^m)\leq i,
\textrm{ for } 0\leq x\leq d_m-1.
\eeqnarray
For $d_m\leq x\leq d_m+B(\dbf_1^{m-1};i-1)+1$,
we have from \reqnarray{proof-OQ-LR-MRI-general-(ii)-222}
that $d_m\leq x\leq d_m+B(\dbf_1^{m-1};i-1)+1
\leq d_m+\sum_{\ell=1}^{m-1}d_{\ell}=\sum_{\ell=1}^{m}d_{\ell}$,
and it is then clear from $\dbf_1^m\in \Acal_m$,
$d_m\leq x\leq \sum_{\ell=1}^{m}d_{\ell}$,
and \rlemma{sum of C-transform}(ii)
(with $\ell'=m-1$ in \rlemma{sum of C-transform}(ii)) that
\beqnarray{proof-OQ-LR-MRI-general-(ii)-444}
\sum_{\ell=1}^{m}I_{\ell}(x;\dbf_1^m)
=\sum_{\ell=1}^{m-1}I_{\ell}(x-d_m;\dbf_1^{m-1})+1,
\textrm{ for } d_m\leq x\leq d_m+B(\dbf_1^{m-1};i-1)+1.
\eeqnarray
From \reqnarray{proof-OQ-LR-MRI-general-(ii)-444},
$B(\dbf_1^{m-1};i-1)<\sum_{\ell=1}^{m-1}d_{\ell}$
in \reqnarray{proof-OQ-LR-MRI-general-(ii)-222},
and the definition of $B(\dbf_1^{m-1};i-1)$
in \reqnarray{maximum representable integer}, we have
\beqnarray{}
\alignspace
\sum_{\ell=1}^{m}I_{\ell}(x;\dbf_1^m)
=\sum_{\ell=1}^{m-1}I_{\ell}(x-d_m;\dbf_1^{m-1})+1\leq (i-1)+1=i, \nn\\
\alignspace \hspace*{1.8in}
\textrm{ for } d_m\leq x\leq d_m+B(\dbf_1^{m-1};i-1),
\label{eqn:proof-OQ-LR-MRI-general-(ii)-555} \\
\alignspace
\sum_{\ell=1}^{m}I_{\ell}(x;\dbf_1^m)
=\sum_{\ell=1}^{m-1}I_{\ell}(x-d_m;\dbf_1^{m-1})+1>(i-1)+1=i, \nn\\
\alignspace \hspace*{1.8in}
\textrm{ for } x=d_m+B(\dbf_1^{m-1};i-1)+1.
\label{eqn:proof-OQ-LR-MRI-general-(ii)-666}
\eeqnarray
Therefore, it follows from \reqnarray{proof-OQ-LR-MRI-general-(ii)-333},
\reqnarray{proof-OQ-LR-MRI-general-(ii)-555},
\reqnarray{proof-OQ-LR-MRI-general-(ii)-666},
$d_m+B(\dbf_1^{m-1};i-1)<d_m+\sum_{\ell=1}^{m-1}d_{\ell}=\sum_{\ell=1}^{m}d_{\ell}$,
and the definition of $B(\dbf_1^m;i)$
in \reqnarray{maximum representable integer} that
\beqnarray{}
B(\dbf_1^m;i)=d_m+B(\dbf_1^{m-1};i-1), \nn
\eeqnarray
which is the desired result in \reqnarray{OQ-LR-MRI-general-2}.

(iii) In \rlemma{OQ-LR-MRI-general}(iii), we have $d_1\leq d_2\leq \cdots\leq d_m$.
If $B(\dbf_1^m;i)\geq d_m$,
then we see from the definition of $\ell'$
in \reqnarray{OQ-LR-MRI-general-3} that $\ell'=m$.
As such, it follows from $\dbf_1^m\in \Acal_m$, $m\geq 1$, $i\geq 1$, $B(\dbf_1^m;i)\geq d_m$,
\rlemma{OQ-LR-MRI-general}(ii), and $\ell'=m$ that
\beqnarray{}
B(\dbf_1^m;i)=d_m+B(\dbf_1^{m-1};i-1)=d_{\ell'}+B(\dbf_1^{\ell'-1};i-1), \nn
\eeqnarray
which is the desired result in \reqnarray{OQ-LR-MRI-general-4}.

On the other hand, if $B(\dbf_1^m;i)<d_m$,
then we see from $d_1=1$, $d_1\leq d_2\leq \cdots\leq d_m$,
and the definition of $\ell'$ in \reqnarray{OQ-LR-MRI-general-3} that
$\ell'$ is the unique integer in $\{1,2,\ldots,m-1\}$ such that
\beqnarray{proof-OQ-LR-MRI-general-(iii)-111}
d_{\ell'}\leq B(\dbf_1^m;i)<d_{\ell'+1}.
\eeqnarray
Thus, it follows from $\dbf_1^m\in \Acal_m$, $m\geq 1$, $i\geq 1$,
$d_{\ell'+1}=\min\{d_{\ell'+1},d_{\ell'+2},\ldots,d_m\}$,
$B(\dbf_1^m;i)<d_{\ell'+1}$ in \reqnarray{proof-OQ-LR-MRI-general-(iii)-111},
$1\leq \ell'\leq m-1$, and \rlemma{OQ-LR-MRI-general}(i) that
\beqnarray{proof-OQ-LR-MRI-general-(iii)-222}
B(\dbf_1^m;i)=B(\dbf_1^{m-1};i)=\cdots=B(\dbf_1^{\ell'};i).
\eeqnarray
We claim that $1\leq i\leq \ell'$.
Suppose on the contrary that $i\geq \ell'+1$.
Then we have
\beqnarray{}
B(\dbf_1^{\ell'};i)=\sum_{\ell=1}^{\ell'}d_{\ell}
<\sum_{\ell=1}^{\ell'+1}d_{\ell}=B(\dbf_1^{\ell'+1};i), \nn
\eeqnarray
contradicting to $B(\dbf_1^{\ell'};i)=B(\dbf_1^{\ell'+1};i)$
in \reqnarray{proof-OQ-LR-MRI-general-(iii)-222}.
From \reqnarray{proof-OQ-LR-MRI-general-(iii)-222}
and \reqnarray{proof-OQ-LR-MRI-general-(iii)-111},
we have $B(\dbf_1^{\ell'};i)=B(\dbf_1^m;i)\geq d_{\ell'}$.
It then follows from
$\dbf_1^{\ell'}\in \Acal_{\ell'}$ (as $\dbf_1^m\in \Acal_m$ and $\ell'<m$),
$B(\dbf_1^{\ell'};i)\geq d_{\ell'}$, $1\leq i\leq \ell'$,
and \rlemma{OQ-LR-MRI-general}(ii) that
\beqnarray{proof-OQ-LR-MRI-general-(iii)-333}
B(\dbf_1^{\ell'};i)=d_{\ell'}+B(\dbf_1^{\ell'-1};i-1).
\eeqnarray
Therefore, we have from \reqnarray{proof-OQ-LR-MRI-general-(iii)-222}
and \reqnarray{proof-OQ-LR-MRI-general-(iii)-333} that
\beqnarray{}
B(\dbf_1^m;i)=B(\dbf_1^{m-1};i)=\cdots=B(\dbf_1^{\ell'};i)=d_{\ell'}+B(\dbf_1^{\ell'-1};i-1), \nn
\eeqnarray
which is the desired result in \reqnarray{OQ-LR-MRI-general-4}.

(iv) In \rlemma{OQ-LR-MRI-general}(iv),
we have $d_{\ell'+1}=\min\{d_{\ell'+1},d_{\ell'+2},\ldots,d_m\}$ for some $1\leq \ell'\leq m-1$.
Suppose that $B(\dbf_1^{\ell'};i)<d_{\ell'+1}-1$.
For $0\leq x\leq B(\dbf_1^{\ell'};i)+1$,
we have from $\dbf_1^m\in \Acal_m$,
$0\leq x\leq B(\dbf_1^{\ell'};i)+1<d_{\ell'+1}=\min\{d_{\ell'+1},d_{\ell'+2},\ldots,d_m\}$,
$1\leq \ell'\leq m-1$, and \rlemma{sum of C-transform}(i) that
\beqnarray{proof-OQ-LR-MRI-general-(iv)-111}
\sum_{\ell=1}^{m}I_{\ell}(x;\dbf_1^m)
=\sum_{\ell=1}^{m-1}I_{\ell}(x;\dbf_1^{m-1})
=\cdots
=\sum_{\ell=1}^{\ell'}I_{\ell}(x;\dbf_1^{\ell'}),
\textrm{ for } 0\leq x\leq B(\dbf_1^{\ell'};i)+1.
\eeqnarray
Thus, it follows from \reqnarray{proof-OQ-LR-MRI-general-(iv)-111},
$B(\dbf_1^{\ell'};i)<d_{\ell'+1}-1\leq \sum_{\ell=1}^{\ell'}d_{\ell}$
(as $\dbf_1^m\in \Acal_m$ and $\ell'\leq m-1$),
and the definition of $B(\dbf_1^{\ell'};i)$
in \reqnarray{maximum representable integer} that
\beqnarray{}
\alignspace
\sum_{\ell=1}^{m}I_{\ell}(x;\dbf_1^m)
=\sum_{\ell=1}^{m-1}I_{\ell}(x;\dbf_1^{m-1})
=\cdots
=\sum_{\ell=1}^{\ell'}I_{\ell}(x;\dbf_1^{\ell'})\leq i,
\textrm{ for } 0\leq x\leq B(\dbf_1^{\ell'};i),
\label{eqn:proof-OQ-LR-MRI-general-(iv)-333} \\
\alignspace
\sum_{\ell=1}^{m}I_{\ell}(x;\dbf_1^m)
=\sum_{\ell=1}^{m-1}I_{\ell}(x;\dbf_1^{m-1})
=\cdots
=\sum_{\ell=1}^{\ell'}I_{\ell}(x;\dbf_1^{\ell'})>i,
\textrm{ for } x=B(\dbf_1^{\ell'};i)+1.
\label{eqn:proof-OQ-LR-MRI-general-(iv)-444}
\eeqnarray
Let $\ell'\leq m'\leq m$.
Then we see from
\reqnarray{proof-OQ-LR-MRI-general-(iv)-333},
\reqnarray{proof-OQ-LR-MRI-general-(iv)-444},
$B(\dbf_1^{\ell'};i)<d_{\ell'+1}-1\leq \sum_{\ell=1}^{\ell'}d_{\ell}
\leq \sum_{\ell=1}^{m'}d_{\ell}$,
and the definition of $B(\dbf_1^{m'};i)$
in \reqnarray{maximum representable integer} that
\beqnarray{}
B(\dbf_1^{m'};i)=B(\dbf_1^{\ell'};i),
\textrm{ for } \ell'\leq m'\leq m, \nn
\eeqnarray
which is the desired result that
$B(\dbf_1^{m};i)=B(\dbf_1^{m-1};i)=\cdots=B(\dbf_1^{\ell'};i)$
in \reqnarray{OQ-LR-MRI-general-5}.

\bappendix{Proof of \rlemma{OQ-LR-MRI-greedy-1}}{OQ-LR-MRI-greedy-1}

In this lemma, we have $\dbf_1^{s_i+j}\in \Acal_{s_i+j}$
for some $1\leq i\leq k-1$ and $0\leq j\leq n_{i+1}$.
For $0\leq x\leq d_{s_i+j}-1$,
we have from $\dbf_1^{s_i+j}\in \Acal_{s_i+j}$,
$x<d_{s_i+j}$, and \rlemma{sum of C-transform}(i)
(with $m=s_i+j\geq 1$ and $\ell'=m-1=s_i+j-1$ in \rlemma{sum of C-transform}(i)) that
\beqnarray{proof-OQ-LR-MRI-greedy-1-111}
\sum_{\ell=1}^{s_i+j}I_{\ell}(x;\dbf_1^{s_i+j})
=\sum_{\ell=1}^{s_i+j-1}I_{\ell}(x;\dbf_1^{s_i+j-1}),
\textrm{ for } 0\leq x\leq d_{s_i+j}-1.
\eeqnarray
If $j=0$, then we have from \reqnarray{OQ-LR-delays-greedy-1} that
\beqnarray{proof-OQ-LR-MRI-greedy-1-222}
d_{s_i+j}=d_{s_i}=d_{s_{i-1}+n_i}=B(\dbf_1^{s_{i-1}+n_i-1};i)+1
=B(\dbf_1^{s_i-1};i)+1=B(\dbf_1^{s_i+j-1};i)+1.
\eeqnarray
Thus, it follows from
\reqnarray{proof-OQ-LR-MRI-greedy-1-111},
$d_{s_i+j}-1=B(\dbf_1^{s_i+j-1};i)$
in \reqnarray{proof-OQ-LR-MRI-greedy-1-222},
and the definition of $B(\dbf_1^{s_i+j-1};i)$
in \reqnarray{maximum representable integer} that
\beqnarray{proof-OQ-LR-MRI-greedy-1-333}
\sum_{\ell=1}^{s_i+j}I_{\ell}(x;\dbf_1^{s_i+j})
=\sum_{\ell=1}^{s_i+j-1}I_{\ell}(x;\dbf_1^{s_i+j-1})\leq i,
\textrm{ for } 0\leq x\leq d_{s_i+j}-1.
\eeqnarray
On the other hand, if $1\leq j\leq n_{i+1}$,
then we have from \reqnarray{OQ-LR-delays-greedy-1} that
\beqnarray{proof-OQ-LR-MRI-greedy-1-444}
d_{s_i+j}=B(\dbf_1^{s_i+j-1};i+1)+1.
\eeqnarray
(We note that $d_{s_i+j}=B(\dbf_1^{s_i+j-1};i)+1$ in \reqnarray{proof-OQ-LR-MRI-greedy-1-222}
is different from $d_{s_i+j}=B(\dbf_1^{s_i+j-1};i+1)+1$ in \reqnarray{proof-OQ-LR-MRI-greedy-1-444},
and that is why we need to consider the two cases $j=0$ and $1\leq j\leq n_{i+1}$ separately.)
Thus, it follows from
\reqnarray{proof-OQ-LR-MRI-greedy-1-111},
$d_{s_i+j}-1=B(\dbf_1^{s_i+j-1};i+1)$
in \reqnarray{proof-OQ-LR-MRI-greedy-1-444},
and the definition of $B(\dbf_1^{s_i+j-1};i+1)$
in \reqnarray{maximum representable integer} that
\beqnarray{proof-OQ-LR-MRI-greedy-1-555}
\sum_{\ell=1}^{s_i+j}I_{\ell}(x;\dbf_1^{s_i+j})
=\sum_{\ell=1}^{s_i+j-1}I_{\ell}(x;\dbf_1^{s_i+j-1})\leq i+1,
\textrm{ for } 0\leq x\leq d_{s_i+j}-1.
\eeqnarray

Furthermore, for $x=d_{s_i+j}$, it is clear from \reqnarray{C-transform} that
$I_{s_i+j}(x;\dbf_1^{s_i+j})=1$ and
$I_{\ell}(x;\dbf_1^{s_i+j})=0$ for $\ell=s_i+j-1,s_i+j-2,\ldots,1$.
It follows that
\beqnarray{proof-OQ-LR-MRI-greedy-1-666}
\sum_{\ell=1}^{s_i+j}I_{\ell}(x;\dbf_1^{s_i+j})=1\leq i+1,
\textrm{ for } x=d_{s_i+j}.
\eeqnarray
As such, we see from
\reqnarray{proof-OQ-LR-MRI-greedy-1-333},
\reqnarray{proof-OQ-LR-MRI-greedy-1-555},
\reqnarray{proof-OQ-LR-MRI-greedy-1-666},
$d_{s_i+j}\leq \sum_{\ell=1}^{s_i+j}d_{\ell}$,
and the definition of $B(\dbf_1^{s_i+j};i+1)$
in \reqnarray{maximum representable integer} that
\beqnarray{proof-OQ-LR-MRI-greedy-1-777}
B(\dbf_1^{s_i+j};i+1)\geq d_{s_i+j}.
\eeqnarray

Since $\nbf_1^k\in \Ncal_{M,k}$,
we have $n_1\geq 2$ and $n_2,n_3,\ldots,n_k\geq 1$.
It is then clear from $i\geq 1$ and $j\geq 0$ that $s_i+j\geq s_i\geq i+1$.
Therefore, we have from $\dbf_1^{s_i+j}\in \Acal_{s_i+j}$, $B(\dbf_1^{s_i+j};i+1)\geq d_{s_i+j}$
in \reqnarray{proof-OQ-LR-MRI-greedy-1-777},
$1<i+1\leq s_i+j$, and \rlemma{OQ-LR-MRI-general}(ii) that
\beqnarray{}
B(\dbf_1^{s_i+j};i+1)=d_{s_i+j}+B(\dbf_1^{s_i+j-1};i), \nn
\eeqnarray
which is the desired result in \reqnarray{OQ-LR-MRI-greedy-1}.

\bappendix{Proof of \rlemma{OQ-LR-MRI-greedy-2}}{OQ-LR-MRI-greedy-2}

In this lemma, the sequence $\dbf_1^{s_i+j}=(d_1,d_2,\ldots,d_{s_i+j})$
is given by \reqnarray{OQ-LR-delays-greedy-2} and \reqnarray{OQ-LR-delays-greedy-3}
for some $1\leq i\leq k-1$ and $1\leq j\leq n_{i+1}$.

(i) To prove $\dbf_1^{s_i+j}\in \Bcal_{s_i+j}$,
we need to show that $d_1=1$ and $d_{\ell}\leq d_{\ell+1}\leq 2d_{\ell}$ for $\ell=1,2,\ldots,s_i+j-1$.
From \reqnarray{OQ-LR-delays-greedy-2}, we have $d_{\ell}=\ell$ for $\ell=1,2,\ldots,s_1$, and it is easy
to verify that $d_1=1$ and $d_{\ell}\leq d_{\ell+1}\leq 2d_{\ell}$ for $\ell=1,2,\ldots,s_1-1$.
In the following two possible cases, we show that $d_{\ell}\leq d_{\ell+1}\leq 2d_{\ell}$
for $\ell=s_1,s_1+1,\ldots,s_i+j-1$, and the proof is completed.

\emph{Case 1: $\ell=s_{i'}$, where $1\leq i'\leq i$.}
In this case, we have from \reqnarray{OQ-LR-delays-greedy-3} that
$d_{\ell+1}=d_{s_{i'}+1}=2d_{s_{i'}}=2d_{\ell}$.
It is clear that $d_{\ell}\leq d_{\ell+1}\leq 2d_{\ell}$.

\emph{Case 2: $\ell=s_{i'}+j'$, where $1\leq i'\leq i-1$ and $1\leq j'\leq n_{i'+1}-1$
or $i'=i$ and $1\leq j'\leq j-1$.}
In this case, we have $1\leq i'\leq i-1$ and $2\leq j'+1\leq n_{i'+1}$
or $i'=i$ and $2\leq j'+1\leq j\leq n_{i+1}$.
Thus, we have from \reqnarray{OQ-LR-delays-greedy-3} that
\beqnarray{}
d_{\ell+1}-d_{\ell}
\aligneq d_{s_{i'}+j'+1}-d_{s_{i'}+j'} \nn\\
\aligneq (2d_{s_{i'}}+j'(d_{s_1}+d_{s_2}+\cdots+d_{s_{i'}}+1)) \nn\\
\alignspace -(2d_{s_{i'}}+(j'-1)(d_{s_1}+d_{s_2}+\cdots+d_{s_{i'}}+1)) \nn\\
\aligneq d_{s_1}+d_{s_2}+\cdots+d_{s_{i'}}+1\geq 0,
\label{eqn:proof-OQ-LR-MRI-greedy-2-(i)-111} \\
d_{\ell+1}-2d_{\ell}
\aligneq d_{s_{i'}+j'+1}-2d_{s_{i'}+j'} \nn\\
\aligneq (2d_{s_{i'}}+j'(d_{s_1}+d_{s_2}+\cdots+d_{s_{i'}}+1)) \nn\\
\alignspace -2(2d_{s_{i'}}+(j'-1)(d_{s_1}+d_{s_2}+\cdots+d_{s_{i'}}+1)) \nn\\
\aligneq -2d_{s_{i'}}-(j'-2)(d_{s_1}+d_{s_2}+\cdots+d_{s_{i'}}+1).
\label{eqn:proof-OQ-LR-MRI-greedy-2-(i)-222}
\eeqnarray

If $j'=1$, then let $a=\max\{1\leq i''\leq i':n_{i''}\geq 2\}$
(note that $a$ is well defined as we have from $\nbf_1^k\in \Ncal_{M,k}$ that $n_1\geq 2$).
Thus, we have from \reqnarray{proof-OQ-LR-MRI-greedy-2-(i)-222}, $j'=1$,
\reqnarray{OQ-LR-delays-greedy-3}, $n_{i'}=n_{i'-1}=\cdots=n_{a+1}=1$, and $n_{a}\geq 2$ that
\beqnarray{proof-OQ-LR-MRI-greedy-2-(i)-333}
d_{\ell+1}-2d_{\ell}
\aligneq -2d_{s_{i'}}-(j'-2)(d_{s_1}+d_{s_2}+\cdots+d_{s_{i'}}+1) \nn\\
\aligneq d_{s_1}+d_{s_2}+\cdots+d_{s_{i'-1}}+1-d_{s_{i'}} \nn\\
\aligneq d_{s_1}+d_{s_2}+\cdots+d_{s_{i'-1}}+1
-(2d_{s_{i'-1}}+(n_{i'}-1)(d_{s_1}+d_{s_2}+\cdots+d_{s_{i'-1}}+1)) \nn\\
\aligneq d_{s_1}+d_{s_2}+\cdots+d_{s_{i'-2}}+1-d_{s_{i'-1}} \nn\\
&\vdots& \nn\\
\aligneq d_{s_1}+d_{s_2}+\cdots+d_{s_{a-1}}+1-d_{s_a} \nn\\
\aligneq d_{s_1}+d_{s_2}+\cdots+d_{s_{a-1}}+1
-(2d_{s_{a-1}}+(n_{a}-1)(d_{s_1}+d_{s_2}+\cdots+d_{s_{a-1}}+1)) \nn\\
\alignleq -2d_{s_{a-1}}\leq 0.
\eeqnarray
On the other hand, if $j'\geq 2$,
then we have from \reqnarray{proof-OQ-LR-MRI-greedy-2-(i)-222} that
\beqnarray{proof-OQ-LR-MRI-greedy-2-(i)-444}
d_{\ell+1}-2d_{\ell}=-2d_{s_{i'}}-(j'-2)(d_{s_1}+d_{s_2}+\cdots+d_{s_{i'}}+1)
\leq -2d_{s_{i'}}\leq 0.
\eeqnarray
Therefore, it follows from \reqnarray{proof-OQ-LR-MRI-greedy-2-(i)-111},
\reqnarray{proof-OQ-LR-MRI-greedy-2-(i)-333},
and \reqnarray{proof-OQ-LR-MRI-greedy-2-(i)-444}
that $d_{\ell}\leq d_{\ell+1}\leq 2d_{\ell}$.

(ii) Note that from \rlemma{OQ-LR-MRI-greedy-2}(i),
we have $\dbf_1^{s_i+j}\in \Bcal_{s_i+j}$ and it follows that
$d_1\leq d_2\leq \cdots\leq d_{s_i+j}$.
In the following, we will show that
\beqnarray{proof-OQ-LR-MRI-greedy-2-(ii)-111}
B(\dbf_1^{s_i};i)<d_{s_i+1}-1.
\eeqnarray
As $\nbf_1^k\in \Ncal_{M,k}$,
we have $n_1\geq 2$ and $n_2,n_3,\ldots,n_k\geq 1$,
and it is clear from $i\geq 1$ and $j\geq 1$ that $s_i+j>s_i>i\geq 1$.
It then follows from $\dbf_1^{s_i+j}\in \Bcal_{s_i+j}\subseteq \Acal_{s_i+j}$,
$s_i+j>1$, $d_{s_i+1}=\min\{d_{s_i+1},d_{s_i+2},\ldots,d_{s_i+j}\}$,
$B(\dbf_1^{s_i};i)<d_{s_i+1}-1$
in \reqnarray{proof-OQ-LR-MRI-greedy-2-(ii)-111},
$1<s_i\leq s_i+j-1$, \rlemma{OQ-LR-MRI-general}(iv)
(with $m=s_i+j$ and $\ell'=s_i$ in \rlemma{OQ-LR-MRI-general}(iv)),
and $B(\dbf_1^{s_i};i)=d_{s_1}+d_{s_2}+\cdots+d_{s_i}$
in \reqnarray{OQ-LR-MRI at break point-greedy} that
\beqnarray{}
B(\dbf_1^{s_i+j};i)=B(\dbf_1^{s_i+j-1};i)=\cdots=B(\dbf_1^{s_i};i)
=d_{s_1}+d_{s_2}+\cdots+d_{s_i}, \nn
\eeqnarray
which is the desired result in \reqnarray{OQ-LR-MRI-greedy-2}.

To prove \reqnarray{proof-OQ-LR-MRI-greedy-2-(ii)-111},
let $a=\max\{1\leq i'\leq i:n_{i'}\geq 2\}$
(note that $a$ is well defined as we have from $\nbf_1^k\in \Ncal_{M,k}$ that $n_1\geq 2$).
It then follows from \reqnarray{OQ-LR-MRI at break point-greedy},
\reqnarray{OQ-LR-delays-greedy-3}, and $n_i=n_{i-1}=\cdots=n_{a+1}=1$ that
\beqnarray{proof-OQ-LR-MRI-greedy-2-(ii)-222}
\alignspace \hspace*{-0.16in} B(\dbf_1^{s_i};i)+1-d_{s_i+1} \nn\\
\aligneq d_{s_1}+d_{s_2}+\cdots+d_{s_i}+1-2d_{s_i} \nn\\
\aligneq d_{s_1}+d_{s_2}+\cdots+d_{s_{i-1}}+1-d_{s_i} \nn\\
\aligneq d_{s_1}+d_{s_2}+\cdots+d_{s_{i-1}}+1
-(2d_{s_{i-1}}+(n_i-1)(d_{s_1}+d_{s_2}+\cdots+d_{s_{i-1}}+1)) \nn\\
\aligneq d_{s_1}+d_{s_2}+\cdots+d_{s_{i-2}}+1-d_{s_{i-1}} \nn\\
&\vdots& \nn\\
\aligneq d_{s_1}+d_{s_2}+\cdots+d_{s_{a-1}}+1-d_{s_a}.
\eeqnarray
If $a=1$, then we see from \reqnarray{proof-OQ-LR-MRI-greedy-2-(ii)-222}
and $d_{s_1}=s_1=n_1\geq 2$ in \reqnarray{OQ-LR-delays-greedy-2} that
\beqnarray{proof-OQ-LR-MRI-greedy-2-(ii)-333}
B(\dbf_1^{s_i};i)+1-d_{s_i+1}
=d_{s_1}+d_{s_2}+\cdots+d_{s_{a-1}}+1-d_{s_a}
=1-d_{s_1}<0.
\eeqnarray
On the other hand, if $a\geq 2$,
then we see from \reqnarray{proof-OQ-LR-MRI-greedy-2-(ii)-222},
\reqnarray{OQ-LR-delays-greedy-3}, $n_a\geq 2$, and $d_{s_{a-1}}\geq d_{s_1}>0$ that
\beqnarray{proof-OQ-LR-MRI-greedy-2-(ii)-444}
\alignspace B(\dbf_1^{s_i};i)+1-d_{s_i+1} \nn\\
\alignspace =d_{s_1}+d_{s_2}+\cdots+d_{s_{a-1}}+1-d_{s_a} \nn\\
\alignspace =d_{s_1}+d_{s_2}+\cdots+d_{s_{a-1}}+1
-(2d_{s_{a-1}}+(n_a-1)(d_{s_1}+d_{s_2}+\cdots+d_{s_{a-1}}+1)) \nn\\
\alignspace \leq -2d_{s_{a-1}}< 0.
\eeqnarray
As such, \reqnarray{proof-OQ-LR-MRI-greedy-2-(ii)-111}
follows from \reqnarray{proof-OQ-LR-MRI-greedy-2-(ii)-333}
and \reqnarray{proof-OQ-LR-MRI-greedy-2-(ii)-444}.

\bappendix{Proof of \rlemma{OQ-LR-MRI-monotone}}{OQ-LR-MRI-monotone}

In this lemmas, we have $\dbf_1^m\in \Acal_m$, $m\geq 1$, and $0\leq i\leq i'$.

(i) \rlemma{OQ-LR-MRI-monotone}(i) follows from the trivial fact that
if there are no more than $i$ 1-entries in the $\Ccal$-transform
of a nonnegative integer, then there must be no more than $i'$ 1-entries
in its $\Ccal$-transform (as we have $i\leq i'$).

(ii) $(\Leftarrow)$
Suppose that $i\geq m$.
Then it is clear that $B(\dbf_1^m;i)=\sum_{\ell=1}^{m}d_{\ell}=B(\dbf_1^m;i+1)$.

$(\Rightarrow)$
Suppose that $B(\dbf_1^m;i)=B(\dbf_1^m;i+1)$.
We will prove that $i\geq m$ by contradiction.
Assume on the contrary that $i<m$.
Let $x=B(\dbf_1^m;i)$ and let
$y_{\ell}=x-\sum_{\ell'=\ell+1}^{m}I_{\ell'}(x;\dbf_1^m)d_{\ell'}$
for $1\leq \ell\leq m$
(note that we have from \reqnarray{C-transform}
that $I_{\ell}(x;\dbf_1^m)=1$ if $y_{\ell}\geq d_{\ell}$
and $I_{\ell}(x;\dbf_1^m)=0$ if $y_{\ell}<d_{\ell}$).
From $x=B(\dbf_1^m;i)$, the definition of $B(\dbf_1^m;i)$
in \reqnarray{maximum representable integer}, and $i<m$,
we see that
\beqnarray{proof-OQ-LR-MRI-monotone-111}
\sum_{\ell=1}^{m}I_{\ell}(x;\dbf_1^m)\leq i<m.
\eeqnarray
It follows from $0\leq I_{\ell}(x;\dbf_1^m)\leq 1$
and $\sum_{\ell=1}^{m}I_{\ell}(x;\dbf_1^m)<m$
in \reqnarray{proof-OQ-LR-MRI-monotone-111} that
there exists $1\leq \ell\leq m$ such that $I_{\ell}(x;\dbf_1^m)=0$.
Let $\ell_1$ be the smallest such positive integer,
i.e., $\ell_1=\min\{1\leq \ell\leq m: I_{\ell}(x;\dbf_1^m)=0\}$.
Clearly, we have $I_1(x;\dbf_1^m)=I_2(x;\dbf_1^m)=\cdots=I_{\ell_1-1}(x;\dbf_1^m)=1$
and $I_{\ell_1}(x;\dbf_1^m)=0$.
From \reqnarray{C-transform} and $I_{\ell_1}(x;\dbf_1^m)=0$,
it is easy to see that $y_{\ell_1}<d_{\ell_1}$.
If $\ell_1=1$, then we see from $y_{\ell_1}\geq 0$, $d_1=1$ (as $\dbf_1^m\in \Acal_m$),
and $y_{\ell_1}<d_{\ell_1}$ that $0\leq y_{\ell_1}<d_{\ell_1}=d_1=1$,
and hence
\beqnarray{proof-OQ-LR-MRI-monotone-222}
y_{\ell_1}=0=d_{\ell_1}-1.
\eeqnarray
On the other hand, if $2\leq \ell_1\leq m$,
then we see from $\dbf_1^m\in \Acal_m$,
$I_1(x;\dbf_1^m)=I_2(x;\dbf_1^m)=\cdots=I_{\ell_1-1}(x;\dbf_1^m)=1$,
$I_{\ell_1}(x;\dbf_1^m)=0$,
and $y_{\ell_1}=x-\sum_{\ell=\ell_1+1}^{m}I_{\ell}(x;\dbf_1^m)d_{\ell}
=\sum_{\ell=1}^{m}I_{\ell}(x;\dbf_1^m)d_{\ell}
-\sum_{\ell=\ell_1+1}^{m}I_{\ell}(x;\dbf_1^m)d_{\ell}
=\sum_{\ell=1}^{\ell_1}I_{\ell}(x;\dbf_1^m)d_{\ell}$ that
\beqnarray{proof-OQ-LR-MRI-monotone-333}
d_{\ell_1}
\leq \sum_{\ell=1}^{\ell_1-1}d_{\ell}+1
=\sum_{\ell=1}^{\ell_1-1}1\cdot d_{\ell}+0\cdot d_{\ell_1}+1
=\sum_{\ell=1}^{\ell_1}I_{\ell}(x;\dbf_1^m)d_{\ell}+1
=y_{\ell_1}+1.
\eeqnarray
As we also have $y_{\ell_1}<d_{\ell_1}$,
it follows from $y_{\ell_1}\geq d_{\ell_1}-1$
in \reqnarray{proof-OQ-LR-MRI-monotone-333} that
\beqnarray{proof-OQ-LR-MRI-monotone-444}
y_{\ell_1}=d_{\ell_1}-1.
\eeqnarray

From \reqnarray{proof-OQ-LR-MRI-monotone-222}
and \reqnarray{proof-OQ-LR-MRI-monotone-444},
we see that there exists $1\leq \ell\leq m$
such that $y_{\ell}=d_{\ell}-1$.
Let $\ell_2$ be the largest such positive integer,
i.e., $\ell_2=\max\{1\leq \ell\leq m: y_{\ell}=d_{\ell}-1\}$.
As we assume that $i<m$, we have $x=B(\dbf_1^m;i)<\sum_{\ell=1}^{m}d_{\ell}$
and $x+1\leq \sum_{\ell=1}^{m}d_{\ell}$.
In the following, we show that
\beqnarray{}
\alignspace
I_{\ell}(x+1;\dbf_1^m)=I_{\ell}(x;\dbf_1^m),
\textrm{ for } \ell=m,m-1,\ldots,\ell_2+1,
\label{eqn:proof-OQ-LR-MRI-monotone-555} \\
\alignspace
I_{\ell_2}(x+1;\dbf_1^m)=1.
\label{eqn:proof-OQ-LR-MRI-monotone-666} \\
\alignspace
I_{\ell}(x+1;\dbf_1^m)=0,
\textrm{ for } \ell=\ell_2-1,\ell_2-2,\ldots,1,
\label{eqn:proof-OQ-LR-MRI-monotone-777}
\eeqnarray
It then follows from \reqnarray{proof-OQ-LR-MRI-monotone-555}--\reqnarray{proof-OQ-LR-MRI-monotone-777},
$x=B(\dbf_1^m;i)$, and the definition of $B(\dbf_1^m;i)$
in \reqnarray{maximum representable integer} that
\beqnarray{proof-OQ-LR-MRI-monotone-888}
\sum_{\ell=1}^{m}I_{\ell}(x+1;\dbf_1^m)
=\sum_{\ell=\ell_2+1}^{m}I_{\ell}(x;\dbf_1^m)+1
\leq \sum_{\ell=1}^{m}I_{\ell}(x;\dbf_1^m)+1
\leq i+1.
\eeqnarray
Since we also have from the definition of $B(\dbf_1^m;i)$
in \reqnarray{maximum representable integer} that
$\sum_{\ell=1}^{m}I_{\ell}(x';\dbf_1^m)\leq i<i+1$
for $0\leq x'\leq x=B(\dbf_1^m;i)$,
it is immediate from \reqnarray{proof-OQ-LR-MRI-monotone-888} that
\beqnarray{proof-OQ-LR-MRI-monotone-999}
\sum_{\ell=1}^{m}I_{\ell}(x';\dbf_1^m)\leq i+1,
\textrm{ for } 0\leq x'\leq x+1.
\eeqnarray
As such, we see from \reqnarray{proof-OQ-LR-MRI-monotone-999},
$x+1\leq \sum_{\ell=1}^{m}d_{\ell}$,
and the definition of $B(\dbf_1^m;i+1)$
in \reqnarray{maximum representable integer} that
$B(\dbf_1^m;i+1)\geq x+1=B(\dbf_1^m;i)+1$,
and we have reached a contradiction to $B(\dbf_1^m;i)=B(\dbf_1^m;i+1)$.

To prove \reqnarray{proof-OQ-LR-MRI-monotone-555}--\reqnarray{proof-OQ-LR-MRI-monotone-777},
let $z_{\ell}=(x+1)-\sum_{\ell'=\ell+1}^{m}I_{\ell'}(x+1;\dbf_1^m)d_{\ell'}$
for $1\leq \ell \leq m$
(note that we have from \reqnarray{C-transform}
that $I_{\ell}(x+1;\dbf_1^m)=1$ if $z_{\ell}\geq d_{\ell}$
and $I_{\ell}(x+1;\dbf_1^m)=0$ if $z_{\ell}<d_{\ell}$).
In the case that $\ell_2=m$, we have $y_m=d_m-1$.
As $x+1=y_m+1=d_m$, it is clear from \reqnarray{C-transform} that
\beqnarray{}
\alignspace
I_m(x+1;\dbf_1^m)=1, \nn\\
\alignspace
I_{\ell}(x+1;\dbf_1^m)=0, \textrm{ for } \ell=m-1,m-2,\ldots,1. \nn
\eeqnarray
Thus, \reqnarray{proof-OQ-LR-MRI-monotone-555}--\reqnarray{proof-OQ-LR-MRI-monotone-777}
are proved in this case.
On the other hand, in the case that $1\leq \ell_2\leq m-1$,
we have $y_{\ell_2}=d_{\ell_2}-1$.
In this case, we first show \reqnarray{proof-OQ-LR-MRI-monotone-555}
by induction on $\ell$.
If $y_m\geq d_m$, then we have $z_m=x+1=y_m+1>d_m$,
and it is clear from \reqnarray{C-transform} that
\beqnarray{proof-OQ-LR-MRI-monotone-aaa}
I_m(x+1;\dbf_1^m)=I_m(x;\dbf_1^m)=1.
\eeqnarray
Otherwise, if $y_m<d_m$, then we see from $m>\ell_2$
and the definition of $\ell_2$ that $y_m<d_m-1$,
and hence we have $z_m=x+1=y_m+1<d_m$.
It is then clear from \reqnarray{C-transform} that
\beqnarray{proof-OQ-LR-MRI-monotone-bbb}
I_m(x+1;\dbf_1^m)=I_m(x;\dbf_1^m)=0.
\eeqnarray
As such, we have from \reqnarray{proof-OQ-LR-MRI-monotone-aaa}
and \reqnarray{proof-OQ-LR-MRI-monotone-bbb} that $I_m(x+1;\dbf_1^m)=I_m(x;\dbf_1^m)$.
Assume as the induction hypothesis that
$I_m(x+1;\dbf_1^m)=I_m(x;\dbf_1^m),I_{m-1}(x+1;\dbf_1^m)=I_{m-1}(x;\dbf_1^m),
\ldots,I_{\ell+1}(x+1;\dbf_1^m)=I_{\ell+1}(x;\dbf_1^m)$
for some $\ell_2+1\leq \ell\leq m-1$.
If $y_{\ell}\geq d_{\ell}$,
then we see from $z_{\ell}=(x+1)-\sum_{\ell'=\ell+1}^{m}I_{\ell'}(x+1;\dbf_1^m)d_{\ell'}$,
the induction hypothesis,
and $y_{\ell}=x-\sum_{\ell'=\ell+1}^{m}I_{\ell'}(x;\dbf_1^m)d_{\ell'}$ that
\beqnarray{proof-OQ-LR-MRI-monotone-ccc}
z_{\ell}=(x+1)-\sum_{\ell'=\ell+1}^{m}I_{\ell'}(x+1;\dbf_1^m)d_{\ell'}
=(x+1)-\sum_{\ell'=\ell+1}^{m}I_{\ell'}(x;\dbf_1^m)d_{\ell'}
=y_{\ell}+1>d_{\ell}.
\eeqnarray
It is clear from \reqnarray{proof-OQ-LR-MRI-monotone-ccc}
and \reqnarray{C-transform} that
\beqnarray{proof-OQ-LR-MRI-monotone-ddd}
I_{\ell}(x+1;\dbf_1^m)=I_{\ell}(x;\dbf_1^m)=1.
\eeqnarray
Otherwise, if $y_{\ell}<d_{\ell}$, then we see from $\ell>\ell_2$
and the definition of $\ell_2$ that $y_{\ell}<d_{\ell}-1$,
and hence we have from $z_{\ell}=(x+1)-\sum_{\ell'=\ell+1}^{m}I_{\ell'}(x+1;\dbf_1^m)d_{\ell'}$,
the induction hypothesis,
and $y_{\ell}=x-\sum_{\ell'=\ell+1}^{m}I_{\ell'}(x;\dbf_1^m)d_{\ell'}$ that
\beqnarray{proof-OQ-LR-MRI-monotone-eee}
z_{\ell}=(x+1)-\sum_{\ell'=\ell+1}^{m}I_{\ell'}(x+1;\dbf_1^m)d_{\ell'}
=(x+1)-\sum_{\ell'=\ell+1}^{m}I_{\ell'}(x;\dbf_1^m)d_{\ell'}
=y_{\ell}+1<d_{\ell}.
\eeqnarray
It is also clear from \reqnarray{proof-OQ-LR-MRI-monotone-eee}
and \reqnarray{C-transform} that
\beqnarray{proof-OQ-LR-MRI-monotone-fff}
I_{\ell}(x+1;\dbf_1^m)=I_{\ell}(x;\dbf_1^m)=0.
\eeqnarray
The induction is completed by combining \reqnarray{proof-OQ-LR-MRI-monotone-ddd}
and \reqnarray{proof-OQ-LR-MRI-monotone-fff}.
Thus, \reqnarray{proof-OQ-LR-MRI-monotone-555} is proved in this case.
From $z_{\ell_2}=(x+1)-\sum_{\ell'=\ell_2+1}^{m}I_{\ell'}(x+1;\dbf_1^m)d_{\ell'}$,
\reqnarray{proof-OQ-LR-MRI-monotone-555},
$y_{\ell_2}=x-\sum_{\ell'=\ell_2+1}^{m}I_{\ell'}(x;\dbf_1^m)d_{\ell'}$,
and $y_{\ell_2}=d_{\ell_2}-1$, we see that
\beqnarray{proof-OQ-LR-MRI-monotone-ggg}
z_{\ell_2}=(x+1)-\sum_{\ell'=\ell_2+1}^{m}I_{\ell'}(x+1;\dbf_1^m)d_{\ell'}
=(x+1)-\sum_{\ell'=\ell_2+1}^{m}I_{\ell'}(x;\dbf_1^m)d_{\ell'}
=y_{\ell_2}+1=d_{\ell_2}.
\eeqnarray
It is then clear from $z_{\ell_2}=d_{\ell_2}$
in \reqnarray{proof-OQ-LR-MRI-monotone-ggg}
and \reqnarray{C-transform} that
\beqnarray{}
\alignspace
I_{\ell_2}(x+1;\dbf_1^m)=1, \nn\\
\alignspace
I_{\ell}(x+1;\dbf_1^m)=0, \textrm{ for } \ell=\ell_2-1,\ell_2-2,\ldots,1. \nn
\eeqnarray
Therefore, \reqnarray{proof-OQ-LR-MRI-monotone-666}
and \reqnarray{proof-OQ-LR-MRI-monotone-777}
are also proved in this case.

\bappendix{Proof of \rlemma{OQ-LR-optimal-is-nondecreasing}}{OQ-LR-optimal-is-nondecreasing}

Note that in this lemma, we have $M\geq 2$, $1\leq k\leq M-1$, ${\dbf^*}_1^M\in \Acal_M$,
and $B({\dbf^*}_1^M;k)=\max_{\dbf_1^M\in \Acal_M}B(\dbf_1^M;k)$.

(i) We will show by contradiction that $B({\dbf^*}_1^M;k)\geq d_M^*$.
Assume on the contrary that $B({\dbf^*}_1^M;k)<d_M^*$.
Let $d'_{\ell}=d_{\ell}^*$ for $\ell=1,2,\ldots,M-1$ and $d'_M=B({\dbf^*}_1^M;k)+1$.
Then it follows from ${\dbf^*}_1^M\in \Acal_M$ and $B({\dbf^*}_1^M;k)<d_M^*$ that
\beqnarray{}
\alignspace
{\dbf'}_1^{M-1}={\dbf^*}_1^{M-1}\in \Acal_{M-1},
\label{eqn:proof-OQ-LR-optimal-is-nondecreasing-(i)-111} \\
\alignspace
1\leq d'_M=B({\dbf^*}_1^M;k)+1\leq d_M^*\leq \sum_{\ell=1}^{M-1}d_{\ell}^*+1
=\sum_{\ell=1}^{M-1}d'_{\ell}+1.
\label{eqn:proof-OQ-LR-optimal-is-nondecreasing-(i)-222}
\eeqnarray
From \reqnarray{proof-OQ-LR-optimal-is-nondecreasing-(i)-111}
and \reqnarray{proof-OQ-LR-optimal-is-nondecreasing-(i)-222},
we immediately see that ${\dbf'}_1^M\in \Acal_M$.
Thus, it is clear that
\beqnarray{proof-OQ-LR-optimal-is-nondecreasing-(i)-333}
B({\dbf'}_1^M;k)\leq \max_{\dbf_1^M\in \Acal_M}B(\dbf_1^M;k)=B({\dbf^*}_1^M;k).
\eeqnarray

Note that from ${\dbf^*}_1^M\in \Acal_M$, $1\leq k<M$, $B({\dbf^*}_1^M;k)<d_M^*$,
and \rlemma{OQ-LR-MRI-general}(i)
(with $m=M$, $i=k$, and $\ell'=m-1=M-1$ in \rlemma{OQ-LR-MRI-general}(i)),
we have $B({\dbf^*}_1^M;k)=B({\dbf^*}_1^{M-1};k)$.
It then follows from ${\dbf'}_1^{M-1}={\dbf^*}_1^{M-1}$,
$B({\dbf^*}_1^M;k)=B({\dbf^*}_1^{M-1};k)$,
and $d'_M=B({\dbf^*}_1^M;k)+1$ that
\beqnarray{proof-OQ-LR-optimal-is-nondecreasing-(i)-444}
B({\dbf'}_1^{M-1};k)=B({\dbf^*}_1^{M-1};k)=B({\dbf^*}_1^M;k)=d'_M-1.
\eeqnarray
As such, we have from ${\dbf'}_1^M\in \Acal_M$, $1\leq k<M$,
$B({\dbf'}_1^{M-1};k)=d'_M-1$ in \reqnarray{proof-OQ-LR-optimal-is-nondecreasing-(i)-444},
\rlemma{OQ-LR-MRI-equivalent conditions}
(with $m=M$, $i=k$, and $\ell'=m-1=M-1$ in \rlemma{OQ-LR-MRI-equivalent conditions}),
and $d'_M=B({\dbf^*}_1^M;k)+1$ that
\beqnarray{}
B({\dbf'}_1^M;k)\geq d'_M=B({\dbf^*}_1^M;k)+1>B({\dbf^*}_1^M;k), \nn
\eeqnarray
contradicting to $B({\dbf'}_1^M;k)\leq B({\dbf^*}_1^M;k)$
in \reqnarray{proof-OQ-LR-optimal-is-nondecreasing-(i)-333}.

(ii) We will show by contradiction that $d_1^*\leq d_2^*\leq \cdots\leq d_M^*$.
Assume on the contrary that there exists $2\leq \ell\leq M-1$
such that $d_{\ell}^*>d_{\ell+1}^*$
(note that as $d_1^*=1$ and $d_2^*\geq 1$, it is impossible that $d_1^*>d_2^*$).
Let $\ell_1$ be the largest such positive integer,
i.e., $\ell_1=\max\{2\leq \ell\leq M-1:d_{\ell}^*>d_{\ell+1}^*\}$.
We consider the two cases $\ell_1=M-1$ and $2\leq \ell_1\leq M-2$ separately.

\emph{Case 1: $\ell_1=M-1$.}
In this case, we see from the definition of $\ell_1$ that $d_{M-1}^*>d_M^*$.
From \rlemma{OQ-LR-optimal-is-nondecreasing}(i),
we know that $B({\dbf^*}_1^M;k)\geq d_M^*$.
Since ${\dbf^*}_1^M\in \Acal_M$ and $1\leq k\leq M-1$,
we also have $B({\dbf^*}_1^M;k)<\sum_{\ell=1}^{M}d_{\ell}^*$.
We then consider the two subcases $d_M^*\leq B({\dbf^*}_1^M;k)<d_{M-1}^*+d_M^*$
and $d_{M-1}^*+d_M^*\leq B({\dbf^*}_1^M;k)<\sum_{\ell=1}^{M}d_{\ell}^*$ separately.

\emph{Subcase 1(a): $d_M^*\leq B({\dbf^*}_1^M;k)<d_{M-1}^*+d_M^*$.}
Let $d'_{\ell}=d_{\ell}^*$ for $\ell=1,2,\ldots,M-2$,
$d'_{M-1}=d_M^*$, and $d'_M=B({\dbf^*}_1^M;k)+1$.
Then it follows from ${\dbf^*}_1^M\in \Acal_M$,
$d_{M-1}^*>d_M^*$, and $B({\dbf^*}_1^M;k)<d_{M-1}^*+d_M^*$ that
\beqnarray{}
\alignspace
{\dbf'}_1^{M-2}={\dbf^*}_1^{M-2}\in \Acal_{M-2},
\label{eqn:proof-OQ-LR-optimal-is-nondecreasing-(ii)-case-1-111} \\
\alignspace
1\leq d'_{M-1}=d_M^*<d_{M-1}^*\leq \sum_{\ell=1}^{M-2}d_{\ell}^*+1
=\sum_{\ell=1}^{M-2}d'_{\ell}+1,
\label{eqn:proof-OQ-LR-optimal-is-nondecreasing-(ii)-case-1-222} \\
\alignspace
1\leq d'_M=B({\dbf^*}_1^M;k)+1\leq d_{M-1}^*+d_M^*
\leq \left(\sum_{\ell=1}^{M-2}d_{\ell}^*+1\right)+d'_{M-1}
=\sum_{\ell=1}^{M-1}d'_{\ell}+1.
\label{eqn:proof-OQ-LR-optimal-is-nondecreasing-(ii)-case-1-333}
\eeqnarray
From \reqnarray{proof-OQ-LR-optimal-is-nondecreasing-(ii)-case-1-111}--\reqnarray{proof-OQ-LR-optimal-is-nondecreasing-(ii)-case-1-333},
we immediately see that ${\dbf'}_1^M\in \Acal_M$.
Thus, it is clear that
\beqnarray{proof-OQ-LR-optimal-is-nondecreasing-(ii)-case-1-444}
B({\dbf'}_1^M;k)\leq \max_{\dbf_1^M\in \Acal_M}B(\dbf_1^M;k)=B({\dbf^*}_1^M;k).
\eeqnarray

For $0\leq x\leq d'_{M-1}-1$,
we have $x<d'_{M-1}=d_M^*\leq B({\dbf^*}_1^M;k)=d'_M-1<d'_M$
and $x<d'_{M-1}=d_M^*<d_{M-1}^*$,
and hence it follows from
${\dbf'}_1^M\in \Acal_M$, $x<\min\{d'_{M-1},d'_M\}$,
${\dbf^*}_1^M\in \Acal_M$,  $x<\min\{d_{M-1}^*,d_M^*\}$,
and \rlemma{sum of C-transform}(i)
(with $m=M>1$ and $\ell'=m-2=M-2$ in \rlemma{sum of C-transform}(i)) that
\beqnarray{}
\alignspace
\sum_{\ell=1}^{M}I_{\ell}(x;{\dbf'}_1^M)=\sum_{\ell=1}^{M-2}I_{\ell}(x;{\dbf'}_1^{M-2}),
\textrm{ for } 0\leq x\leq d'_{M-1}-1,
\label{eqn:proof-OQ-LR-optimal-is-nondecreasing-(ii)-case-1-555} \\
\alignspace
\sum_{\ell=1}^{M}I_{\ell}(x;{\dbf^*}_1^M)=\sum_{\ell=1}^{M-2}I_{\ell}(x;{\dbf^*}_1^{M-2}),
\textrm{ for } 0\leq x\leq d'_{M-1}-1.
\label{eqn:proof-OQ-LR-optimal-is-nondecreasing-(ii)-case-1-666}
\eeqnarray
From \reqnarray{proof-OQ-LR-optimal-is-nondecreasing-(ii)-case-1-555},
${\dbf'}_1^{M-2}={\dbf^*}_1^{M-2}$,
\reqnarray{proof-OQ-LR-optimal-is-nondecreasing-(ii)-case-1-666},
$d'_{M-1}-1=d_M^*-1<B({\dbf^*}_1^M;k)$,
and the definition of $B({\dbf^*}_1^M;k)$
in \reqnarray{maximum representable integer},
we have
\beqnarray{proof-OQ-LR-optimal-is-nondecreasing-(ii)-case-1-777}
\sum_{\ell=1}^{M}I_{\ell}(x;{\dbf'}_1^M)=\sum_{\ell=1}^{M-2}I_{\ell}(x;{\dbf'}_1^{M-2})
=\sum_{\ell=1}^{M-2}I_{\ell}(x;{\dbf^*}_1^{M-2})=\sum_{\ell=1}^{M}I_{\ell}(x;{\dbf^*}_1^M)
\leq k, \nn\\
\textrm{ for } 0\leq x\leq d'_{M-1}-1.
\eeqnarray
Furthermore, for $d'_{M-1}\leq x\leq B({\dbf^*}_1^M;k)$,
we have $x\leq B({\dbf^*}_1^M;k)=d'_M-1<d'_M$, $x\geq d'_{M-1}=d_M^*$,
and $x-d_M^*\leq B({\dbf^*}_1^M;k)-d_M^*<d_{M-1}^*$.
Thus, it follows from
${\dbf'}_1^M\in \Acal_M$, $x<d'_M$,
\rlemma{sum of C-transform}(i)
(with $m=M>1$ and $\ell'=m-1=M-1$ in \rlemma{sum of C-transform}(i)),
${\dbf'}_1^{M-1}\in \Acal_{M-1}$, $x\geq d'_{M-1}$,
and \rlemma{sum of C-transform}(ii)
(with $m=M-1\geq 1$ and $\ell'=m-1=M-2$ in \rlemma{sum of C-transform}(ii)) that
\beqnarray{proof-OQ-LR-optimal-is-nondecreasing-(ii)-case-1-888}
\sum_{\ell=1}^{M}I_{\ell}(x;{\dbf'}_1^M)=\sum_{\ell=1}^{M-1}I_{\ell}(x;{\dbf'}_1^{M-1})
=\sum_{\ell=1}^{M-2}I_{\ell}(x-d'_{M-1};{\dbf'}_1^{M-2})+1, \nn\\
\textrm{ for } d'_{M-1}\leq x\leq B({\dbf^*}_1^M;k).
\eeqnarray
It also follows from
${\dbf^*}_1^M\in \Acal_M$, $x\geq d_M^*$,
\rlemma{sum of C-transform}(ii)
(with $m=M>1$ and $\ell'=m-1=M-1$ in \rlemma{sum of C-transform}(ii)),
${\dbf^*}_1^{M-1}\in \Acal_{M-1}$, $x-d_M^*<d_{M-1}^*$,
and \rlemma{sum of C-transform}(i)
(with $m=M-1\geq 1$ and $\ell'=m-1=M-2$ in \rlemma{sum of C-transform}(i)) that
\beqnarray{proof-OQ-LR-optimal-is-nondecreasing-(ii)-case-1-999}
\sum_{\ell=1}^{M}I_{\ell}(x;{\dbf^*}_1^M)=\sum_{\ell=1}^{M-1}I_{\ell}(x-d_M^*;{\dbf^*}_1^{M-1})+1
=\sum_{\ell=1}^{M-2}I_{\ell}(x-d_M^*;{\dbf^*}_1^{M-2})+1, \nn\\
\textrm{ for } d'_{M-1}\leq x\leq B({\dbf^*}_1^M;k).
\eeqnarray
From \reqnarray{proof-OQ-LR-optimal-is-nondecreasing-(ii)-case-1-888},
$d'_{M-1}=d_M^*$, ${\dbf'}_1^{M-2}={\dbf^*}_1^{M-2}$,
\reqnarray{proof-OQ-LR-optimal-is-nondecreasing-(ii)-case-1-999},
and the definition of $B({\dbf^*}_1^M;k)$
in \reqnarray{maximum representable integer},
we have
\beqnarray{proof-OQ-LR-optimal-is-nondecreasing-(ii)-case-1-aaa}
\sum_{\ell=1}^{M}I_{\ell}(x;{\dbf'}_1^M)=\sum_{\ell=1}^{M-2}I_{\ell}(x-d'_{M-1};{\dbf'}_1^{M-2})+1
=\sum_{\ell=1}^{M-2}I_{\ell}(x-d_M^*;{\dbf^*}_1^{M-2})+1 \nn\\
=\sum_{\ell=1}^{M}I_{\ell}(x;{\dbf^*}_1^M)\leq k,
\textrm{ for } d'_{M-1}\leq x\leq B({\dbf^*}_1^M;k).
\eeqnarray
Finally, for $x=B({\dbf^*}_1^M;k)+1=d'_M$,
it is clear from \reqnarray{C-transform} that $I_M(x;{\dbf'}_1^M)=1$
and $I_{\ell}(x;{\dbf'}_1^M)=0$ for $\ell=M-1,M-2,\ldots,1$.
Thus, it follows from $k\geq 1$ that
\beqnarray{proof-OQ-LR-optimal-is-nondecreasing-(ii)-case-1-bbb}
\sum_{\ell=1}^{M}I_{\ell}(x;{\dbf'}_1^M)=1\leq k,
\textrm{ for } x=B({\dbf^*}_1^M;k)+1.
\eeqnarray
As $B({\dbf^*}_1^M;k)+1=d'_M\leq \sum_{\ell=1}^{M}d'_{\ell}$,
we see from \reqnarray{proof-OQ-LR-optimal-is-nondecreasing-(ii)-case-1-777},
\reqnarray{proof-OQ-LR-optimal-is-nondecreasing-(ii)-case-1-aaa},
\reqnarray{proof-OQ-LR-optimal-is-nondecreasing-(ii)-case-1-bbb},
and the definition of $B({\dbf'}_1^M;k)$
in \reqnarray{maximum representable integer} that
\beqnarray{}
B({\dbf'}_1^M;k)\geq B({\dbf^*}_1^M;k)+1>B({\dbf^*}_1^M;k), \nn
\eeqnarray
contradicting to $B({\dbf'}_1^M;k)\leq B({\dbf^*}_1^M;k)$
in \reqnarray{proof-OQ-LR-optimal-is-nondecreasing-(ii)-case-1-444}.

\emph{Subcase 1(b): $d_{M-1}^*+d_M^*\leq B({\dbf^*}_1^M;k)<\sum_{\ell=1}^{M}d_{\ell}^*$.}
Let $d'_{\ell}=d_{\ell}^*$ for $\ell=1,2,\ldots,M-1$ and $d'_M=d_M^*+1$.
Then it follows from ${\dbf^*}_1^M\in \Acal_M$ and $d_{M-1}^*>d_M^*$ that
\beqnarray{}
\alignspace
{\dbf'}_1^{M-1}={\dbf^*}_1^{M-1}\in \Acal_{M-1},
\label{eqn:proof-OQ-LR-optimal-is-nondecreasing-(ii)-case-1-ccc} \\
\alignspace
1\leq d'_M=d_M^*+1\leq d_{M-1}^*=d'_{M-1}\leq \sum_{\ell=1}^{M-1}d'_{\ell}+1.
\label{eqn:proof-OQ-LR-optimal-is-nondecreasing-(ii)-case-1-ddd}
\eeqnarray
From \reqnarray{proof-OQ-LR-optimal-is-nondecreasing-(ii)-case-1-ccc}
and \reqnarray{proof-OQ-LR-optimal-is-nondecreasing-(ii)-case-1-ddd},
we immediately see that ${\dbf'}_1^M\in \Acal_M$.
Thus, it is clear that
\beqnarray{proof-OQ-LR-optimal-is-nondecreasing-(ii)-case-1-eee}
B({\dbf'}_1^M;k)\leq \max_{\dbf_1^M\in \Acal_M}B(\dbf_1^M;k)=B({\dbf^*}_1^M;k).
\eeqnarray

From ${\dbf^*}_1^M\in \Acal_M$, $1\leq k<M$,
$B({\dbf^*}_1^M;k)\geq d_{M-1}^*+d_M^*\geq d_M^*$,
and \rlemma{OQ-LR-MRI-general}(ii)
(with $m=M$ and $i=k$ in \rlemma{OQ-LR-MRI-general}(ii)),
we have
\beqnarray{proof-OQ-LR-optimal-is-nondecreasing-(ii)-case-1-fff}
B({\dbf^*}_1^M;k)=d_M^*+B({\dbf^*}_1^{M-1};k-1).
\eeqnarray
As $B({\dbf^*}_1^M;k)\geq d_{M-1}^*+d_M^*$,
we see from \reqnarray{proof-OQ-LR-optimal-is-nondecreasing-(ii)-case-1-fff} that
\beqnarray{proof-OQ-LR-optimal-is-nondecreasing-(ii)-case-1-ggg}
B({\dbf^*}_1^{M-1};k-1)=B({\dbf^*}_1^M;k)-d_M^* \geq d_{M-1}^*.
\eeqnarray
From ${\dbf'}_1^{M-1}\in \Acal_{M-1}$, $0\leq k-1<M-1$,
and \rlemma{OQ-LR-MRI-monotone}(i)
(with $m=M-1$, $i=k-1$, and $i'=k$ in \rlemma{OQ-LR-MRI-monotone}(i)),
we have $B({\dbf'}_1^{M-1};k-1)\leq B({\dbf'}_1^{M-1};k)$.
It then follows from $B({\dbf'}_1^{M-1};k)\geq B({\dbf'}_1^{M-1};k-1)$,
${\dbf'}_1^{M-1}={\dbf^*}_1^{M-1}$, $B({\dbf^*}_1^{M-1};k-1)\geq d_{M-1}^*$
in \reqnarray{proof-OQ-LR-optimal-is-nondecreasing-(ii)-case-1-ggg},
$d_{M-1}^*>d_M^*$, and $d'_M=d_M^*+1$ that
\beqnarray{proof-OQ-LR-optimal-is-nondecreasing-(ii)-case-1-hhh}
B({\dbf'}_1^{M-1};k)\geq B({\dbf'}_1^{M-1};k-1)
=B({\dbf^*}_1^{M-1};k-1)\geq d_{M-1}^*\geq d_M^*+1=d'_M.
\eeqnarray
From ${\dbf'}_1^M\in \Acal_M$, $1\leq k<M$,
$B({\dbf'}_1^{M-1};k)\geq d'_M$
in \reqnarray{proof-OQ-LR-optimal-is-nondecreasing-(ii)-case-1-hhh},
and \rlemma{OQ-LR-MRI-general}(ii)
(with $m=M$ and $i=k$ in \rlemma{OQ-LR-MRI-general}(ii)),
we have
\beqnarray{proof-OQ-LR-optimal-is-nondecreasing-(ii)-case-1-iii}
B({\dbf'}_1^M;k)=d'_M+B({\dbf'}_1^{M-1};k-1).
\eeqnarray
Therefore, we see from \reqnarray{proof-OQ-LR-optimal-is-nondecreasing-(ii)-case-1-iii},
$d'_M=d_M^*+1$, ${\dbf'}_1^{M-1}={\dbf^*}_1^{M-1}$,
and \reqnarray{proof-OQ-LR-optimal-is-nondecreasing-(ii)-case-1-fff} that
\beqnarray{}
B({\dbf'}_1^M;k)=d'_M+B({\dbf'}_1^{M-1};k-1)
=d_M^*+1+B({\dbf^*}_1^{M-1};k-1)=B({\dbf^*}_1^M;k)+1,
\eeqnarray
contradicting to $B({\dbf'}_1^M;k)\leq B({\dbf^*}_1^M;k)$
in \reqnarray{proof-OQ-LR-optimal-is-nondecreasing-(ii)-case-1-eee}.

\emph{Case 2: $2\leq \ell_1\leq M-2$.}
In this case, we see from the definition of $\ell_1$ that $d_{\ell_1}^*>d_{\ell_1+1}^*$
and $d_{\ell_1+1}^*\leq d_{\ell_1+2}^*\leq \cdots\leq d_M^*$.
We claim that
\beqnarray{proof-OQ-LR-optimal-is-nondecreasing-(ii)-case-2-111}
d_M^*=B({\dbf^*}_1^{M-1};k)+1.
\eeqnarray
From \rlemma{OQ-LR-optimal-is-nondecreasing}(i),
we know that $B({\dbf^*}_1^M;k)\geq d_M^*$.
It then follows from ${\dbf^*}_1^M\in \Acal_M$, $1\leq k<M$,
$B({\dbf^*}_1^M;k)\geq d_M^*$,
\rlemma{OQ-LR-MRI-equivalent conditions}
(with $m=M$, $i=k$, and $\ell'=m-1=M-1$ in \rlemma{OQ-LR-MRI-equivalent conditions}) that
\beqnarray{proof-OQ-LR-optimal-is-nondecreasing-(ii)-case-2-222}
B({\dbf^*}_1^{M-1};k)\geq d_M^*-1.
\eeqnarray
To prove \reqnarray{proof-OQ-LR-optimal-is-nondecreasing-(ii)-case-2-111},
we see from \reqnarray{proof-OQ-LR-optimal-is-nondecreasing-(ii)-case-2-222}
that it suffices to show that it cannot be the case that $B({\dbf^*}_1^{M-1};k)>d_M^*-1$.
Assume on the contrary that $B({\dbf^*}_1^{M-1};k)>d_M^*-1$.
Let $d'_{\ell}=d_{\ell}^*$ for $\ell=1,2,\ldots,M-1$ and $d'_M=B({\dbf^*}_1^{M-1};k)+1$.
Then it follows from ${\dbf^*}_1^M\in \Acal_M$
and $B({\dbf^*}_1^{M-1};k)\leq \sum_{\ell=1}^{M-1}d_{\ell}^*$ that
\beqnarray{}
\alignspace
{\dbf'}_1^{M-1}={\dbf^*}_1^{M-1}\in \Acal_{M-1},
\label{eqn:proof-OQ-LR-optimal-is-nondecreasing-(ii)-case-2-333} \\
\alignspace
1\leq d'_M=B({\dbf^*}_1^{M-1};k)+1\leq \sum_{\ell=1}^{M-1}d_{\ell}^*+1
=\sum_{\ell=1}^{M-1}d'_{\ell}+1.
\label{eqn:proof-OQ-LR-optimal-is-nondecreasing-(ii)-case-2-444}
\eeqnarray
From \reqnarray{proof-OQ-LR-optimal-is-nondecreasing-(ii)-case-2-333}
and \reqnarray{proof-OQ-LR-optimal-is-nondecreasing-(ii)-case-2-444},
we immediately see that ${\dbf'}_1^M\in \Acal_M$.
Thus, it is clear that
\beqnarray{proof-OQ-LR-optimal-is-nondecreasing-(ii)-case-2-555}
B({\dbf'}_1^M;k)\leq \max_{\dbf_1^M\in \Acal_M}B(\dbf_1^M;k)=B({\dbf^*}_1^M;k).
\eeqnarray

Note that from ${\dbf^*}_1^M\in \Acal_M$, $1\leq k<M$,
$B({\dbf^*}_1^M;k)\geq d_M^*$,
and \rlemma{OQ-LR-MRI-general}(ii)
(with $m=M$ and $i=k$ in \rlemma{OQ-LR-MRI-general}(ii)),
we have
\beqnarray{proof-OQ-LR-optimal-is-nondecreasing-(ii)-case-2-666}
B({\dbf^*}_1^M;k)=d_M^*+B({\dbf^*}_1^{M-1};k-1).
\eeqnarray
As we have $B({\dbf'}_1^{M-1};k)=B({\dbf^*}_1^{M-1};k)=d'_M-1$,
it follows from ${\dbf'}_1^M\in \Acal_M$, $1\leq k<M$,
and \rlemma{OQ-LR-MRI-general}(ii)
(with $m=M$ and $i=k$ in \rlemma{OQ-LR-MRI-general}(ii)) that
\beqnarray{proof-OQ-LR-optimal-is-nondecreasing-(ii)-case-2-777}
B({\dbf'}_1^M;k)=d'_M+B({\dbf'}_1^{M-1};k-1).
\eeqnarray
Therefore, we have from
\reqnarray{proof-OQ-LR-optimal-is-nondecreasing-(ii)-case-2-777},
$d'_M=B({\dbf^*}_1^{M-1};k)+1>d_M^*$, ${\dbf'}_1^{M-1}={\dbf^*}_1^{M-1}$,
and \reqnarray{proof-OQ-LR-optimal-is-nondecreasing-(ii)-case-2-666} that
\beqnarray{}
B({\dbf'}_1^M;k)=d'_M+B({\dbf'}_1^{M-1};k-1)
>d_M^*+B({\dbf^*}_1^{M-1};k-1)=B({\dbf^*}_1^M;k),
\eeqnarray
and we have reached a contradiction to $B({\dbf'}_1^M;k)\leq B({\dbf^*}_1^M;k)$
in \reqnarray{proof-OQ-LR-optimal-is-nondecreasing-(ii)-case-2-555}.

Define $k_M=k$.
Then we have from \reqnarray{proof-OQ-LR-optimal-is-nondecreasing-(ii)-case-2-111} that
\beqnarray{proof-OQ-LR-optimal-is-nondecreasing-(ii)-case-2-888}
d_M^*=B({\dbf^*}_1^{M-1};k_M)+1.
\eeqnarray
Let $E_{M-1}=\{\max\{k_M-1,0\}\leq i\leq k_M:d_{M-1}^*=B({\dbf^*}_1^{M-2};i)+1\}$.
If $E_{M-1}\neq \emptyset$,
then we define $k_{M-1}=\min\{i: i\in E_{M-1}\}$
so that we have
\beqnarray{proof-OQ-LR-optimal-is-nondecreasing-(ii)-case-2-999}
\max\{k_M-1,0\}\leq k_{M-1}\leq k_M
\textrm{ and } d_{M-1}^*=B({\dbf^*}_1^{M-2};k_{M-1})+1.
\eeqnarray
Otherwise, if $E_{M-1}=\emptyset$,
then we define $\ell_2=M-1$.
In the case that $E_{M-1}\neq \emptyset$ and $M-1>\ell_1+1$,
we continue the above process and
let $E_{M-2}=\{\max\{k_{M-1}-1,0\}\leq i\leq k_{M-1}:d_{M-2}^*=B({\dbf^*}_1^{M-3};i)+1\}$.
If $E_{M-2}\neq \emptyset$,
then we define $k_{M-2}=\min\{i: i\in E_{M-2}\}$
so that we have
\beqnarray{proof-OQ-LR-optimal-is-nondecreasing-(ii)-case-2-aaa}
\max\{k_{M-1}-1,0\}\leq k_{M-2}\leq k_{M-1}
\textrm{ and } d_{M-2}^*=B({\dbf^*}_1^{M-3};k_{M-2})+1.
\eeqnarray
Otherwise, if $E_{M-2}=\emptyset$,
then we define $\ell_2=M-2$.
Again, in the case that $E_{M-2}\neq \emptyset$ and $M-2>\ell_1+1$,
we continue the above process and
let $E_{M-3}=\{\max\{k_{M-2}-1,0\}\leq i\leq k_{M-2}:d_{M-3}^*=B({\dbf^*}_1^{M-4};i)+1\}$.
If $E_{M-3}\neq \emptyset$,
then we define $k_{M-3}=\min\{i: i\in E_{M-3}\}$
so that we have
\beqnarray{proof-OQ-LR-optimal-is-nondecreasing-(ii)-case-2-bbb}
\max\{k_{M-2}-1,0\}\leq k_{M-3}\leq k_{M-2}
\textrm{ and } d_{M-3}^*=B({\dbf^*}_1^{M-4};k_{M-3})+1.
\eeqnarray
Otherwise, if $E_{M-3}=\emptyset$,
then we define $\ell_2=M-3$.
Clearly, we can continue the above process until
either $E_{M-1}\neq \emptyset,E_{M-2}\neq \emptyset,\ldots,E_{\ell_1+1}\neq \emptyset$
(in this scenario, $k_M,k_{M-1},\ldots,k_{\ell_1+1}$ are defined)
or $E_{M-1}\neq \emptyset,E_{M-2}\neq \emptyset,\ldots,E_{\ell_2+1}\neq \emptyset$,
and $E_{\ell_2}=\emptyset$ for some $\ell_1+1\leq \ell_2\leq M-1$
(in this scenario, $k_M,k_{M-1},\ldots,k_{\ell_2+1}$ are defined).

It is clear that for the scenario that
$E_{M-1}\neq \emptyset,E_{M-2}\neq \emptyset,\ldots,E_{\ell_1+1}\neq \emptyset$,
we have
\beqnarray{}
\alignspace
E_{\ell}=\{\max\{k_{\ell+1}-1,0\}\leq i\leq k_{\ell+1}:d_{\ell}^*=B({\dbf^*}_1^{\ell-1};i)+1\}, \nn\\
\alignspace \hspace*{1.72in} \textrm{ for } \ell=M-1,M-2,\ldots,\ell_1+1,
\label{eqn:proof-OQ-LR-optimal-is-nondecreasing-(ii)-case-2-ccc} \\
\alignspace
k_{\ell}=\min\{i: i\in E_{\ell}\},
\textrm{ for } \ell=M-1,M-2,\ldots,\ell_1+1.
\label{eqn:proof-OQ-LR-optimal-is-nondecreasing-(ii)-case-2-ddd}
\eeqnarray
It follows from \reqnarray{proof-OQ-LR-optimal-is-nondecreasing-(ii)-case-2-888},
\reqnarray{proof-OQ-LR-optimal-is-nondecreasing-(ii)-case-2-ccc},
and \reqnarray{proof-OQ-LR-optimal-is-nondecreasing-(ii)-case-2-ddd} that
\beqnarray{}
\alignspace
\max\{k_{\ell+1}-1,0\}\leq k_{\ell}\leq k_{\ell+1},
\textrm{ for } \ell=M-1,M-2,\ldots,\ell_1+1,
\label{eqn:proof-OQ-LR-optimal-is-nondecreasing-(ii)-case-2-eee} \\
\alignspace
d_{\ell}^*=B({\dbf^*}_1^{\ell-1};k_{\ell})+1,
\textrm{ for } \ell=M,M-1,\ldots,\ell_1+1.
\label{eqn:proof-OQ-LR-optimal-is-nondecreasing-(ii)-case-2-fff}
\eeqnarray
For example, if $M=12$, $k=5$, and ${\dbf^*}_1^M=(1,2,3,5,12,1,1,2,4,8,16,32)$,
then we have $\ell_1=\max\{2\leq \ell\leq M-1:d_{\ell}^*>d_{\ell+1}^*\}=5$
and $k_{12}=k_M=k=5$.
From the values of $B({\dbf^*}_1^{\ell};i)$ for $\ell_1\leq \ell\leq M$ and $0\leq i\leq k$
in \rtable{OQ-LR-optimal-is-nondecreasing-1},
it is easy to see that
\beqnarray{}
k_{11}\aligneq 4 \textrm{ (as } d_{11}^*=16=B({\dbf^*}_1^{10};4)+1 \textrm{ so that } E_{11}=\{4\}), \nn\\
k_{10}\aligneq 3 \textrm{ (as } d_{10}^*=8=B({\dbf^*}_1^{9};3)+1 \textrm{ so that } E_{10}=\{3\}), \nn\\
k_{9}\aligneq 2 \textrm{ (as } d_{9}^*=4=B({\dbf^*}_1^{8};2)+1 \textrm{ so that } E_{9}=\{2\}), \nn\\
k_{8}\aligneq 1 \textrm{ (as } d_{8}^*=2=B({\dbf^*}_1^{7};1)+1 \textrm{ so that } E_{8}=\{1\}), \nn\\
k_{7}\aligneq 0 \textrm{ (as } d_{7}^*=1=B({\dbf^*}_1^{6};0)+1 \textrm{ so that } E_{7}=\{0\}), \nn\\
k_{6}\aligneq 0 \textrm{ (as } d_{6}^*=1=B({\dbf^*}_1^{5};0)+1 \textrm{ so that } E_{6}=\{0\}). \nn
\eeqnarray
{\tiny
\btable{htbp}{||c||c|c|c|c|c|c||}
\hline
$\ell \verb+\+ i$ & 0 & 1 &  2 &  3 &  4 &  5 \\
\hline
5                 & 0 & 3 &  8 & 10 & 22 & 23 \\
\hline
6                 & 0 & 1 &  4 &  9 & 11 & 23 \\
\hline
7                 & 0 & 1 &  2 &  5 & 10 & 12 \\
\hline
8                 & 0 & 2 &  3 &  4 &  7 & 12 \\
\hline
9                 & 0 & 2 &  6 &  7 &  8 & 11 \\
\hline
10                & 0 & 2 &  6 & 14 & 15 & 16 \\
\hline
11                & 0 & 2 &  6 & 14 & 30 & 31 \\
\hline
12                & 0 & 2 &  6 & 14 & 30 & 62 \\
\hline
\etable{OQ-LR-optimal-is-nondecreasing-1}
{The values of $B({\dbf^*}_1^{\ell};i)$ for $\ell_1\leq \ell\leq M$ and $0\leq i\leq k$,
where $M=12$, $k=5$, ${\dbf^*}_1^M=(1,2,3,5,12,1,1,2,4,8,16,32)$, and $\ell_1=5$.}}

It is also clear that for the scenario that
$E_{M-1}\neq \emptyset,E_{M-2}\neq \emptyset,\ldots,E_{\ell_2+1}\neq \emptyset$,
and $E_{\ell_2}=\emptyset$ for some $\ell_1+1\leq \ell_2\leq M-1$,
we have
\beqnarray{}
\alignspace
E_{\ell}=\{\max\{k_{\ell+1}-1,0\}\leq i\leq k_{\ell+1}:d_{\ell}^*=B({\dbf^*}_1^{\ell-1};i)+1\}, \nn\\
\alignspace \hspace*{2.02in} \textrm{ for } \ell=M-1,M-2,\ldots,\ell_2,
\label{eqn:proof-OQ-LR-optimal-is-nondecreasing-(ii)-case-2-ggg} \\
\alignspace
k_{\ell}=\min\{i: i\in E_{\ell}\},
\textrm{ for } \ell=M-1,M-2,\ldots,\ell_2+1.
\label{eqn:proof-OQ-LR-optimal-is-nondecreasing-(ii)-case-2-hhh}
\eeqnarray
It follows from \reqnarray{proof-OQ-LR-optimal-is-nondecreasing-(ii)-case-2-888},
\reqnarray{proof-OQ-LR-optimal-is-nondecreasing-(ii)-case-2-ggg},
and \reqnarray{proof-OQ-LR-optimal-is-nondecreasing-(ii)-case-2-hhh} that
\beqnarray{}
\alignspace
\max\{k_{\ell+1}-1,0\}\leq k_{\ell}\leq k_{\ell+1},
\textrm{ for } \ell=M-1,M-2,\ldots,\ell_2+1,
\label{eqn:proof-OQ-LR-optimal-is-nondecreasing-(ii)-case-2-iii} \\
\alignspace
d_{\ell}^*=B({\dbf^*}_1^{\ell-1};k_{\ell})+1,
\textrm{ for } \ell=M,M-1,\ldots,\ell_2+1.
\label{eqn:proof-OQ-LR-optimal-is-nondecreasing-(ii)-case-2-jjj}
\eeqnarray
For example, if $M=9$, $k=8$, and ${\dbf^*}_1^M=(1,2,3,5,2,8,22,44,88)$,
then we have $\ell_1=\max\{2\leq \ell\leq M-1:d_{\ell}^*>d_{\ell+1}^*\}=4$
and $k_{9}=k_M=k=8$.
From the values of $B({\dbf^*}_1^{\ell};i)$ for $\ell_1\leq \ell\leq M$ and $0\leq i\leq k$
in \rtable{OQ-LR-optimal-is-nondecreasing-2},
it is easy to see that
\beqnarray{}
k_{8}\aligneq 7 \textrm{ (as } d_{8}^*=44=B({\dbf^*}_1^{7};7)+1=B({\dbf^*}_1^{7};8)+1 \textrm{ so that } E_{8}=\{7,8\}), \nn\\
k_{7}\aligneq 6 \textrm{ (as } d_{7}^*=22=B({\dbf^*}_1^{6};6)+1=B({\dbf^*}_1^{6};7)+1 \textrm{ so that } E_{7}=\{6,7\}), \nn\\
\ell_2\aligneq 6 \textrm{ (as } d_{6}^*=8<B({\dbf^*}_1^{5};5)+1=B({\dbf^*}_1^{5};6)+1 \textrm{ so that } E_{6}=\emptyset). \nn
\eeqnarray
As another example, if $M=9$, $k=5$, and ${\dbf^*}_1^M=(1,2,3,5,2,8,14,28,56)$,
then we have $\ell_1=\max\{2\leq \ell\leq M-1:d_{\ell}^*>d_{\ell+1}^*\}=4$
and $k_{9}=k_M=k=5$.
From the values of $B({\dbf^*}_1^{\ell};i)$ for $\ell_1\leq \ell\leq M$ and $0\leq i\leq k$
in \rtable{OQ-LR-optimal-is-nondecreasing-3},
it is easy to see that
\beqnarray{}
k_{8}\aligneq 4 \textrm{ (as } d_{8}^*=28=B({\dbf^*}_1^{7};4)+1 \textrm{ so that } E_{8}=\{4\}), \nn\\
k_{7}\aligneq 3 \textrm{ (as } d_{7}^*=14=B({\dbf^*}_1^{6};3)+1 \textrm{ so that } E_{7}=\{3\}), \nn\\
\ell_2\aligneq 6 \textrm{ (as } B({\dbf^*}_1^{5};2)+1<d_{6}^*=8<B({\dbf^*}_1^{5};3)+1 \textrm{ so that } E_{6}=\emptyset). \nn
\eeqnarray
{\tiny
\btable{htbp}{||c||c|c|c|c|c|c|c|c|c||}
\hline
$\ell \verb+\+ i$ & 0 & 1 &  2 &  3 &  4 &  5 &  6 &  7 &   8 \\
\hline
4                 & 0 & 3 &  8 & 10 & 11 & 11 & 11 & 11 &  11 \\
\hline
5                 & 0 & 2 &  5 & 10 & 12 & 13 & 13 & 13 &  13 \\
\hline
6                 & 0 & 2 &  5 & 13 & 18 & 20 & 21 & 21 &  21 \\
\hline
7                 & 0 & 2 &  5 & 13 & 18 & 20 & 42 & 43 &  43 \\
\hline
8                 & 0 & 2 &  5 & 13 & 18 & 20 & 42 & 86 &  87 \\
\hline
9                 & 0 & 2 &  5 & 13 & 18 & 20 & 42 & 86 & 174 \\
\hline
\etable{OQ-LR-optimal-is-nondecreasing-2}
{The values of $B({\dbf^*}_1^{\ell};i)$ for $\ell_1\leq \ell\leq M$ and $0\leq i\leq k$,
where $M=9$, $k=8$, ${\dbf^*}_1^M=(1,2,3,5,2,8,22,44,88)$, and $\ell_1=4$.}}
{\tiny
\btable{htbp}{||c||c|c|c|c|c|c|c|c|c||}
\hline
$\ell \verb+\+ i$ & 0 & 1 &  2 &  3 &  4 &   5 \\
\hline
4                 & 0 & 3 &  8 & 10 & 11 &  11 \\
\hline
5                 & 0 & 2 &  5 & 10 & 12 &  13 \\
\hline
6                 & 0 & 2 &  5 & 13 & 18 &  20 \\
\hline
7                 & 0 & 2 &  5 & 19 & 27 &  32 \\
\hline
8                 & 0 & 2 &  5 & 19 & 47 &  55 \\
\hline
9                 & 0 & 2 &  5 & 19 & 47 & 103 \\
\hline
\etable{OQ-LR-optimal-is-nondecreasing-3}
{The values of $B({\dbf^*}_1^{\ell};i)$ for $\ell_1\leq \ell\leq M$ and $0\leq i\leq k$,
where $M=9$, $k=5$, ${\dbf^*}_1^M=(1,2,3,5,2,8,14,28,56)$, and $\ell_1=4$.}}

We need to consider the above two scenarios separately.

\emph{Subcase 2(a): $E_{M-1}\neq \emptyset,E_{M-2}\neq \emptyset,\ldots,E_{\ell_1+1}\neq \emptyset$.}
In this subcase, we let $d'_{\ell}=d_{\ell}^*$ for $\ell=1,2,\ldots,\ell_1-1$,
$d'_{\ell}=d_{\ell+1}^*$ for $\ell=\ell_1,\ell_1+1,\ldots,M-1$,
and $d'_M=B({{\dbf}^*}_1^M;k)+1$.
We will show that
\beqnarray{proof-OQ-LR-optimal-is-nondecreasing-(ii)-subcase-2(a)-111}
{\dbf'}_1^M\in \Acal_M \textrm{ and } B({\dbf'}_1^M;k)>B({\dbf^*}_1^M;k).
\eeqnarray
As it is clear from ${\dbf'}_1^M\in \Acal_M$ that
$B({\dbf'}_1^M;k)\leq \max_{\dbf_1^M\in \Acal_M}B(\dbf_1^M;k)=B({\dbf^*}_1^M;k)$,
we have reached a contradiction to $B({\dbf'}_1^M;k)>B({\dbf^*}_1^M;k)$
in \reqnarray{proof-OQ-LR-optimal-is-nondecreasing-(ii)-subcase-2(a)-111}.
Note that for the example
that $M=12$, $k=5$, and ${\dbf^*}_1^M=(1,2,3,5,12,1,1,2,4,8,16,32)$
as shown in \rtable{OQ-LR-optimal-is-nondecreasing-1},
we have $B({\dbf^*}_1^M;k)=62$.
If we let ${\dbf'}_1^M=(1,2,3,5,1,1,2,4,8,16,32,63)$,
then it can be seen that $B({\dbf'}_1^M;k)=93>B({\dbf^*}_1^M;k)=62$
and \reqnarray{proof-OQ-LR-optimal-is-nondecreasing-(ii)-subcase-2(a)-111} is satisfied.

We first show by induction on $\ell$ that
\beqnarray{proof-OQ-LR-optimal-is-nondecreasing-(ii)-subcase-2(a)-222}
\sum_{\ell'=\ell_1}^{\ell}d_{\ell'}^*>d_{\ell+1}^*,
\textrm{ for } \ell_1\leq \ell\leq M-1.
\eeqnarray
As we already have $d_{\ell_1}^*>d_{\ell_1+1}^*$,
we assume as the induction hypothesis that $\sum_{\ell'=\ell_1}^{\ell-1}d_{\ell'}^*>d_{\ell}^*$
for some $\ell_1+1\leq \ell\leq M-1$.
From $d_{\ell}^*=B({\dbf^*}_1^{\ell-1};k_{\ell})+1$
in \reqnarray{proof-OQ-LR-optimal-is-nondecreasing-(ii)-case-2-fff}
(as $\ell_1+1\leq \ell\leq M-1$),
$\ell-1\geq \ell_1>1$,
$0\leq \max\{k_{\ell+1}-1,0\}\leq k_{\ell}\leq k_{\ell+1}$
in \reqnarray{proof-OQ-LR-optimal-is-nondecreasing-(ii)-case-2-eee},
and \rlemma{OQ-LR-MRI-monotone}(i)
(with $m=\ell-1$, $i=k_{\ell}$, and $i'=k_{\ell+1}$ in \rlemma{OQ-LR-MRI-monotone}(i)),
we have
\beqnarray{proof-OQ-LR-optimal-is-nondecreasing-(ii)-subcase-2(a)-333}
d_{\ell}^*=B({\dbf^*}_1^{\ell-1};k_{\ell})+1\leq B({\dbf^*}_1^{\ell-1};k_{\ell+1})+1.
\eeqnarray
Note that $k_{\ell+1}\geq 0$
(as we have from \reqnarray{proof-OQ-LR-optimal-is-nondecreasing-(ii)-case-2-eee}
that $k_{\ell+1}\geq \max\{k_{\ell+2}-1,0\}\geq 0$
in the case that $\ell_1+1\leq \ell\leq M-2$,
and we have $k_{\ell+1}=k_M=k\geq 1$ in the case that $\ell=M-1$).
If $k_{\ell+1}\geq 1$,
then we see from ${\dbf^*}_1^{\ell}\in \Acal_{\ell}$
(as ${\dbf^*}_1^M\in \Acal_M$ and $\ell<M$),
$B({\dbf^*}_1^{\ell-1};k_{\ell+1})\geq d_{\ell}^*-1$
in \reqnarray{proof-OQ-LR-optimal-is-nondecreasing-(ii)-subcase-2(a)-333},
and \rlemma{OQ-LR-MRI-general}(ii)
(with $m=\ell\geq \ell_1+1>1$ and $i=k_{\ell+1}\geq 1$ in \rlemma{OQ-LR-MRI-general}(ii)) that
\beqnarray{proof-OQ-LR-optimal-is-nondecreasing-(ii)-subcase-2(a)-444}
B({\dbf^*}_1^{\ell};k_{\ell+1})=d_{\ell}^*+B({\dbf^*}_1^{\ell-1};k_{\ell+1}-1).
\eeqnarray
As such, it follows from $d_{\ell+1}^*=B({\dbf^*}_1^{\ell};k_{\ell+1})+1$
in \reqnarray{proof-OQ-LR-optimal-is-nondecreasing-(ii)-case-2-fff}
(as $\ell_1+2\leq \ell+1\leq M$),
\reqnarray{proof-OQ-LR-optimal-is-nondecreasing-(ii)-subcase-2(a)-444},
$\ell-1\geq \ell_1>1$,
$0\leq k_{\ell+1}-1\leq \max\{k_{\ell+1}-1,0\}\leq k_{\ell}$
in \reqnarray{proof-OQ-LR-optimal-is-nondecreasing-(ii)-case-2-eee},
\rlemma{OQ-LR-MRI-monotone}(i)
(with $m=\ell-1$, $i=k_{\ell+1}-1$, and $i'=k_{\ell}$ in \rlemma{OQ-LR-MRI-monotone}(i)),
$d_{\ell}^*=B({\dbf^*}_1^{\ell-1};k_{\ell})+1$
in \reqnarray{proof-OQ-LR-optimal-is-nondecreasing-(ii)-case-2-fff}
(as $\ell_1+1\leq \ell\leq M-1$),
and $d_{\ell}^*<\sum_{\ell'=\ell_1}^{\ell-1}d_{\ell'}^*$ in the induction hypothesis that
\beqnarray{proof-OQ-LR-optimal-is-nondecreasing-(ii)-subcase-2(a)-555}
d_{\ell+1}^*
\aligneq B({\dbf^*}_1^{\ell};k_{\ell+1})+1
=d_{\ell}^*+B({\dbf^*}_1^{\ell-1};k_{\ell+1}-1)+1 \nn\\
\alignleq d_{\ell}^*+B({\dbf^*}_1^{\ell-1};k_{\ell})+1=d_{\ell}^*+d_{\ell}^* \nn\\
\alignless d_{\ell}^*+\sum_{\ell'=\ell_1}^{\ell-1}d_{\ell'}^*=\sum_{\ell'=\ell_1}^{\ell}d_{\ell'}^*.
\eeqnarray
On the other hand, if $k_{\ell+1}=0$,
then we see from $d_{\ell+1}^*=B({\dbf^*}_1^{\ell};k_{\ell+1})+1$
in \reqnarray{proof-OQ-LR-optimal-is-nondecreasing-(ii)-case-2-fff}
(as $\ell_1+2\leq \ell+1\leq M$),
$d_{\ell_1}^*\geq 1$, $d_{\ell_1+1}^*\geq 1$,
and $\ell\geq \ell_1+1$ that
\beqnarray{proof-OQ-LR-optimal-is-nondecreasing-(ii)-subcase-2(a)-666}
d_{\ell+1}^*=B({\dbf^*}_1^{\ell};k_{\ell+1})+1=B({\dbf^*}_1^{\ell};0)+1=1
<d_{\ell_1}^*+d_{\ell_1+1}^*\leq \sum_{\ell'=\ell_1}^{\ell}d_{\ell'}^*.
\eeqnarray
The induction is completed by combining
\reqnarray{proof-OQ-LR-optimal-is-nondecreasing-(ii)-subcase-2(a)-555}
and \reqnarray{proof-OQ-LR-optimal-is-nondecreasing-(ii)-subcase-2(a)-666}.

Now we use \reqnarray{proof-OQ-LR-optimal-is-nondecreasing-(ii)-subcase-2(a)-222}
to show that
\beqnarray{proof-OQ-LR-optimal-is-nondecreasing-(ii)-subcase-2(a)-777}
I_{\ell_1}(x;{\dbf^*}_1^M)=0, \textrm{ for all } 0\leq x\leq B({\dbf^*}_1^M;k).
\eeqnarray
We prove \reqnarray{proof-OQ-LR-optimal-is-nondecreasing-(ii)-subcase-2(a)-777} by contradiction.
Suppose on the contrary that $I_{\ell_1}(x;{\dbf^*}_1^M)=1$
for some $0\leq x\leq B({\dbf^*}_1^M;k)$.
In the following, we show by induction on $\ell$ that
$I_{\ell}(x;{\dbf^*}_1^M)=1$ for $\ell=\ell_1,\ell_1+1,\ldots,M$.
As we have already assumed that $I_{\ell_1}(x;{\dbf^*}_1^M)=1$,
we assume as the induction hypothesis that
$I_{\ell_1}(x;{\dbf^*}_1^M)=I_{\ell_1+1}(x;{\dbf^*}_1^M)=\cdots=I_{\ell}(x;{\dbf^*}_1^M)=1$
for some $\ell_1\leq \ell\leq M-1$.
From $x=\sum_{\ell'=1}^{M}I_{\ell'}(x;{\dbf^*}_1^M)d_{\ell'}^*$
(the unique representation property of the $\Ccal$-transform),
the induction hypothesis, $\ell_1+1\leq \ell\leq M-1$,
and \reqnarray{proof-OQ-LR-optimal-is-nondecreasing-(ii)-subcase-2(a)-222},
we see that
\beqnarray{proof-OQ-LR-optimal-is-nondecreasing-(ii)-subcase-2(a)-888}
x-\sum_{\ell'=\ell+2}^{M}I_{\ell'}(x;{\dbf^*}_1^M)d_{\ell'}^*
\aligneq \sum_{\ell'=1}^{M}I_{\ell'}(x;{\dbf^*}_1^M)d_{\ell'}^*
-\sum_{\ell'=\ell+2}^{M}I_{\ell'}(x;{\dbf^*}_1^M)d_{\ell'}^*
=\sum_{\ell'=1}^{\ell+1}I_{\ell'}(x;{\dbf^*}_1^M)d_{\ell'}^* \nn\\
\aligngeq \sum_{\ell'=\ell_1}^{\ell}I_{\ell'}(x;{\dbf^*}_1^M)d_{\ell'}^*
=\sum_{\ell'=\ell_1}^{\ell}d_{\ell'}^*>d_{\ell+1}^*.
\eeqnarray
It then follows from \reqnarray{proof-OQ-LR-optimal-is-nondecreasing-(ii)-subcase-2(a)-888}
and \reqnarray{C-transform} that $I_{\ell+1}(x;{\dbf^*}_1^M)=1$,
and the induction is completed.
To continue with the proof of \reqnarray{proof-OQ-LR-optimal-is-nondecreasing-(ii)-subcase-2(a)-777},
let $x'=d_{\ell_1+1}^*+\sum_{\ell=\ell_1+1}^{M}d_{\ell}^*$.
From $d_{\ell_1}^*>d_{\ell_1+1}^*$,
$I_{\ell}(x;{\dbf^*}_1^M)=1$ for $\ell=\ell_1,\ell_1+1,\ldots,M$,
and $\sum_{\ell=1}^{M}I_{\ell}(x;{\dbf^*}_1^M)d_{\ell}^*=x\leq B({\dbf^*}_1^M;k)$,
we have
\beqnarray{proof-OQ-LR-optimal-is-nondecreasing-(ii)-subcase-2(a)-999}
x'
\aligneq d_{\ell_1+1}^*+\sum_{\ell=\ell_1+1}^{M}d_{\ell}^*
<d_{\ell_1}^*+\sum_{\ell=\ell_1+1}^{M}d_{\ell}^*
=\sum_{\ell=\ell_1}^{M}I_{\ell}(x;{\dbf^*}_1^M)d_{\ell}^* \nn\\
\alignleq \sum_{\ell=1}^{M}I_{\ell}(x;{\dbf^*}_1^M)d_{\ell}^*
=x\leq B({\dbf^*}_1^M;k).
\eeqnarray
It is clear from \reqnarray{proof-OQ-LR-optimal-is-nondecreasing-(ii)-subcase-2(a)-999}
and the definition of $B({\dbf^*}_1^M;k)$
in \reqnarray{maximum representable integer} that
\beqnarray{proof-OQ-LR-optimal-is-nondecreasing-(ii)-subcase-2(a)-aaa}
\sum_{\ell=1}^{M}I_{\ell}(x';{\dbf^*}_1^M)\leq k.
\eeqnarray
Furthermore, we see from ${\dbf^*}_1^M \in \Acal_M$,
$\sum_{\ell=\ell_1+1}^{M}d_{\ell}^*\leq x'\leq B({\dbf^*}_1^M;k)\leq \sum_{\ell=1}^{M}d_{\ell}^*$
in \reqnarray{proof-OQ-LR-optimal-is-nondecreasing-(ii)-subcase-2(a)-999},
$2\leq \ell_1\leq M-2$, and \rlemma{sum of C-transform}(ii)
(with $m=M$ and $\ell'=\ell_1$ in \rlemma{sum of C-transform}(ii)) that
\beqnarray{proof-OQ-LR-optimal-is-nondecreasing-(ii)-subcase-2(a)-bbb}
\sum_{\ell=1}^{M}I_{\ell}(x';{\dbf^*}_1^M)
=\sum_{\ell=1}^{\ell_1}I_{\ell}(d_{\ell_1+1}^*;{\dbf^*}_1^{\ell_1})+M-\ell_1.
\eeqnarray
As we have $d_{\ell_1+1}^*=B({\dbf^*}_1^{\ell_1};k_{\ell_1+1})+1$
in \reqnarray{proof-OQ-LR-optimal-is-nondecreasing-(ii)-case-2-fff}
and $d_{\ell_1+1}^*<d_{\ell_1}^*\leq \sum_{\ell=1}^{\ell_1}d_{\ell}^*$,
it is clear from the definition of $B({\dbf^*}_1^{\ell_1};k_{\ell_1+1})$
in \reqnarray{maximum representable integer} that
\beqnarray{proof-OQ-LR-optimal-is-nondecreasing-(ii)-subcase-2(a)-ccc}
\sum_{\ell=1}^{\ell_1}I_{\ell}(d_{\ell_1+1}^*;{\dbf^*}_1^{\ell_1})
\geq k_{\ell_1+1}+1.
\eeqnarray
It is also easy to see from $k_M=k$ and $k_{\ell}\geq \max\{k_{\ell+1}-1,0\}\geq k_{\ell+1}-1$
for $\ell_1+1\leq \ell\leq M-1$
in \reqnarray{proof-OQ-LR-optimal-is-nondecreasing-(ii)-case-2-bbb} that
\beqnarray{proof-OQ-LR-optimal-is-nondecreasing-(ii)-subcase-2(a)-ddd}
k_{\ell_1+1}\geq k_M-(M-\ell_1-1)=k-M+\ell_1+1.
\eeqnarray
As such, we have from
\reqnarray{proof-OQ-LR-optimal-is-nondecreasing-(ii)-subcase-2(a)-bbb},
\reqnarray{proof-OQ-LR-optimal-is-nondecreasing-(ii)-subcase-2(a)-ccc},
and \reqnarray{proof-OQ-LR-optimal-is-nondecreasing-(ii)-subcase-2(a)-ddd} that
\beqnarray{}
\sum_{\ell=1}^{M}I_{\ell}(x';{\dbf^*}_1^M)
\geq k_{\ell_1+1}+1+M-\ell_1
\geq (k-M+\ell_1+1)+1+M-\ell_1=k+2>k, \nn
\eeqnarray
contradicting to $\sum_{\ell=1}^{M}I_{\ell}(x';{\dbf^*}_1^M)\leq k$
in \reqnarray{proof-OQ-LR-optimal-is-nondecreasing-(ii)-subcase-2(a)-aaa}.

To prove ${\dbf'}_1^M\in \Acal_M$
in \reqnarray{proof-OQ-LR-optimal-is-nondecreasing-(ii)-subcase-2(a)-111},
we see from ${\dbf^*}_1^M\in \Acal_M$, $d'_{\ell}=d_{\ell}^*$ for $1\leq \ell\leq \ell_1-1$,
$d'_{\ell}=d_{\ell+1}^*$ for $\ell_1\leq \ell\leq M-1$, $d'_M=B({{\dbf}^*}_1^M;k)+1$,
$\sum_{\ell'=\ell_1}^{\ell}d_{\ell'}^*>d_{\ell+1}^*$ for $\ell_1\leq \ell\leq M-1$
in \reqnarray{proof-OQ-LR-optimal-is-nondecreasing-(ii)-subcase-2(a)-222},
and $I_{\ell_1}(B({\dbf^*}_1^M;k);{\dbf^*}_1^M)=0$
in \reqnarray{proof-OQ-LR-optimal-is-nondecreasing-(ii)-subcase-2(a)-777} that
\beqnarray{}
\alignspace
{\dbf'}_1^{\ell_1-1}={\dbf^*}_1^{\ell_1-1}\in \Acal_{\ell_1-1},
\label{eqn:proof-OQ-LR-optimal-is-nondecreasing-(ii)-subcase-2(a)-eee} \\
\alignspace
1\leq d'_{\ell}=d_{\ell+1}^*<\sum_{\ell'=\ell_1}^{\ell}d_{\ell'}^*
=d_{\ell_1}^*+\sum_{\ell'=\ell_1+1}^{\ell}d_{\ell'}^*
\leq \left(\sum_{\ell'=1}^{\ell_1-1}d_{\ell'}^*+1\right)
+\sum_{\ell'=\ell_1+1}^{\ell}d_{\ell'}^* \nn\\
\alignspace \hspace*{0.5in}
=\sum_{\ell'=1}^{\ell_1-1}d'_{\ell'}+1+\sum_{\ell'=\ell_1}^{\ell-1}d'_{\ell'}
=\sum_{\ell=1}^{\ell-1}d'_{\ell}+1,
\textrm{ for } \ell_1\leq \ell\leq M-1,
\label{eqn:proof-OQ-LR-optimal-is-nondecreasing-(ii)-subcase-2(a)-fff} \\
\alignspace
1\leq d'_M=B({{\dbf}^*}_1^M;k)+1
=\sum_{\ell=1}^{M}I_{\ell}(B({\dbf^*}_1^M;k);{\dbf^*}_1^M)d_{\ell}^*+1 \nn\\
\alignspace \hspace*{0.58in}
=\sum_{\ell=1}^{\ell_1-1}I_{\ell}(B({\dbf^*}_1^M;k);{\dbf^*}_1^M)d_{\ell}^*
+\sum_{\ell=\ell_1+1}^{M}I_{\ell}(B({\dbf^*}_1^M;k);{\dbf^*}_1^M)d_{\ell}^*+1 \nn\\
\alignspace \hspace*{0.58in}
\leq \sum_{\ell=1}^{\ell_1-1}d_{\ell}^*+\sum_{\ell=\ell_1+1}^{M}d_{\ell}^*+1
=\sum_{\ell'=1}^{\ell_1-1}d'_{\ell'}+\sum_{\ell'=\ell_1}^{M-1}d'_{\ell'}+1
=\sum_{\ell=1}^{M-1}d'_{\ell}+1.
\label{eqn:proof-OQ-LR-optimal-is-nondecreasing-(ii)-subcase-2(a)-ggg}
\eeqnarray
From \reqnarray{proof-OQ-LR-optimal-is-nondecreasing-(ii)-subcase-2(a)-eee}--\reqnarray{proof-OQ-LR-optimal-is-nondecreasing-(ii)-subcase-2(a)-ggg},
we immediately obtain ${\dbf'}_1^M\in \Acal_M$.

To prove $B({\dbf'}_1^M;k)>B({\dbf^*}_1^M;k)$
in \reqnarray{proof-OQ-LR-optimal-is-nondecreasing-(ii)-subcase-2(a)-111},
we need to show that for $0\leq x\leq B({\dbf^*}_1^M;k)$, we have
\beqnarray{}
\alignspace
I_{\ell}(x;{\dbf'}_1^{M-1})=I_{\ell+1}(x;{\dbf^*}_1^M),
\textrm{ for } \ell=M-1,M-2,\ldots,\ell_1,
\label{eqn:proof-OQ-LR-optimal-is-nondecreasing-(ii)-subcase-2(a)-hhh} \\
\alignspace
I_{\ell}(x;{\dbf'}_1^{M-1})=I_{\ell}(x;{\dbf^*}_1^M),
\textrm{ for } \ell=\ell_1-1,\ell_1-2,\ldots,1,
\label{eqn:proof-OQ-LR-optimal-is-nondecreasing-(ii)-subcase-2(a)-iii}
\eeqnarray
Let $0\leq x\leq B({\dbf^*}_1^M;k)$,
let $y_{\ell}=x-\sum_{\ell'=\ell+1}^{M}I_{\ell'}(x;{\dbf^*}_1^M)d_{\ell'}^*$
for $1\leq \ell\leq M$,
and let $z_{\ell}=x-\sum_{\ell'=\ell+1}^{M-1}I_{\ell'}(x;{\dbf'}_1^{M-1})d'_{\ell'}$
for $1\leq \ell\leq M-1$.
Note that we have from \reqnarray{C-transform} that
\beqnarray{}
\alignspace
I_{\ell}(x;{\dbf^*}_1^M)=1 \textrm{ if and only if } y_{\ell}\geq d_{\ell}^*,
\textrm{ for } 1\leq \ell\leq M,
\label{eqn:proof-OQ-LR-optimal-is-nondecreasing-(ii)-subcase-2(a)-jjj} \\
\alignspace
I_{\ell}(x;{\dbf'}_1^{M-1})=1 \textrm{ if and only if } z_{\ell}\geq d'_{\ell},
\textrm{ for } 1\leq \ell\leq M-1,
\label{eqn:proof-OQ-LR-optimal-is-nondecreasing-(ii)-subcase-2(a)-kkk}
\eeqnarray
We first prove \reqnarray{proof-OQ-LR-optimal-is-nondecreasing-(ii)-subcase-2(a)-hhh}
by induction on $\ell$.
As $z_{M-1}=y_M=x$ and $d'_{M-1}=d_M^*$,
we have $z_{M-1}\geq d'_{M-1}$ if and only if $y_M\geq d_M^*$,
and it follows from \reqnarray{proof-OQ-LR-optimal-is-nondecreasing-(ii)-subcase-2(a)-jjj}
and \reqnarray{proof-OQ-LR-optimal-is-nondecreasing-(ii)-subcase-2(a)-kkk} that
\beqnarray{}
I_{M-1}(x;{\dbf'}_1^{M-1})=I_M(x;{\dbf^*}_1^M). \nn
\eeqnarray
Assume as the induction hypothesis that
$I_{M-1}(x;{\dbf'}_1^{M-1})=I_M(x;{\dbf^*}_1^M),
I_{M-2}(x;{\dbf'}_1^{M-1})=I_{M-1}(x;{\dbf^*}_1^M),\ldots,
I_{\ell+1}(x;{\dbf'}_1^{M-1})=I_{\ell+2}(x;{\dbf^*}_1^M)$
for some $\ell_1\leq \ell\leq M-2$.
Then we have from
$z_{\ell}=x-\sum_{\ell'=\ell+1}^{M-1}I_{\ell'}(x;{\dbf'}_1^{M-1})d'_{\ell'}$,
the induction hypothesis,
$d'_{\ell'}=d_{\ell'+1}^*$ for $\ell_1\leq \ell'\leq M-1$,
$\ell_1\leq \ell\leq M-2$,
and $y_{\ell+1}=x-\sum_{\ell'=\ell+2}^{M}I_{\ell'}(x;{\dbf^*}_1^M)d_{\ell'}^*$ that
\beqnarray{proof-OQ-LR-optimal-is-nondecreasing-(ii)-subcase-2(a)-1111}
z_{\ell}
\aligneq x-\sum_{\ell'=\ell+1}^{M-1}I_{\ell'}(x;{\dbf'}_1^{M-1})d'_{\ell'}
=x-\sum_{\ell'=\ell+1}^{M-1}I_{\ell'+1}(x;{\dbf^*}_1^M)d_{\ell'+1}^* \nn\\
\aligneq x-\sum_{\ell'=\ell+2}^{M}I_{\ell'}(x;{\dbf^*}_1^M)d_{\ell'}^*
=y_{\ell+1}.
\eeqnarray
As such, we see from $z_{\ell}=y_{\ell+1}$
in \reqnarray{proof-OQ-LR-optimal-is-nondecreasing-(ii)-subcase-2(a)-1111}
and $d'_{\ell}=d_{\ell+1}^*$ (as $\ell_1\leq \ell\leq M-2$)
that $z_{\ell}\geq d'_{\ell}$ if and only if $y_{\ell+1}\geq d_{\ell+1}^*$,
and it follows from
\reqnarray{proof-OQ-LR-optimal-is-nondecreasing-(ii)-subcase-2(a)-jjj}
and \reqnarray{proof-OQ-LR-optimal-is-nondecreasing-(ii)-subcase-2(a)-kkk} that
\beqnarray{}
I_{\ell}(x;{\dbf'}_1^{M-1})=I_{\ell+1}(x;{\dbf^*}_1^M). \nn
\eeqnarray
Therefore, the induction is completed
and \reqnarray{proof-OQ-LR-optimal-is-nondecreasing-(ii)-subcase-2(a)-hhh} is proved.
Now we prove \reqnarray{proof-OQ-LR-optimal-is-nondecreasing-(ii)-subcase-2(a)-iii}
by induction on $\ell$.
From \reqnarray{proof-OQ-LR-optimal-is-nondecreasing-(ii)-subcase-2(a)-hhh},
$d'_{\ell'}=d_{\ell'+1}^*$ for $\ell_1\leq \ell'\leq M-1$,
and $I_{\ell_1}(x;{\dbf^*}_1^M)=0$
in \reqnarray{proof-OQ-LR-optimal-is-nondecreasing-(ii)-subcase-2(a)-777},
we see that
\beqnarray{proof-OQ-LR-optimal-is-nondecreasing-(ii)-subcase-2(a)-2222}
z_{\ell_1-1}
\aligneq x-\sum_{\ell'=\ell_1}^{M-1}I_{\ell'}(x;{\dbf'}_1^{M-1})d'_{\ell'}
=x-\sum_{\ell'=\ell_1}^{M-1}I_{\ell'+1}(x;{\dbf^*}_1^M)d_{\ell'+1}^* \nn\\
\aligneq x-0\cdot d_{\ell_1}^*
-\sum_{\ell'=\ell_1+1}^{M}I_{\ell'}(x;{\dbf^*}_1^M)d_{\ell'}^*
=x-\sum_{\ell'=\ell_1}^{M}I_{\ell'}(x;{\dbf^*}_1^M)d_{\ell'}^*
=y_{\ell_1-1}.
\eeqnarray
As we have from $z_{\ell_1-1}=y_{\ell_1-1}$
in \reqnarray{proof-OQ-LR-optimal-is-nondecreasing-(ii)-subcase-2(a)-2222}
and $d'_{\ell_1-1}=d_{\ell_1-1}^*$
that $z_{\ell_1-1}\geq d'_{\ell_1-1}$ if and only if $y_{\ell_1-1}\geq d_{\ell_1-1}^*$,
it follows from
\reqnarray{proof-OQ-LR-optimal-is-nondecreasing-(ii)-subcase-2(a)-jjj}
and \reqnarray{proof-OQ-LR-optimal-is-nondecreasing-(ii)-subcase-2(a)-kkk} that
\beqnarray{}
I_{\ell_1-1}(x;{\dbf'}_1^{M-1})=I_{\ell_1-1}(x;{\dbf^*}_1^M). \nn
\eeqnarray
Assume as the induction hypothesis that
$I_{\ell_1-1}(x;{\dbf'}_1^{M-1})=I_{\ell_1-1}(x;{\dbf^*}_1^M),
I_{\ell_1-2}(x;{\dbf'}_1^{M-1})=I_{\ell_1-2}(x;{\dbf^*}_1^M),\ldots,
I_{\ell+1}(x;{\dbf'}_1^{M-1})=I_{\ell+1}(x;{\dbf^*}_1^M)$
for some $1\leq \ell\leq \ell_1-2$.
Then we have from the induction hypothesis,
\reqnarray{proof-OQ-LR-optimal-is-nondecreasing-(ii)-subcase-2(a)-hhh},
$d'_{\ell'}=d_{\ell'}^*$ for $1\leq \ell'\leq \ell_1-1$,
$d'_{\ell'}=d_{\ell'+1}^*$ for $\ell_1\leq \ell'\leq M-1$,
and $I_{\ell_1}(x;{\dbf^*}_1^M)=0$
in \reqnarray{proof-OQ-LR-optimal-is-nondecreasing-(ii)-subcase-2(a)-777} that
\beqnarray{proof-OQ-LR-optimal-is-nondecreasing-(ii)-subcase-2(a)-3333}
z_{\ell}
\aligneq x-\sum_{\ell'=\ell+1}^{M-1}I_{\ell'}(x;{\dbf'}_1^{M-1})d'_{\ell'}
=x-\sum_{\ell'=\ell+1}^{\ell_1-1}I_{\ell'}(x;{\dbf'}_1^{M-1})d'_{\ell'}
-\sum_{\ell'=\ell_1}^{M-1}I_{\ell'}(x;{\dbf'}_1^{M-1})d'_{\ell'} \nn\\
\aligneq x-\sum_{\ell'=\ell+1}^{\ell_1-1}I_{\ell'}(x;{\dbf^*}_1^M)d_{\ell'}^*
-\sum_{\ell'=\ell_1}^{M-1}I_{\ell'+1}(x;{\dbf^*}_1^M)d_{\ell'+1}^* \nn\\
\aligneq x-\sum_{\ell'=\ell+1}^{\ell_1-1}I_{\ell'}(x;{\dbf^*}_1^M)d_{\ell'}^*
-0\cdot d_{\ell_1}^*-\sum_{\ell'=\ell_1+1}^{M}I_{\ell'}(x;{\dbf^*}_1^M)d_{\ell'}^* \nn\\
\aligneq x-\sum_{\ell'=\ell+1}^{M}I_{\ell'}(x;{\dbf^*}_1^M)d_{\ell'}^*
=y_{\ell}.
\eeqnarray
From $z_{\ell}=y_{\ell}$
in \reqnarray{proof-OQ-LR-optimal-is-nondecreasing-(ii)-subcase-2(a)-3333}
and $d'_{\ell}=d_{\ell}^*$ (as $1\leq \ell\leq \ell_1-2$),
we see that $z_{\ell}\geq d'_{\ell}$ if and only if $y_{\ell}\geq d_{\ell}^*$,
and hence it follows from
\reqnarray{proof-OQ-LR-optimal-is-nondecreasing-(ii)-subcase-2(a)-jjj}
and \reqnarray{proof-OQ-LR-optimal-is-nondecreasing-(ii)-subcase-2(a)-kkk} that
\beqnarray{}
I_{\ell}(x;{\dbf'}_1^{M-1})=I_{\ell}(x;{\dbf^*}_1^M). \nn
\eeqnarray
Therefore, the induction is completed
and \reqnarray{proof-OQ-LR-optimal-is-nondecreasing-(ii)-subcase-2(a)-iii} is proved.

For $0\leq x\leq B({\dbf^*}_1^M;k)$,
it is clear from
\reqnarray{proof-OQ-LR-optimal-is-nondecreasing-(ii)-subcase-2(a)-iii},
\reqnarray{proof-OQ-LR-optimal-is-nondecreasing-(ii)-subcase-2(a)-hhh},
$d'_{\ell}=d_{\ell}^*$ for $1\leq \ell\leq \ell_1-1$,
$d'_{\ell}=d_{\ell+1}^*$ for $\ell_1\leq \ell\leq M-1$,
$I_{\ell_1}(x;{\dbf^*}_1^M)=0$
in \reqnarray{proof-OQ-LR-optimal-is-nondecreasing-(ii)-subcase-2(a)-777},
and the definition of $B({\dbf^*}_1^M;k)$
in \reqnarray{maximum representable integer} that
\beqnarray{proof-OQ-LR-optimal-is-nondecreasing-(ii)-subcase-2(a)-4444}
\sum_{\ell=1}^{M-1}I_{\ell}(x;{\dbf'}_1^{M-1})d'_{\ell}
\aligneq
\sum_{\ell=1}^{\ell_1-1}I_{\ell}(x;{\dbf'}_1^{M-1})d'_{\ell}
+\sum_{\ell=\ell_1}^{M-1}I_{\ell}(x;{\dbf'}_1^{M-1})d'_{\ell} \nn\\
\aligneq
\sum_{\ell=1}^{\ell_1-1}I_{\ell}(x;{\dbf^*}_1^M)d_{\ell}^*
+\sum_{\ell=\ell_1}^{M-1}I_{\ell+1}(x;{\dbf^*}_1^M)d_{\ell+1}^* \nn\\
\aligneq
\sum_{\ell=1}^{\ell_1-1}I_{\ell}(x;{\dbf^*}_1^M)d_{\ell}^*
+0\cdot d_{\ell_1}^*
+\sum_{\ell=\ell_1+1}^{M}I_{\ell}(x;{\dbf^*}_1^M)d_{\ell}^* \nn\\
\aligneq
\sum_{\ell=1}^{M}I_{\ell}(x;{\dbf^*}_1^M)d_{\ell}^*\leq k,
\textrm{ for } 0\leq x\leq B({\dbf^*}_1^M;k).
\eeqnarray
It then follows from \reqnarray{proof-OQ-LR-optimal-is-nondecreasing-(ii)-subcase-2(a)-4444},
$B({\dbf^*}_1^M;k)\leq \sum_{\ell=1}^{M-1}d'_{\ell}$
in \reqnarray{proof-OQ-LR-optimal-is-nondecreasing-(ii)-subcase-2(a)-ggg},
the definition of $B({\dbf'}_1^{M-1};k)$
in \reqnarray{maximum representable integer},
and $d'_M=B({\dbf^*}_1^M;k)+1$ that
\beqnarray{proof-OQ-LR-optimal-is-nondecreasing-(ii)-subcase-2(a)-5555}
B({\dbf'}_1^{M-1};k)\geq B({\dbf^*}_1^M;k)=d'_M-1.
\eeqnarray
As such, we have from ${\dbf'}_1^M\in \Acal_M$,
$B({\dbf'}_1^{M-1};k)\geq d'_M-1$
in \reqnarray{proof-OQ-LR-optimal-is-nondecreasing-(ii)-subcase-2(a)-5555},
$1\leq k<M$, \rlemma{OQ-LR-MRI-general}(ii)
(with $m=M$ and $i=k$ in \rlemma{OQ-LR-MRI-general}(ii)),
and $d'_M=B({\dbf^*}_1^M;k)+1$ that
\beqnarray{}
B({\dbf'}_1^M;k)=d'_M+B({\dbf'}_1^{M-1};k-1)
=B({\dbf^*}_1^M;k)+1+B({\dbf'}_1^{M-1};k-1)
>B({\dbf^*}_1^M;k). \nn
\eeqnarray
Therefore, we have proved that $B({\dbf'}_1^M;k)>B({\dbf^*}_1^M;k)$
in \reqnarray{proof-OQ-LR-optimal-is-nondecreasing-(ii)-subcase-2(a)-111}

\emph{Subcase 2(b): $E_{M-1}\neq \emptyset,E_{M-2}\neq \emptyset,\ldots,E_{\ell_2+1}\neq \emptyset$,
and $E_{\ell_2}=\emptyset$ for some $\ell_1+1\leq \ell_2\leq M-1$.}
In this subcase, we let $d'_{\ell}=d_{\ell}^*$ for $\ell=1,2,\ldots,\ell_2-1$
and $d'_{\ell}=d_{\ell}^*+1$ for $\ell=\ell_2,\ell_2+1,\ldots,M$.
We will show that
\beqnarray{proof-OQ-LR-optimal-is-nondecreasing-(ii)-subcase-2(b)-111}
{\dbf'}_1^M\in \Acal_M \textrm{ and } B({\dbf'}_1^M;k)>B({\dbf^*}_1^M;k).
\eeqnarray
As it is clear from ${\dbf'}_1^M\in \Acal_M$ that
$B({\dbf'}_1^M;k)\leq \max_{\dbf_1^M\in \Acal_M}B(\dbf_1^M;k)=B({\dbf^*}_1^M;k)$,
we have reached a contradiction to $B({\dbf'}_1^M;k)>B({\dbf^*}_1^M;k)$
in \reqnarray{proof-OQ-LR-optimal-is-nondecreasing-(ii)-subcase-2(b)-111}.
Note that for the example
that $M=9$, $k=8$, and ${\dbf^*}_1^M=(1,2,3,5,2,8,22,44,88)$
as shown in \rtable{OQ-LR-optimal-is-nondecreasing-2},
we have $B({\dbf^*}_1^M;k)=174$.
If we let ${\dbf'}_1^M=(1,2,3,5,2,9,23,45,89)$,
then it can be seen that $B({\dbf'}_1^M;k)=178>B({\dbf^*}_1^M;k)=174$
and \reqnarray{proof-OQ-LR-optimal-is-nondecreasing-(ii)-subcase-2(b)-111} is satisfied.
Also note that for the example
that $M=9$, $k=5$, and ${\dbf^*}_1^M=(1,2,3,5,2,8,14,28,56)$
as shown in \rtable{OQ-LR-optimal-is-nondecreasing-3},
we have $B({\dbf^*}_1^M;k)=103$.
If we let ${\dbf'}_1^M=(1,2,3,5,2,9,15,29,57)$,
then it can be seen that $B({\dbf'}_1^M;k)=106>B({\dbf^*}_1^M;k)=103$
and \reqnarray{proof-OQ-LR-optimal-is-nondecreasing-(ii)-subcase-2(b)-111} is satisfied.

We first show that
\beqnarray{proof-OQ-LR-optimal-is-nondecreasing-(ii)-subcase-2(b)-222}
k_{\ell_2+1}\geq 1.
\eeqnarray
Suppose on the contrary that $k_{\ell_2+1}=0$.
Then we have from \reqnarray{proof-OQ-LR-optimal-is-nondecreasing-(ii)-case-2-jjj} that
\beqnarray{proof-OQ-LR-optimal-is-nondecreasing-(ii)-subcase-2(b)-333}
d_{\ell_2+1}^*=B({\dbf^*}_1^{\ell_2};k_{\ell_2+1})+1=B({\dbf^*}_1^{\ell_2};0)+1=1.
\eeqnarray
From $d_{\ell_1+1}^*\leq d_{\ell_1+2}^*\leq \cdots\leq d_M^*$,
$\ell_1+1\leq \ell_2\leq M-1$, and $d_{\ell_2+1}^*=1$
in \reqnarray{proof-OQ-LR-optimal-is-nondecreasing-(ii)-subcase-2(b)-333},
it is clear that $1\leq d_{\ell_2}^*\leq d_{\ell_2+1}^*=1$,
which implies that $d_{\ell_2}^*=1$.
Thus, it follows from $d_{\ell_2}^*=1=B({\dbf^*}_1^{\ell_2-1};0)+1$,
$k_{\ell_2+1}=\max\{k_{\ell_2+1}-1,0\}=0$,
and \reqnarray{proof-OQ-LR-optimal-is-nondecreasing-(ii)-case-2-ggg}
that $E_{\ell_2}=\{0\}$, contradicting to $E_{\ell_2}=\emptyset$.

Furthermore, we show that
\beqnarray{proof-OQ-LR-optimal-is-nondecreasing-(ii)-subcase-2(b)-444}
d_{\ell_2}^*<B({\dbf^*}_1^{\ell_2-1};k_{\ell_2+1})+1.
\eeqnarray
Suppose on the contrary that $d_{\ell_2}^*\geq B({\dbf^*}_1^{\ell_2-1};k_{\ell_2+1})+1$.
As $k_{\ell_2+1}\geq 1$ and
$E_{\ell_2}=\{\max\{k_{\ell_2+1}-1,0\}\leq i\leq k_{\ell_2+1}:d_{\ell_2}^*
=B({\dbf^*}_1^{\ell_2-1};i)+1\}=\emptyset$, we have
\beqnarray{proof-OQ-LR-optimal-is-nondecreasing-(ii)-subcase-2(b)-555}
d_{\ell_2}^*\neq B({\dbf^*}_1^{\ell_2-1};k_{\ell_2+1}-1)+1
\textrm{ and } d_{\ell_2}^*\neq B({\dbf^*}_1^{\ell_2-1};k_{\ell_2+1})+1.
\eeqnarray
It follows from $d_{\ell_2}^*\geq B({\dbf^*}_1^{\ell_2-1};k_{\ell_2+1})+1$
and $d_{\ell_2}^*\neq B({\dbf^*}_1^{\ell_2-1};k_{\ell_2+1})+1$
in \reqnarray{proof-OQ-LR-optimal-is-nondecreasing-(ii)-subcase-2(b)-555}
that $d_{\ell_2}^*>B({\dbf^*}_1^{\ell_2-1};k_{\ell_2+1})+1$.
Thus, we have from ${\dbf^*}_1^{\ell_2}\in \Acal_{\ell_2}$
(as ${\dbf^*}_1^M\in \Acal_M$ and $\ell_2<M$),
$\ell_2\geq \ell_1+1>1$, $k_{\ell_2+1}\geq 1$
in \reqnarray{proof-OQ-LR-optimal-is-nondecreasing-(ii)-subcase-2(b)-222},
$B({\dbf^*}_1^{\ell_2-1};k_{\ell_2+1})<d_{\ell_2}^*-1$,
and \rlemma{OQ-LR-MRI-general}(i) (with $m=\ell_2$, $i=k_{\ell_2+1}$,
and $\ell'=m-1=\ell_2-1$ in \rlemma{OQ-LR-MRI-general}(i)) that
\beqnarray{proof-OQ-LR-optimal-is-nondecreasing-(ii)-subcase-2(b)-666}
B({\dbf^*}_1^{\ell_2};k_{\ell_2+1})=B({\dbf^*}_1^{\ell_2-1};k_{\ell_2+1}).
\eeqnarray
As such, it follows from
\reqnarray{proof-OQ-LR-optimal-is-nondecreasing-(ii)-case-2-jjj},
\reqnarray{proof-OQ-LR-optimal-is-nondecreasing-(ii)-subcase-2(b)-666},
and $d_{\ell_2}^*>B({\dbf^*}_1^{\ell_2-1};k_{\ell_2+1})+1$ that
\beqnarray{proof-OQ-LR-optimal-is-nondecreasing-(ii)-subcase-2(b)-777}
d_{\ell_2+1}^*=B({\dbf^*}_1^{\ell_2};k_{\ell_2+1})+1
=B({\dbf^*}_1^{\ell_2-1};k_{\ell_2+1})+1<d_{\ell_2}^*.
\eeqnarray
As  $\ell_1=\max\{2\leq \ell\leq M-1:d_{\ell}^*>d_{\ell+1}^*\}$,
it is clear from
\reqnarray{proof-OQ-LR-optimal-is-nondecreasing-(ii)-subcase-2(b)-777}
that $\ell_2\leq \ell_1$,
and we have reached a contradiction to $\ell_2\geq \ell_1+1$.

To prove ${\dbf'}_1^M\in \Acal_M$
in \reqnarray{proof-OQ-LR-optimal-is-nondecreasing-(ii)-subcase-2(b)-111},
we see from ${\dbf^*}_1^M\in \Acal_M$, $d'_{\ell}=d_{\ell}^*$ for $1\leq \ell\leq \ell_2-1$,
$d'_{\ell}=d_{\ell}^*+1$ for $\ell=\ell_2,\ell_2+1,\ldots,M$,
$d_{\ell_2}^*<B({\dbf^*}_1^{\ell_2-1};k_{\ell_2+1})+1$
in \reqnarray{proof-OQ-LR-optimal-is-nondecreasing-(ii)-subcase-2(b)-444},
and $B({\dbf^*}_1^{\ell_2-1};k_{\ell_2+1})\leq \sum_{\ell'=1}^{\ell_2-1}d_{\ell'}^*$
that
\beqnarray{}
\alignspace
{\dbf'}_1^{\ell_2-1}={\dbf^*}_1^{\ell_2-1}\in \Acal_{\ell_2-1},
\label{eqn:proof-OQ-LR-optimal-is-nondecreasing-(ii)-subcase-2(b)-888} \\
\alignspace
1\leq d'_{\ell_2}=d_{\ell_2}^*+1\leq B({\dbf^*}_1^{\ell_2-1};k_{\ell_2+1})+1
\leq \sum_{\ell'=1}^{\ell_2-1}d_{\ell'}^*+1,
\label{eqn:proof-OQ-LR-optimal-is-nondecreasing-(ii)-subcase-2(b)-999} \\
\alignspace
1\leq d'_{\ell}=d_{\ell}^*+1
\leq \left(\sum_{\ell'=1}^{\ell-1}d_{\ell'}^*+1\right)+1
=\sum_{\ell'=1}^{\ell_2-1}d_{\ell'}^*+\sum_{\ell'=\ell_2}^{\ell-1}d_{\ell'}^*+2 \nn\\
\alignspace \hspace*{0.49in}
=\sum_{\ell'=1}^{\ell_2-1}d'_{\ell'}+\sum_{\ell'=\ell_2}^{\ell-1}(d'_{\ell'}-1)+2
=\sum_{\ell'=1}^{\ell-1}d'_{\ell'}-(\ell-\ell_2)+2 \nn\\
\alignspace \hspace*{0.49in}
\leq \sum_{\ell'=1}^{\ell-1}d'_{\ell'}+1, \textrm{ for } \ell_2+1\leq \ell \leq M.
\label{eqn:proof-OQ-LR-optimal-is-nondecreasing-(ii)-subcase-2(b)-aaa}
\eeqnarray
From \reqnarray{proof-OQ-LR-optimal-is-nondecreasing-(ii)-subcase-2(b)-888}--\reqnarray{proof-OQ-LR-optimal-is-nondecreasing-(ii)-subcase-2(b)-aaa},
we immediately obtain ${\dbf'}_1^M\in \Acal_M$.

To prove $B({\dbf'}_1^M;k)>B({\dbf^*}_1^M;k)$
in \reqnarray{proof-OQ-LR-optimal-is-nondecreasing-(ii)-subcase-2(b)-111},
we show that
\beqnarray{}
\alignspace
B({\dbf'}_1^{\ell};k_{\ell_2+1}-1)\geq B({\dbf^*}_1^{\ell};k_{\ell_2+1}-1),
\textrm{ for } \ell_2\leq \ell\leq M,
\label{eqn:proof-OQ-LR-optimal-is-nondecreasing-(ii)-subcase-2(b)-bbb} \\
\alignspace
B({\dbf'}_1^{\ell};i)\geq B({\dbf^*}_1^{\ell};i)+1,
\textrm{ for } \ell_2\leq \ell\leq M \textrm{ and } k_{\ell_2+1}\leq i\leq k.
\label{eqn:proof-OQ-LR-optimal-is-nondecreasing-(ii)-subcase-2(b)-ccc}
\eeqnarray
It is clear that with $\ell=M$ and $i=k$
in \reqnarray{proof-OQ-LR-optimal-is-nondecreasing-(ii)-subcase-2(b)-ccc},
we have $B({\dbf'}_1^M;k)>B({\dbf^*}_1^M;k)$
in \reqnarray{proof-OQ-LR-optimal-is-nondecreasing-(ii)-subcase-2(b)-111}.
Note that for the example
that $M=9$, $k=8$, and ${\dbf^*}_1^M=(1,2,3,5,2,8,22,44,88)$
as shown in \rtable{OQ-LR-optimal-is-nondecreasing-2},
we have $\ell_1=4$, $\ell_2=6$, and $k_{\ell_2+1}=k_7=6$.
If we let ${\dbf'}_1^M=(1,2,3,5,2,9,23,45,89)$,
then it can be seen from the values of $B({\dbf^*}_1^{\ell};i)$
for $\ell_1\leq \ell\leq M$ and $0\leq i\leq k$
in \rtable{OQ-LR-optimal-is-nondecreasing-2}
and the values of $B({\dbf'}_1^{\ell};i)$
for $\ell_1\leq \ell\leq M$ and $0\leq i\leq k$
in \rtable{OQ-LR-optimal-is-nondecreasing-2'}
that \reqnarray{proof-OQ-LR-optimal-is-nondecreasing-(ii)-subcase-2(b)-bbb}
and \reqnarray{proof-OQ-LR-optimal-is-nondecreasing-(ii)-subcase-2(b)-ccc} are satisfied.
Also note that for the example
that $M=9$, $k=5$, and ${\dbf^*}_1^M=(1,2,3,5,2,8,14,28,56)$
as shown in \rtable{OQ-LR-optimal-is-nondecreasing-3},
we have $\ell_1=4$, $\ell_2=6$, and $k_{\ell_2+1}=k_7=3$.
If we let ${\dbf'}_1^M=(1,2,3,5,2,9,15,29,57)$,
then it can be seen from the values of $B({\dbf^*}_1^{\ell};i)$
for $\ell_1\leq \ell\leq M$ and $0\leq i\leq k$
in \rtable{OQ-LR-optimal-is-nondecreasing-3}
and the values of $B({\dbf'}_1^{\ell};i)$
for $\ell_1\leq \ell\leq M$ and $0\leq i\leq k$
in \rtable{OQ-LR-optimal-is-nondecreasing-3'}
that \reqnarray{proof-OQ-LR-optimal-is-nondecreasing-(ii)-subcase-2(b)-bbb}
and \reqnarray{proof-OQ-LR-optimal-is-nondecreasing-(ii)-subcase-2(b)-ccc} are satisfied.

{\tiny
\btable{htbp}{||c||c|c|c|c|c|c|c|c|c||}
\hline
$\ell \verb+\+ i$ & 0 & 1 &  2 &  3 &  4 &  5 &  6 &   7 &   8 \\
\hline
4                 & 0 & 3 &  8 & 10 & 11 & 11 & 11 &  11 &  11 \\
\hline
5                 & 0 & 2 &  5 & 10 & 12 & 13 & 13 &  13 &  13 \\
\hline
6                 & 0 & 2 &  5 & 14 & 19 & \textbf{21} & \textbf{22} &  \textbf{22} &  \textbf{22} \\
\hline
7                 & 0 & 2 &  5 & 14 & 19 & \textbf{21} & \textbf{44} &  \textbf{45} &  \textbf{45} \\
\hline
8                 & 0 & 2 &  5 & 14 & 19 & \textbf{21} & \textbf{66} &  \textbf{89} &  \textbf{90} \\
\hline
9                 & 0 & 2 &  5 & 14 & 19 & \textbf{21} & \textbf{66} & \textbf{155} & \textbf{178} \\
\hline
\etable{OQ-LR-optimal-is-nondecreasing-2'}
{The values of $B({\dbf'}_1^{\ell};i)$ for $\ell_1\leq \ell\leq M$ and $0\leq i\leq k$,
where $M=9$, $k=8$, ${\dbf'}_1^M=(1,2,3,5,2,9,23,45,89)$,
$\ell_1=4$, $\ell_2=6$, and $k_{\ell_2+1}=6$
(note that the values of $B({\dbf'}_1^{\ell};i)$
for $\ell_2\leq \ell\leq M$ and $k_{\ell_2+1}-1\leq i\leq k$
in \reqnarray{proof-OQ-LR-optimal-is-nondecreasing-(ii)-subcase-2(b)-bbb}
and \reqnarray{proof-OQ-LR-optimal-is-nondecreasing-(ii)-subcase-2(b)-ccc}
are in boldface).}}
{\tiny
\btable{htbp}{||c||c|c|c|c|c|c|c|c|c||}
\hline
$\ell \verb+\+ i$ & 0 & 1 &  2 &  3 &  4 &   5 \\
\hline
4                 & 0 & 3 &  8 & 10 & 11 &  11 \\
\hline
5                 & 0 & 2 &  5 & 10 & 12 &  13 \\
\hline
6                 & 0 & 2 &  \textbf{5} & \textbf{14} & \textbf{19} &  \textbf{21} \\
\hline
7                 & 0 & 2 &  \textbf{5} & \textbf{20} & \textbf{29} &  \textbf{34} \\
\hline
8                 & 0 & 2 &  \textbf{5} & \textbf{20} & \textbf{49} &  \textbf{58} \\
\hline
9                 & 0 & 2 &  \textbf{5} & \textbf{20} & \textbf{49} & \textbf{106} \\
\hline
\etable{OQ-LR-optimal-is-nondecreasing-3'}
{The values of $B({\dbf'}_1^{\ell};i)$ for $\ell_1\leq \ell\leq M$ and $0\leq i\leq k$,
where $M=9$, $k=5$, ${\dbf'}_1^M=(1,2,3,5,2,9,15,29,57)$,
$\ell_1=4$, $\ell_2=6$, and $k_{\ell_2+1}=3$
(note that the values of $B({\dbf'}_1^{\ell};i)$
for $\ell_2\leq \ell\leq M$ and $k_{\ell_2+1}-1\leq i\leq k$
in \reqnarray{proof-OQ-LR-optimal-is-nondecreasing-(ii)-subcase-2(b)-bbb}
and \reqnarray{proof-OQ-LR-optimal-is-nondecreasing-(ii)-subcase-2(b)-ccc}
are in boldface).}}

To prove \reqnarray{proof-OQ-LR-optimal-is-nondecreasing-(ii)-subcase-2(b)-bbb},
we first show that if $B({\dbf^*}_1^{\ell};k_{\ell+1}-1)=B({\dbf^*}_1^{\ell};k_{\ell+1})$
for some $\ell_2\leq \ell\leq M-2$
(note that it is easy to see from
\reqnarray{proof-OQ-LR-optimal-is-nondecreasing-(ii)-case-2-iii},
\reqnarray{proof-OQ-LR-optimal-is-nondecreasing-(ii)-subcase-2(b)-222},
and $\ell_2\leq \ell\leq M-2$ that $k_{\ell+1}\geq k_{\ell_2+1}\geq 1$),
then we have
\beqnarray{proof-OQ-LR-optimal-is-nondecreasing-(ii)-subcase-2(b)-ddd}
k_{\ell+2}=k_{\ell+1}+1 \textrm{ and }
B({\dbf^*}_1^{\ell+1};k_{\ell+2}-1)=B({\dbf^*}_1^{\ell+1};k_{\ell+2}).
\eeqnarray
So suppose that $B({\dbf^*}_1^{\ell};k_{\ell+1}-1)=B({\dbf^*}_1^{\ell};k_{\ell+1})$
for some $\ell_2\leq \ell\leq M-2$.
From $\ell_2\leq \ell\leq M-2$,
\reqnarray{proof-OQ-LR-optimal-is-nondecreasing-(ii)-case-2-jjj},
and $B({\dbf^*}_1^{\ell};k_{\ell+1}-1)=B({\dbf^*}_1^{\ell};k_{\ell+1})$,
we have
\beqnarray{proof-OQ-LR-optimal-is-nondecreasing-(ii)-subcase-2(b)-eee}
d_{\ell+1}^*=B({\dbf^*}_1^{\ell};k_{\ell+1})+1
=B({\dbf^*}_1^{\ell};k_{\ell+1}-1)+1.
\eeqnarray
As $\ell_2\leq \ell\leq M-2$,
it is clear from \reqnarray{proof-OQ-LR-optimal-is-nondecreasing-(ii)-case-2-iii}
and \reqnarray{proof-OQ-LR-optimal-is-nondecreasing-(ii)-subcase-2(b)-222}
that $k_{\ell+2}\geq k_{\ell+1}\geq \cdots \geq k_{\ell_2+1}\geq 1$
and $k_{\ell+2}-1=\max\{k_{\ell+2}-1,0\}\leq k_{\ell+1}\leq k_{\ell+2}$,
i.e., either $k_{\ell+1}=k_{\ell+2}$ or $k_{\ell+1}=k_{\ell+2}-1$.
If $k_{\ell+1}=k_{\ell+2}$,
then we immediately see from
\reqnarray{proof-OQ-LR-optimal-is-nondecreasing-(ii)-subcase-2(b)-eee}
and $k_{\ell+2}\geq 1$
that $d_{\ell+1}^*=B({\dbf^*}_1^{\ell};k_{\ell+2}-1)+1
=B({\dbf^*}_1^{\ell};k_{\ell+2})+1$.
Thus, it follows from
\reqnarray{proof-OQ-LR-optimal-is-nondecreasing-(ii)-case-2-ggg},
\reqnarray{proof-OQ-LR-optimal-is-nondecreasing-(ii)-case-2-hhh},
and $\ell_2\leq \ell\leq M-2$
that $E_{\ell+1}=\{k_{\ell+2}-1,k_{\ell+2}\}$
and $k_{\ell+1}=\min\{i:i\in E_{\ell+1}\}=k_{\ell+2}-1$,
and this contradicts to $k_{\ell+1}=k_{\ell+2}$.
Thus, we must have $k_{\ell+1}=k_{\ell+2}-1$,
i.e., $k_{\ell+2}=k_{\ell+1}+1$
in \reqnarray{proof-OQ-LR-optimal-is-nondecreasing-(ii)-subcase-2(b)-ddd}.
From $d_{\ell+1}^*=B({\dbf^*}_1^{\ell};k_{\ell+1})+1$
in \reqnarray{proof-OQ-LR-optimal-is-nondecreasing-(ii)-subcase-2(b)-eee},
$\ell\geq \ell_2\geq \ell_1+1>1$, $k_{\ell+2}\geq k_{\ell+1}\geq 1$,
and \rlemma{OQ-LR-MRI-monotone}(i)
(with $m=\ell$, $i=k_{\ell+1}$, and $i'=k_{\ell+2}$ in \rlemma{OQ-LR-MRI-monotone}(i)),
we have
\beqnarray{proof-OQ-LR-optimal-is-nondecreasing-(ii)-subcase-2(b)-fff}
d_{\ell+1}^*=B({\dbf^*}_1^{\ell};k_{\ell+1})+1\leq B({\dbf^*}_1^{\ell};k_{\ell+2})+1.
\eeqnarray
As such, we see from ${\dbf^*}_1^{\ell+1}\in \Acal_{\ell+1}$
(as ${\dbf^*}_1^M\in \Acal_M$ and $\ell+1<M$),
$B({\dbf^*}_1^{\ell};k_{\ell+2})\geq d_{\ell+1}^*-1$
in \reqnarray{proof-OQ-LR-optimal-is-nondecreasing-(ii)-subcase-2(b)-fff},
and \rlemma{OQ-LR-MRI-general}(ii)
(with $m=\ell+1>1$ and $i=k_{\ell+2}\geq 1$ in \rlemma{OQ-LR-MRI-general}(ii)) that
\beqnarray{proof-OQ-LR-optimal-is-nondecreasing-(ii)-subcase-2(b)-ggg}
B({\dbf^*}_1^{\ell+1};k_{\ell+2})=d_{\ell+1}^*+B({\dbf^*}_1^{\ell};k_{\ell+2}-1).
\eeqnarray
Similarly, we see from ${\dbf^*}_1^{\ell+1}\in \Acal_{\ell+1}$,
$B({\dbf^*}_1^{\ell};k_{\ell+1})=d_{\ell+1}^*-1$
in \reqnarray{proof-OQ-LR-optimal-is-nondecreasing-(ii)-subcase-2(b)-fff},
and \rlemma{OQ-LR-MRI-general}(ii)
(with $m=\ell+1>1$ and $i=k_{\ell+1}\geq 1$ in \rlemma{OQ-LR-MRI-general}(ii)) that
\beqnarray{proof-OQ-LR-optimal-is-nondecreasing-(ii)-subcase-2(b)-hhh}
B({\dbf^*}_1^{\ell+1};k_{\ell+1})=d_{\ell+1}^*+B({\dbf^*}_1^{\ell};k_{\ell+1}-1).
\eeqnarray
Therefore, we have from
$k_{\ell+2}=k_{\ell+1}+1$,
\reqnarray{proof-OQ-LR-optimal-is-nondecreasing-(ii)-subcase-2(b)-hhh},
$B({\dbf^*}_1^{\ell};k_{\ell+1}-1)=B({\dbf^*}_1^{\ell};k_{\ell+1})$,
and \reqnarray{proof-OQ-LR-optimal-is-nondecreasing-(ii)-subcase-2(b)-ggg} that
\beqnarray{}
B({\dbf^*}_1^{\ell+1};k_{\ell+2}-1)
\aligneq B({\dbf^*}_1^{\ell+1};k_{\ell+1})
=d_{\ell+1}^*+B({\dbf^*}_1^{\ell};k_{\ell+1}-1)
=d_{\ell+1}^*+B({\dbf^*}_1^{\ell};k_{\ell+1}) \nn\\
\aligneq d_{\ell+1}^*+B({\dbf^*}_1^{\ell};k_{\ell+2}-1)
=B({\dbf^*}_1^{\ell+1};k_{\ell+2}), \nn
\eeqnarray
which is the desired result that
$B({\dbf^*}_1^{\ell+1};k_{\ell+2}-1)=B({\dbf^*}_1^{\ell+1};k_{\ell+2})$
in \reqnarray{proof-OQ-LR-optimal-is-nondecreasing-(ii)-subcase-2(b)-ddd}.

To continue with the proof of
\reqnarray{proof-OQ-LR-optimal-is-nondecreasing-(ii)-subcase-2(b)-bbb},
note that from ${\dbf^*}_1^{\ell_2}\in \Acal_{\ell_2}$
(as ${\dbf^*}_1^M\in \Acal_M$ and $\ell_2<M$),
$\ell_2\geq \ell_1+1>1$, $k_{\ell_2+1}-1\geq 0$
in \reqnarray{proof-OQ-LR-optimal-is-nondecreasing-(ii)-subcase-2(b)-222},
and \rlemma{OQ-LR-MRI-monotone}(i)
(with $m=\ell_2$, $i=k_{\ell_2+1}-1$, and $i'=k_{\ell_2+1}$ in \rlemma{OQ-LR-MRI-monotone}(i)),
we have $B({\dbf^*}_1^{\ell_2};k_{\ell_2+1}-1)\leq B({\dbf^*}_1^{\ell_2};k_{\ell_2+1})$.
We further show that
\beqnarray{proof-OQ-LR-optimal-is-nondecreasing-(ii)-subcase-2(b)-iii}
B({\dbf^*}_1^{\ell_2};k_{\ell_2+1}-1)<B({\dbf^*}_1^{\ell_2};k_{\ell_2+1}).
\eeqnarray
Assume on the contrary that
$B({\dbf^*}_1^{\ell_2};k_{\ell_2+1}-1)\geq B({\dbf^*}_1^{\ell_2};k_{\ell_2+1})$,
i.e., $B({\dbf^*}_1^{\ell_2};k_{\ell_2+1}-1)=B({\dbf^*}_1^{\ell_2};k_{\ell_2+1})$
(as we have already shown that
$B({\dbf^*}_1^{\ell_2};k_{\ell_2+1}-1)\leq B({\dbf^*}_1^{\ell_2};k_{\ell_2+1})$).
Starting with $B({\dbf^*}_1^{\ell_2};k_{\ell_2+1}-1)=B({\dbf^*}_1^{\ell_2};k_{\ell_2+1})$,
it is easy to see that we can apply
\reqnarray{proof-OQ-LR-optimal-is-nondecreasing-(ii)-subcase-2(b)-ddd}
for $M-\ell_2-1$ times and obtain
\beqnarray{proof-OQ-LR-optimal-is-nondecreasing-(ii)-subcase-2(b)-jjj}
k_{\ell+2}=k_{\ell+1}+1 \textrm{ and }
B({\dbf^*}_1^{\ell+1};k_{\ell+2}-1)=B({\dbf^*}_1^{\ell+1};k_{\ell+2}),
\textrm{ for } \ell_2\leq \ell\leq M-2.
\eeqnarray
In particular, we have $B({\dbf^*}_1^{M-1};k_M-1)=B({\dbf^*}_1^{M-1};k_M)$
(with $\ell=M-2$ in \reqnarray{proof-OQ-LR-optimal-is-nondecreasing-(ii)-subcase-2(b)-jjj}).
As $k_M=k$, it is clear that $B({\dbf^*}_1^{M-1};k-1)=B({\dbf^*}_1^{M-1};k)$.
It then follows from \rlemma{OQ-LR-MRI-monotone}(ii)
(with $m=M-1\geq 1$ and $i=k-1\geq 0$ in \rlemma{OQ-LR-MRI-monotone}(ii))
that $k-1\geq M-1$, i.e., $k\geq M$, and we have reached a contradiction to $k\leq M-1$.
Therefore, we see from
$d_{\ell_2}^*<B({\dbf^*}_1^{\ell_2-1};k_{\ell_2+1})+1$
in \reqnarray{proof-OQ-LR-optimal-is-nondecreasing-(ii)-subcase-2(b)-444},
$B({\dbf^*}_1^{\ell_2-1};k_{\ell_2+1}-1)<B({\dbf^*}_1^{\ell_2-1};k_{\ell_2+1})$
in \reqnarray{proof-OQ-LR-optimal-is-nondecreasing-(ii)-subcase-2(b)-iii},
and $d_{\ell_2}^*\neq B({\dbf^*}_1^{\ell_2-1};k_{\ell_2+1}-1)+1$
in \reqnarray{proof-OQ-LR-optimal-is-nondecreasing-(ii)-subcase-2(b)-555}
that we have either $d_{\ell_2}^*<B({\dbf^*}_1^{\ell_2-1};k_{\ell_2+1}-1)+1$
or $B({\dbf^*}_1^{\ell_2-1};k_{\ell_2+1}-1)+1<d_{\ell_2}^*<B({\dbf^*}_1^{\ell_2-1};k_{\ell_2+1})+1$.

If $d_{\ell_2}^*<B({\dbf^*}_1^{\ell_2-1};k_{\ell_2+1}-1)+1$,
then we have $k_{\ell_2+1}-1\geq 1$
(otherwise, if $k_{\ell_2+1}-1=0$,
then $d_{\ell_2}^*<B({\dbf^*}_1^{\ell_2-1};k_{\ell_2+1}-1)+1=B({\dbf^*}_1^{\ell_2-1};0)+1=1$,
contradicting to $d_{\ell_2}^*\geq 1$).
From ${\dbf^*}_1^{\ell_2}\in \Acal_{\ell_2}$, $\ell_2\geq \ell_1+1>1$,
$k_{\ell_2+1}-1\geq 1$, $B({\dbf^*}_1^{\ell_2-1};k_{\ell_2+1}-1)>d_{\ell_2}^*-1$,
and \rlemma{OQ-LR-MRI-general}(ii) (with $m=\ell_2$ and $i=k_{\ell_2+1}-1$
in \rlemma{OQ-LR-MRI-general}(ii)), we have
\beqnarray{proof-OQ-LR-optimal-is-nondecreasing-(ii)-subcase-2(b)-kkk}
B({\dbf^*}_1^{\ell_2};k_{\ell_2+1}-1)=d_{\ell_2}^*+B({\dbf^*}_1^{\ell_2-1};k_{\ell_2+1}-2).
\eeqnarray
As it is easy to see from $d'_{\ell_2}=d_{\ell_2}^*+1$,
$d_{\ell_2}^*<B({\dbf^*}_1^{\ell_2-1};k_{\ell_2+1}-1)+1$,
and ${\dbf'}_1^{\ell_2-1}={\dbf^*}_1^{\ell_2-1}$ that
\beqnarray{proof-OQ-LR-optimal-is-nondecreasing-(ii)-subcase-2(b)-1111}
d'_{\ell_2}=d_{\ell_2}^*+1
\leq B({\dbf^*}_1^{\ell_2-1};k_{\ell_2+1}-1)+1
=B({\dbf'}_1^{\ell_2-1};k_{\ell_2+1}-1)+1,
\eeqnarray
it then follows from ${\dbf'}_1^{\ell_2}\in \Acal_{\ell_2}$, $\ell_2\geq \ell_1+1>1$,
$k_{\ell_2+1}-1\geq 1$, $B({\dbf'}_1^{\ell_2-1};k_{\ell_2+1}-1)\geq d'_{\ell_2}-1$
in \reqnarray{proof-OQ-LR-optimal-is-nondecreasing-(ii)-subcase-2(b)-1111},
and \rlemma{OQ-LR-MRI-general}(ii) (with $m=\ell_2$ and $i=k_{\ell_2+1}-1$
in \rlemma{OQ-LR-MRI-general}(ii)) that
\beqnarray{proof-OQ-LR-optimal-is-nondecreasing-(ii)-subcase-2(b)-2222}
B({\dbf'}_1^{\ell_2};k_{\ell_2+1}-1)=d'_{\ell_2}+B({\dbf'}_1^{\ell_2-1};k_{\ell_2+1}-2).
\eeqnarray
Thus, we see from
\reqnarray{proof-OQ-LR-optimal-is-nondecreasing-(ii)-subcase-2(b)-2222},
$d'_{\ell_2}=d_{\ell_2}^*+1$, ${\dbf'}_1^{\ell_2-1}={\dbf^*}_1^{\ell_2-1}$,
and \reqnarray{proof-OQ-LR-optimal-is-nondecreasing-(ii)-subcase-2(b)-kkk} that
\beqnarray{proof-OQ-LR-optimal-is-nondecreasing-(ii)-subcase-2(b)-3333}
B({\dbf'}_1^{\ell_2};k_{\ell_2+1}-1)
\aligneq d'_{\ell_2}+B({\dbf'}_1^{\ell_2-1};k_{\ell_2+1}-2) \nn\\
\aligneq d_{\ell_2}^*+1+B({\dbf^*}_1^{\ell_2-1};k_{\ell_2+1}-2) \nn\\
\aligneq B({\dbf^*}_1^{\ell_2};k_{\ell_2+1}-1)+1.
\eeqnarray
In the following, we show by induction on $\ell$ that
\beqnarray{}
\alignspace
B({\dbf^*}_1^{\ell};k_{\ell_2+1}-1)=B({\dbf^*}_1^{\ell_2};k_{\ell_2+1}-1),
\textrm{ for } \ell_2\leq \ell\leq M,
\label{eqn:proof-OQ-LR-optimal-is-nondecreasing-(ii)-subcase-2(b)-4444} \\
\alignspace
B({\dbf'}_1^{\ell};k_{\ell_2+1}-1)=B({\dbf'}_1^{\ell_2};k_{\ell_2+1}-1),
\textrm{ for } \ell_2\leq \ell\leq M.
\label{eqn:proof-OQ-LR-optimal-is-nondecreasing-(ii)-subcase-2(b)-5555}
\eeqnarray
As such, follows from
\reqnarray{proof-OQ-LR-optimal-is-nondecreasing-(ii)-subcase-2(b)-5555},
\reqnarray{proof-OQ-LR-optimal-is-nondecreasing-(ii)-subcase-2(b)-3333},
and \reqnarray{proof-OQ-LR-optimal-is-nondecreasing-(ii)-subcase-2(b)-4444} that
\beqnarray{}
B({\dbf'}_1^{\ell};k_{\ell_2+1}-1)
\aligneq B({\dbf'}_1^{\ell_2};k_{\ell_2+1}-1)
=B({\dbf^*}_1^{\ell_2};k_{\ell_2+1}-1)+1 \nn\\
\aligneq B({\dbf^*}_1^{\ell};k_{\ell_2+1}-1)+1,
\textrm{ for } \ell_2\leq \ell\leq M.
\eeqnarray
Therefore, \reqnarray{proof-OQ-LR-optimal-is-nondecreasing-(ii)-subcase-2(b)-bbb}
is proved in this case.
Since \reqnarray{proof-OQ-LR-optimal-is-nondecreasing-(ii)-subcase-2(b)-4444}
and \reqnarray{proof-OQ-LR-optimal-is-nondecreasing-(ii)-subcase-2(b)-5555}
hold trivially for $\ell=\ell_2$,
we assume as the induction hypothesis that
\reqnarray{proof-OQ-LR-optimal-is-nondecreasing-(ii)-subcase-2(b)-4444}
and \reqnarray{proof-OQ-LR-optimal-is-nondecreasing-(ii)-subcase-2(b)-5555}
hold for some $\ell_2\leq \ell\leq M-1$.
From $d_{\ell_1+1}^*\leq d_{\ell_1+2}^*\leq \cdots\leq d_M^*$,
$\ell\geq \ell_2\geq \ell_1+1$,
$d_{\ell_2+1}^*=B({\dbf^*}_1^{\ell_2};k_{\ell_2+1})+1$
in \reqnarray{proof-OQ-LR-optimal-is-nondecreasing-(ii)-case-2-jjj}
$B({\dbf^*}_1^{\ell_2};k_{\ell_2+1}-1)<B({\dbf^*}_1^{\ell_2};k_{\ell_2+1})$
in \reqnarray{proof-OQ-LR-optimal-is-nondecreasing-(ii)-subcase-2(b)-iii},
and $B({\dbf^*}_1^{\ell};k_{\ell_2+1}-1)=B({\dbf^*}_1^{\ell_2};k_{\ell_2+1}-1)$
in the induction hypothesis, we have
\beqnarray{proof-OQ-LR-optimal-is-nondecreasing-(ii)-subcase-2(b)-6666}
d_{\ell+1}^*\geq d_{\ell_2+1}^*=B({\dbf^*}_1^{\ell_2};k_{\ell_2+1})+1
>B({\dbf^*}_1^{\ell_2};k_{\ell_2+1}-1)+1
=B({\dbf^*}_1^{\ell};k_{\ell_2+1}-1)+1.
\eeqnarray
It follows from ${\dbf^*}_1^{\ell+1}\in \Acal_{\ell+1}$, $\ell+1>1$,
$k_{\ell_2+1}-1\geq 1$, $B({\dbf^*}_1^{\ell};k_{\ell_2+1}-1)<d_{\ell+1}^*-1$
in \reqnarray{proof-OQ-LR-optimal-is-nondecreasing-(ii)-subcase-2(b)-6666},
\rlemma{OQ-LR-MRI-general}(i)
(with $m=\ell+1$, $i=k_{\ell_2+1}-1$, and $\ell'=m-1=\ell$ in \rlemma{OQ-LR-MRI-general}(i)),
and $B({\dbf^*}_1^{\ell};k_{\ell_2+1}-1)=B({\dbf^*}_1^{\ell_2};k_{\ell_2+1}-1)$
in the induction hypothesis that
\beqnarray{proof-OQ-LR-optimal-is-nondecreasing-(ii)-subcase-2(b)-7777}
B({\dbf^*}_1^{\ell+1};k_{\ell_2+1}-1)
=B({\dbf^*}_1^{\ell};k_{\ell_2+1}-1)
=B({\dbf^*}_1^{\ell_2};k_{\ell_2+1}-1).
\eeqnarray
Similarly, we have from $d'_{\ell+1}=d_{\ell+1}^*+1$ (as $\ell+1>\ell_2$),
$d_{\ell+1}^*\geq d_{\ell_2+1}^*$ (as $\ell+1\geq \ell_2+1>\ell_1+1$),
\reqnarray{proof-OQ-LR-optimal-is-nondecreasing-(ii)-case-2-jjj},
\reqnarray{proof-OQ-LR-optimal-is-nondecreasing-(ii)-subcase-2(b)-iii},
\reqnarray{proof-OQ-LR-optimal-is-nondecreasing-(ii)-subcase-2(b)-3333},
and $B({\dbf'}_1^{\ell};k_{\ell_2+1}-1)=B({\dbf'}_1^{\ell_2};k_{\ell_2+1}-1)$
in the induction hypothesis that
\beqnarray{proof-OQ-LR-optimal-is-nondecreasing-(ii)-subcase-2(b)-8888}
d'_{\ell+1}
\aligneq d_{\ell+1}^*+1
\geq d_{\ell_2+1}^*+1=(B({\dbf^*}_1^{\ell_2};k_{\ell_2+1})+1)+1 \nn\\
\aligngreater B({\dbf^*}_1^{\ell_2};k_{\ell_2+1}-1)+2
=B({\dbf'}_1^{\ell_2};k_{\ell_2+1}-1)+1 \nn\\
\aligneq B({\dbf'}_1^{\ell};k_{\ell_2+1}-1)+1.
\eeqnarray
It then follows from ${\dbf'}_1^{\ell+1}\in \Acal_{\ell+1}$, $\ell+1>1$,
$k_{\ell_2+1}-1\geq 1$, $B({\dbf'}_1^{\ell};k_{\ell_2+1}-1)<d'_{\ell+1}-1$
in \reqnarray{proof-OQ-LR-optimal-is-nondecreasing-(ii)-subcase-2(b)-8888},
\rlemma{OQ-LR-MRI-general}(i)
(with $m=\ell+1$, $i=k_{\ell_2+1}-1$, and $\ell'=m-1=\ell$ in \rlemma{OQ-LR-MRI-general}(i)),
and $B({\dbf'}_1^{\ell};k_{\ell_2+1}-1)=B({\dbf'}_1^{\ell_2};k_{\ell_2+1}-1)$
in the induction hypothesis that
\beqnarray{proof-OQ-LR-optimal-is-nondecreasing-(ii)-subcase-2(b)-9999}
B({\dbf'}_1^{\ell+1};k_{\ell_2+1}-1)
=B({\dbf'}_1^{\ell};k_{\ell_2+1}-1)
=B({\dbf'}_1^{\ell_2};k_{\ell_2+1}-1).
\eeqnarray
The induction is completed by combining
\reqnarray{proof-OQ-LR-optimal-is-nondecreasing-(ii)-subcase-2(b)-7777}
and \reqnarray{proof-OQ-LR-optimal-is-nondecreasing-(ii)-subcase-2(b)-9999}.

On the other hand,
if $B({\dbf^*}_1^{\ell_2-1};k_{\ell_2+1}-1)+1<d_{\ell_2}^*<B({\dbf^*}_1^{\ell_2-1};k_{\ell_2+1})+1$,
then we show by induction on $\ell$ that
\beqnarray{proof-OQ-LR-optimal-is-nondecreasing-(ii)-subcase-2(b)-aaaa}
\alignspace
B({\dbf'}_1^{\ell};k_{\ell_2+1}-1)
=B({\dbf^*}_1^{\ell};k_{\ell_2+1}-1)
=B({\dbf^*}_1^{\ell_2-1};k_{\ell_2+1}-1),
\textrm{ for } \ell_2-1\leq \ell\leq M.
\eeqnarray
Therefore, \reqnarray{proof-OQ-LR-optimal-is-nondecreasing-(ii)-subcase-2(b)-bbb}
is also proved in this case.
As ${\dbf'}_1^{\ell_2-1}={\dbf^*}_1^{\ell_2-1}$,
it is clear that
\reqnarray{proof-OQ-LR-optimal-is-nondecreasing-(ii)-subcase-2(b)-aaaa}
holds for $\ell=\ell_2-1$.
Assume as the induction hypothesis that
\reqnarray{proof-OQ-LR-optimal-is-nondecreasing-(ii)-subcase-2(b)-aaaa}
holds for some $\ell_2-1\leq \ell\leq M-1$.
From $d_{\ell+1}^*\geq d_{\ell_2}^*$ (as $\ell+1\geq \ell_2\geq \ell_1+1$),
$d_{\ell_2}^*>B({\dbf^*}_1^{\ell_2-1};k_{\ell_2+1}-1)+1$,
and $B({\dbf^*}_1^{\ell};k_{\ell_2+1}-1)=B({\dbf^*}_1^{\ell_2-1};k_{\ell_2+1}-1)$
in the induction hypothesis, we have
\beqnarray{proof-OQ-LR-optimal-is-nondecreasing-(ii)-subcase-2(b)-bbbb}
d_{\ell+1}^*\geq d_{\ell_2}^*
>B({\dbf^*}_1^{\ell_2-1};k_{\ell_2+1}-1)+1
=B({\dbf^*}_1^{\ell};k_{\ell_2+1}-1)+1.
\eeqnarray
It follows from ${\dbf^*}_1^{\ell+1}\in \Acal_{\ell+1}$, $\ell+1>1$,
$k_{\ell_2+1}-1\geq 1$, $B({\dbf^*}_1^{\ell};k_{\ell_2+1}-1)<d_{\ell+1}^*-1$
in \reqnarray{proof-OQ-LR-optimal-is-nondecreasing-(ii)-subcase-2(b)-bbbb},
\rlemma{OQ-LR-MRI-general}(i)
(with $m=\ell+1$, $i=k_{\ell_2+1}-1$, and $\ell'=m-1=\ell$ in \rlemma{OQ-LR-MRI-general}(i)),
and $B({\dbf^*}_1^{\ell};k_{\ell_2+1}-1)=B({\dbf^*}_1^{\ell_2-1};k_{\ell_2+1}-1)$
in the induction hypothesis that
\beqnarray{proof-OQ-LR-optimal-is-nondecreasing-(ii)-subcase-2(b)-cccc}
B({\dbf^*}_1^{\ell+1};k_{\ell_2+1}-1)
=B({\dbf^*}_1^{\ell};k_{\ell_2+1}-1)
=B({\dbf^*}_1^{\ell_2-1};k_{\ell_2+1}-1).
\eeqnarray
Similarly, we have from $d'_{\ell+1}=d_{\ell+1}^*+1$ (as $\ell+1\geq \ell_2$),
$d_{\ell+1}^*\geq d_{\ell_2}^*$ (as $\ell+1\geq \ell_2\geq \ell_1+1$),
$d_{\ell_2}^*>B({\dbf^*}_1^{\ell_2-1};k_{\ell_2+1}-1)+1$,
and $B({\dbf'}_1^{\ell};k_{\ell_2+1}-1)=B({\dbf^*}_1^{\ell_2-1};k_{\ell_2+1}-1)$
in the induction hypothesis that
\beqnarray{proof-OQ-LR-optimal-is-nondecreasing-(ii)-subcase-2(b)-dddd}
d'_{\ell+1}=d_{\ell+1}^*+1
>d_{\ell+1}^*\geq d_{\ell_2}^*
>B({\dbf^*}_1^{\ell_2-1};k_{\ell_2+1}-1)+1
=B({\dbf'}_1^{\ell};k_{\ell_2+1}-1)+1.
\eeqnarray
It then follows from ${\dbf'}_1^{\ell+1}\in \Acal_{\ell+1}$, $\ell+1>1$,
$k_{\ell_2+1}-1\geq 1$, $B({\dbf'}_1^{\ell};k_{\ell_2+1}-1)<d'_{\ell+1}-1$
in \reqnarray{proof-OQ-LR-optimal-is-nondecreasing-(ii)-subcase-2(b)-dddd},
\rlemma{OQ-LR-MRI-general}(i)
(with $m=\ell+1$, $i=k_{\ell_2+1}-1$, and $\ell'=m-1=\ell$ in \rlemma{OQ-LR-MRI-general}(i)),
and $B({\dbf'}_1^{\ell};k_{\ell_2+1}-1)=B({\dbf^*}_1^{\ell_2-1};k_{\ell_2+1}-1)$
in the induction hypothesis that
\beqnarray{proof-OQ-LR-optimal-is-nondecreasing-(ii)-subcase-2(b)-eeee}
B({\dbf'}_1^{\ell+1};k_{\ell_2+1}-1)
=B({\dbf'}_1^{\ell};k_{\ell_2+1}-1)
=B({\dbf^*}_1^{\ell_2-1};k_{\ell_2+1}-1).
\eeqnarray
The induction is completed by combining
\reqnarray{proof-OQ-LR-optimal-is-nondecreasing-(ii)-subcase-2(b)-cccc}
and \reqnarray{proof-OQ-LR-optimal-is-nondecreasing-(ii)-subcase-2(b)-eeee}.

Finally, we prove $B({\dbf'}_1^{\ell};i)\geq B({\dbf^*}_1^{\ell};i)+1$
for $\ell_2\leq \ell\leq M$ and $k_{\ell_2+1}\leq i\leq k$
in \reqnarray{proof-OQ-LR-optimal-is-nondecreasing-(ii)-subcase-2(b)-ccc}.
by induction.
Let $k_{\ell_2+1}\leq i\leq k$.
From $d_{\ell_2}^*<B({\dbf^*}_1^{\ell_2-1};k_{\ell_2+1})+1$
in \reqnarray{proof-OQ-LR-optimal-is-nondecreasing-(ii)-subcase-2(b)-444},
${\dbf^*}_1^{\ell_2-1}\in \Acal_{\ell_2-1}$, $\ell_2-1\geq \ell_1>1$,
$i\geq k_{\ell_2+1}\geq 1$, and \rlemma{OQ-LR-MRI-monotone}(i), we have
\beqnarray{proof-OQ-LR-optimal-is-nondecreasing-(ii)-subcase-2(b)-ffff}
d_{\ell_2}^*
<B({\dbf^*}_1^{\ell_2-1};k_{\ell_2+1})+1
\leq B({\dbf^*}_1^{\ell_2-1};i)+1.
\eeqnarray
It follows from ${\dbf^*}_1^{\ell_2}\in \Acal_{\ell_2}$, $\ell_2\geq \ell_1+1>1$,
$i\geq k_{\ell_2+1}\geq 1$, $B({\dbf^*}_1^{\ell_2-1};i)>d_{\ell_2}^*-1$
in \reqnarray{proof-OQ-LR-optimal-is-nondecreasing-(ii)-subcase-2(b)-ffff},
and \rlemma{OQ-LR-MRI-general}(ii) that
\beqnarray{proof-OQ-LR-optimal-is-nondecreasing-(ii)-subcase-2(b)-gggg}
B({\dbf^*}_1^{\ell_2};i)=d_{\ell_2}^*+B({\dbf^*}_1^{\ell_2-1};i-1).
\eeqnarray
Similarly, we have from $d'_{\ell_2}=d_{\ell_2}^*+1$,
\reqnarray{proof-OQ-LR-optimal-is-nondecreasing-(ii)-subcase-2(b)-444},
${\dbf'}_1^{\ell_2-1}={\dbf^*}_1^{\ell_2-1}$,
${\dbf'}_1^{\ell_2-1}\in \Acal_{\ell_2-1}$,
$\ell_2-1\geq \ell_1>1$, $i\geq k_{\ell_2+1}\geq 1$,
and \rlemma{OQ-LR-MRI-monotone}(i) that
\beqnarray{proof-OQ-LR-optimal-is-nondecreasing-(ii)-subcase-2(b)-hhhh}
d'_{\ell_2}=d_{\ell_2}^*+1
\leq B({\dbf^*}_1^{\ell_2-1};k_{\ell_2+1})+1
=B({\dbf'}_1^{\ell_2-1};k_{\ell_2+1})+1
\leq B({\dbf'}_1^{\ell_2-1};i)+1.
\eeqnarray
It then follows from ${\dbf'}_1^{\ell_2}\in \Acal_{\ell_2}$, $\ell_2\geq \ell_1+1>1$,
$i\geq k_{\ell_2+1}\geq 1$, $B({\dbf'}_1^{\ell_2-1};i)\geq d'_{\ell_2}-1$
in \reqnarray{proof-OQ-LR-optimal-is-nondecreasing-(ii)-subcase-2(b)-hhhh},
\rlemma{OQ-LR-MRI-general}(ii), $d'_{\ell_2}=d_{\ell_2}^*+1$,
${\dbf'}_1^{\ell_2-1}={\dbf^*}_1^{\ell_2-1}$,
and \reqnarray{proof-OQ-LR-optimal-is-nondecreasing-(ii)-subcase-2(b)-gggg} that
\beqnarray{proof-OQ-LR-optimal-is-nondecreasing-(ii)-subcase-2(b)-iiii}
B({\dbf'}_1^{\ell_2};i)
=d'_{\ell_2}+B({\dbf'}_1^{\ell_2-1};i-1)
=d_{\ell_2}^*+1+B({\dbf^*}_1^{\ell_2-1};i-1)
=B({\dbf^*}_1^{\ell_2};i)+1.
\eeqnarray
It is immediate from
\reqnarray{proof-OQ-LR-optimal-is-nondecreasing-(ii)-subcase-2(b)-iiii}
that \reqnarray{proof-OQ-LR-optimal-is-nondecreasing-(ii)-subcase-2(b)-ccc}
holds for $\ell=\ell_2$ and all $k_{\ell_2+1}\leq i\leq k$.

Assume as the induction hypothesis that
\reqnarray{proof-OQ-LR-optimal-is-nondecreasing-(ii)-subcase-2(b)-ccc}
holds for some $\ell_2\leq \ell \leq M-1$ and all $k_{\ell_2+1}\leq i\leq k$.
In the following, we let $k_{\ell_2+1}\leq i\leq k$.
If $d_{\ell+1}^*\leq B({\dbf^*}_1^{\ell};i)+1$,
then we have from ${\dbf^*}_1^{\ell+1}\in \Acal_{\ell+1}$, $\ell+1>1$,
$i\geq k_{\ell_2+1}\geq 1$, and \rlemma{OQ-LR-MRI-general}(ii) that
\beqnarray{proof-OQ-LR-optimal-is-nondecreasing-(ii)-subcase-2(b)-jjjj}
B({\dbf^*}_1^{\ell+1};i)=d_{\ell+1}^*+B({\dbf^*}_1^{\ell};i-1).
\eeqnarray
Similarly, we have from $d'_{\ell+1}=d_{\ell+1}^*+1$ (as $\ell+1>\ell_2$),
$d_{\ell+1}^*\leq B({\dbf^*}_1^{\ell};i)+1$,
and $B({\dbf'}_1^{\ell};i)\geq B({\dbf^*}_1^{\ell};i)+1$
in the induction hypothesis that
\beqnarray{proof-OQ-LR-optimal-is-nondecreasing-(ii)-subcase-2(b)-kkkk}
d'_{\ell+1}=d_{\ell+1}^*+1
\leq (B({\dbf^*}_1^{\ell};i)+1)+1
\leq B({\dbf'}_1^{\ell};i)+1.
\eeqnarray
It then follows from ${\dbf'}_1^{\ell+1}\in \Acal_{\ell+1}$, $\ell+1>1$,
$i\geq k_{\ell_2+1}\geq 1$,
$B({\dbf'}_1^{\ell};i)\geq d'_{\ell+1}-1$
in \reqnarray{proof-OQ-LR-optimal-is-nondecreasing-(ii)-subcase-2(b)-kkkk},
\rlemma{OQ-LR-MRI-general}(ii), and $d'_{\ell+1}=d_{\ell+1}^*+1$ that
\beqnarray{proof-OQ-LR-optimal-is-nondecreasing-(ii)-subcase-2(b)-11111}
B({\dbf'}_1^{\ell+1};i)
=d'_{\ell+1}+B({\dbf'}_1^{\ell};i-1)
=d_{\ell+1}^*+1+B({\dbf'}_1^{\ell};i-1).
\eeqnarray
For the case that $i=k_{\ell_2+1}$,
we have from \reqnarray{proof-OQ-LR-optimal-is-nondecreasing-(ii)-subcase-2(b)-bbb} that
\beqnarray{proof-OQ-LR-optimal-is-nondecreasing-(ii)-subcase-2(b)-22222}
B({\dbf'}_1^{\ell};i-1)=B({\dbf'}_1^{\ell};k_{\ell_2+1}-1)
\geq B({\dbf^*}_1^{\ell};k_{\ell_2+1}-1)=B({\dbf^*}_1^{\ell};i-1).
\eeqnarray
For the case that $k_{\ell_2+1}+1\leq i\leq k$,
we have $k_{\ell_2+1}\leq i-1\leq k-1$
and it follows from the induction hypothesis that
\beqnarray{proof-OQ-LR-optimal-is-nondecreasing-(ii)-subcase-2(b)-33333}
B({\dbf'}_1^{\ell};i-1)\geq B({\dbf^*}_1^{\ell};i-1)+1>B({\dbf^*}_1^{\ell};i-1).
\eeqnarray
As such, we see from
\reqnarray{proof-OQ-LR-optimal-is-nondecreasing-(ii)-subcase-2(b)-11111},
\reqnarray{proof-OQ-LR-optimal-is-nondecreasing-(ii)-subcase-2(b)-22222},
\reqnarray{proof-OQ-LR-optimal-is-nondecreasing-(ii)-subcase-2(b)-33333}
and \reqnarray{proof-OQ-LR-optimal-is-nondecreasing-(ii)-subcase-2(b)-jjjj} that
\beqnarray{proof-OQ-LR-optimal-is-nondecreasing-(ii)-subcase-2(b)-44444}
B({\dbf'}_1^{\ell+1};i)
=d_{\ell+1}^*+1+B({\dbf'}_1^{\ell};i-1)
\geq d_{\ell+1}^*+1+B({\dbf^*}_1^{\ell};i-1)
=B({\dbf^*}_1^{\ell+1};i)+1.
\eeqnarray
On the other hand, if $d_{\ell+1}^*>B({\dbf^*}_1^{\ell};i)+1$,
then we have from ${\dbf^*}_1^{\ell+1}\in \Acal_{\ell+1}$, $\ell+1>1$,
$i\geq k_{\ell_2+1}\geq 1$, and \rlemma{OQ-LR-MRI-general}(i) that
\beqnarray{proof-OQ-LR-optimal-is-nondecreasing-(ii)-subcase-2(b)-55555}
B({\dbf^*}_1^{\ell+1};i)=B({\dbf^*}_1^{\ell};i).
\eeqnarray
For the case that $d'_{\ell+1}\leq B({\dbf'}_1^{\ell};i)+1$,
we have from ${\dbf'}_1^{\ell+1}\in \Acal_{\ell+1}$, $\ell+1>1$,
$i\geq k_{\ell_2+1}\geq 1$, \rlemma{OQ-LR-MRI-general}(ii),
$d'_{\ell+1}=d_{\ell+1}^*+1$, $d_{\ell+1}^*>B({\dbf^*}_1^{\ell};i)+1$,
and \reqnarray{proof-OQ-LR-optimal-is-nondecreasing-(ii)-subcase-2(b)-55555} that
\beqnarray{proof-OQ-LR-optimal-is-nondecreasing-(ii)-subcase-2(b)-66666}
B({\dbf'}_1^{\ell+1};i)
\aligneq d'_{\ell+1}+B({\dbf'}_1^{\ell};i-1)\geq d'_{\ell+1}=d_{\ell+1}^*+1 \nn\\
\aligngreater (B({\dbf^*}_1^{\ell};i)+1)+1=B({\dbf^*}_1^{\ell+1};i)+2.
\eeqnarray
For the case that $d'_{\ell+1}>B({\dbf'}_1^{\ell};i)+1$,
we have from ${\dbf'}_1^{\ell+1}\in \Acal_{\ell+1}$, $\ell+1>1$,
$i\geq k_{\ell_2+1}\geq 1$, \rlemma{OQ-LR-MRI-general}(i),
$B({\dbf'}_1^{\ell};i)\geq B({\dbf^*}_1^{\ell};i)+1$ in the induction hypothesis,
and \reqnarray{proof-OQ-LR-optimal-is-nondecreasing-(ii)-subcase-2(b)-55555} that
\beqnarray{proof-OQ-LR-optimal-is-nondecreasing-(ii)-subcase-2(b)-77777}
B({\dbf'}_1^{\ell+1};i)=B({\dbf'}_1^{\ell};i)
\geq B({\dbf^*}_1^{\ell};i)+1=B({\dbf^*}_1^{\ell+1};i)+1.
\eeqnarray
The induction is completed by combining
\reqnarray{proof-OQ-LR-optimal-is-nondecreasing-(ii)-subcase-2(b)-44444},
\reqnarray{proof-OQ-LR-optimal-is-nondecreasing-(ii)-subcase-2(b)-66666},
and \reqnarray{proof-OQ-LR-optimal-is-nondecreasing-(ii)-subcase-2(b)-77777}.

\bappendix{Proof of \rlemma{OQ-LR-optimal-properties}}{OQ-LR-optimal-properties}

Note that from \rlemma{OQ-LR-optimal-is-nondecreasing},
we have $B({\dbf^*}_1^M;k)\geq d_M^*$
and $d_1^*\leq d_2^*\leq \cdots\leq d_M^*$.

(i) It is clear from $B({\dbf^*}_1^M;k)\geq d_M^*$
and the definition of $s_k$ in \reqnarray{OQ-LR-optimal-properties-1}
that $s_k$ is well defined and $s_k=M$.
Furthermore, we have from $s_k=M$, ${\dbf^*}_1^M\in \Acal_M$, $1\leq k<M$,
the definition of $s_k$ in \reqnarray{OQ-LR-optimal-properties-1},
and \rlemma{OQ-LR-MRI-general}(iii) that
\beqnarray{}
B({\dbf^*}_1^{s_k};k)=B({\dbf^*}_1^M;k)=d_{s_k}^*+B({\dbf^*}_1^{s_k-1};k-1). \nn
\eeqnarray
Therefore, \reqnarray{OQ-LR-optimal-properties-3} is proved.

(ii) As $1\leq k\leq M-1$, we have $s_k=M\geq k+1$.
We first show by induction on $i$ that $s_i$ is well defined and $s_i\geq i+1$
for $i=k,k-1,\ldots,1$ as in \reqnarray{OQ-LR-optimal-properties-4}.
From \reqnarray{OQ-LR-optimal-properties-3},
we know that $s_k$ is well defined and $s_k=M\geq k+1$.
Assume as the induction hypothesis that $s_k,s_{k-1},\ldots,s_{i+1}$ are well defined
and $s_k\geq k+1,s_{k-1}\geq k,\ldots,s_{i+1}\geq i+2$ for some $1\leq i\leq k-1$.
Since we have from ${\dbf^*}_1^{s_{i+1}-1}\in \Acal_{s_{i+1}-1}$ and $i\geq 1$
that $d_1^*=1\leq B({\dbf^*}_1^{s_{i+1}-1};i)$
and we have from the induction hypothesis that $1<i+1\leq s_{i+1}-1$,
it is also clear from the definition of $s_i$
in \reqnarray{OQ-LR-optimal-properties-2} that $s_i$ is well defined.

To complete the induction, it remains to show that $s_i\geq i+1$.
Suppose on the contrary that $s_i\leq i$.
Then we have from $s_i<i+1\leq s_{i+1}-1$ (by the induction hypothesis)
and the definition of $s_i$ in \reqnarray{OQ-LR-optimal-properties-2} that
\beqnarray{proof-OQ-LR-optimal-properties-(ii)-111}
d_{i+1}^*>B({\dbf^*}_1^{s_{i+1}-1};i).
\eeqnarray
We also have from ${\dbf^*}_1^{s_{i+1}-1}\in \Acal_{s_{i+1}-1}$,
$B({\dbf^*}_1^{s_{i+1}-1};i)<d_{i+1}^*$
in \reqnarray{proof-OQ-LR-optimal-properties-(ii)-111},
$1\leq i\leq (s_{i+1}-1)-1$ (by the induction hypothesis),
and \rlemma{OQ-LR-MRI-general}(i)
(with $m=s_{i+1}-1$ and $\ell'=i$ in \rlemma{OQ-LR-MRI-general}(i)) that
\beqnarray{proof-OQ-LR-optimal-properties-(ii)-222}
B({\dbf^*}_1^{s_{i+1}-1};i)=B({\dbf^*}_1^{s_{i+1}-2};i)
=\cdots=B({\dbf^*}_1^{i+1};i)=B({\dbf^*}_1^i;i).
\eeqnarray
Note that as $B({\dbf^*}_1^i;i)=\sum_{\ell=1}^{i}d_{\ell}^*$,
it is clear from ${\dbf^*}_1^M\in \Acal_M$ that
$B({\dbf^*}_1^i;i)=\sum_{\ell=1}^{i}d_{\ell}^*\geq d_{i+1}^*-1$.
It follows from $B({\dbf^*}_1^i;i)\geq d_{i+1}^*-1$
and \rlemma{OQ-LR-MRI-equivalent conditions}
(with $m=i+1$ and $\ell'=m-1=i$ in \rlemma{OQ-LR-MRI-equivalent conditions}) that
\beqnarray{proof-OQ-LR-optimal-properties-(ii)-333}
B({\dbf^*}_1^{i+1};i)\geq d_{i+1}^*.
\eeqnarray
Therefore, we have from \reqnarray{proof-OQ-LR-optimal-properties-(ii)-222}
and \reqnarray{proof-OQ-LR-optimal-properties-(ii)-333} that
\beqnarray{}
B({\dbf^*}_1^{s_{i+1}-1};i)=B({\dbf^*}_1^{i+1};i)\geq d_{i+1}^*, \nn
\eeqnarray
and we have reached a contradiction to $B({\dbf^*}_1^{s_{i+1}-1};i)<d_{i+1}^*$
in \reqnarray{proof-OQ-LR-optimal-properties-(ii)-111}.

Now we prove \reqnarray{OQ-LR-optimal-properties-5}
and \reqnarray{OQ-LR-optimal-properties-6}.
From ${\dbf^*}_1^{s_{i+1}-1}\in \Acal_{s_{i+1}-1}$,
$1\leq i<s_{i+1}-1$ in \reqnarray{OQ-LR-optimal-properties-4},
the definition of $s_i$ in \reqnarray{OQ-LR-optimal-properties-2}
for $i=1,2,\ldots,k-1$, and \rlemma{OQ-LR-MRI-general}(iii), we have
\beqnarray{}
B({\dbf^*}_1^{s_{i+1}-1};i)=B({\dbf^*}_1^{s_{i+1}-2};i)=\cdots
=B({\dbf^*}_1^{s_i};i)=d_{s_i}^*+B({\dbf^*}_1^{s_i-1};i-1), \nn\\
\textrm{ for } i=1,2,\ldots,k-1, \nn
\eeqnarray
which is the desired result in \reqnarray{OQ-LR-optimal-properties-5}.
Furthermore, it is easy to see from \reqnarray{OQ-LR-optimal-properties-3}
and \reqnarray{OQ-LR-optimal-properties-5} that
\beqnarray{}
B({\dbf^*}_1^{s_i};i)
\aligneq d_{s_i}^*+B({\dbf^*}_1^{s_i-1};i-1)
=d_{s_i}^*+d_{s_{i-1}}^*+B({\dbf^*}_1^{s_{i-1}-1};i-2)
=\cdots \nn\\
\aligneq d_{s_i}^*+d_{s_{i-1}}^*+\cdots +d_{s_1}^*+B({\dbf^*}_1^{s_1-1};0)
=d_{s_1}^*+d_{s_2}^*+\cdots +d_{s_i}^*,\nn\\
\alignspace \hspace*{3.04in} \textrm{ for } i=1,2,\ldots,k, \nn
\eeqnarray
which is the desired result in \reqnarray{OQ-LR-optimal-properties-6}.

(iii) First we show that $B({\dbf^*}_1^{s_i};i+1)\geq d_{s_i+1}^*-1$
for $i=1,2,\ldots,k-1$ in \reqnarray{OQ-LR-optimal-properties-7}.
Assume on the contrary that $B({\dbf^*}_1^{s_i};i+1)<d_{s_i+1}^*-1$
for some $1\leq i\leq k-1$.
Then we have from ${\dbf^*}_1^{s_{i+1}}\in \Acal_{s_{i+1}}$,
$s_{i+1}>i+1>1$ in \reqnarray{OQ-LR-optimal-properties-4},
$d_{s_i+1}^*=\min\{d_{s_i+1}^*,d_{s_i+2}^*,\ldots,d_{s_{i+1}}^*\}$
(as $d_1^*\leq d_2^*\leq \cdots\leq d_M^*$),
$B({\dbf^*}_1^{s_i};i+1)<d_{s_i+1}^*-1$, $1\leq s_i\leq s_{i+1}-1$,
\rlemma{OQ-LR-MRI-general}(iv)
(with $m=s_{i+1}$ and $\ell'=s_i$ in \rlemma{OQ-LR-MRI-general}(iv)),
and $d_{s_i+1}^*\leq d_{s_{i+1}}^*$ that
\beqnarray{}
B({\dbf^*}_1^{s_{i+1}};i+1)=B({\dbf^*}_1^{s_{i+1}-1};i+1)
=\cdots=B({\dbf^*}_1^{s_i};i+1)<d_{s_i+1}^*-1\leq d_{s_{i+1}}^*-1, \nn
\eeqnarray
contradicting to $B({\dbf^*}_1^{s_{i+1}};i+1)=d_{s_1}^*+d_{s_2}^*
+\cdots +d_{s_{i+1}}^*\geq d_{s_{i+1}}^*$ in \reqnarray{OQ-LR-optimal-properties-6}.

Now we show that $B({\dbf^*}_1^{s_i-1};i)\geq d_{s_i}^*-1$
for $i=1,2,\ldots,k$ in \reqnarray{OQ-LR-optimal-properties-8}.
Assume on the contrary that $B({\dbf^*}_1^{s_i-1};i)<d_{s_i}^*-1$
for some $1\leq i\leq k$.
Then we have from ${\dbf^*}_1^{s_i}\in \Acal_{s_i}$,
$s_i>i\geq 1$ in \reqnarray{OQ-LR-optimal-properties-4},
$B({\dbf^*}_1^{s_i-1};i)<d_{s_i}^*-1$,
and \rlemma{OQ-LR-MRI-general}(iv)
(with $m=s_i$ and $\ell'=m-1=s_i-1$ in \rlemma{OQ-LR-MRI-general}(iv)) that
\beqnarray{}
B({\dbf^*}_1^{s_i};i)=B({\dbf^*}_1^{s_i-1};i)<d_{s_i}^*-1, \nn
\eeqnarray
contradicting to $B({\dbf^*}_1^{s_{i}};i)=d_{s_1}^*+d_{s_2}^*
+\cdots +d_{s_i}^*\geq d_{s_i}^*$ in \reqnarray{OQ-LR-optimal-properties-6}.

Finally, we prove \reqnarray{OQ-LR-optimal-properties-9}
and \reqnarray{OQ-LR-optimal-properties-10}.
From ${\dbf^*}_1^{s_i}\in \Acal_{s_i}$,
$s_i\geq i+1>1$ in \reqnarray{OQ-LR-optimal-properties-4},
$B({\dbf^*}_1^{s_i};i+1)\geq d_{s_i+1}^*-1\geq d_{s_i}^*-1$
in \reqnarray{OQ-LR-optimal-properties-7},
and \rlemma{OQ-LR-MRI-general}(ii), we have
\beqnarray{}
B({\dbf^*}_1^{s_i};i+1)=d_{s_i}^*+B({\dbf^*}_1^{s_i-1};i),
\textrm{ for } i=1,2,\ldots,k-1, \nn
\eeqnarray
which is the desired result in \reqnarray{OQ-LR-optimal-properties-9}.
From ${\dbf^*}_1^{s_i-1}\in \Acal_{s_i-1}$,
$s_i-1\geq i\geq 1$ in \reqnarray{OQ-LR-optimal-properties-4},
$B({\dbf^*}_1^{s_i-1};i)\geq d_{s_i}^*-1\geq d_{s_i-1}^*-1$
in \reqnarray{OQ-LR-optimal-properties-8},
and \rlemma{OQ-LR-MRI-general}(ii), we also have
\beqnarray{}
B({\dbf^*}_1^{s_i-1};i)=d_{s_i-1}^*+B({\dbf^*}_1^{s_i-2};i-1),
\textrm{ for } i=1,2,\ldots,k, \nn
\eeqnarray
which is the desired result in \reqnarray{OQ-LR-optimal-properties-10}.



\begin{thebibliography}{99}

\baselineskip12pt

\bibitem{PSS86}
P. R. Prucnal, M. A. Santoro, and S. K. Sehgal,
``Ultrafast all-optical synchronous multiple access fiber networks''
\emph{IEEE Journal on Selected Areas in Communications}\/,
vol.~4, pp.~1484--1493, December~1986.

\bibitem{Karol93}
M. J. Karol,
``Shared-memory optical packet (ATM) switch,''
in \emph{Proceedings SPIE : Multigigabit Fiber Communication Systems (1993)}\/,
vol.~2024, pp.~212--222, October~1993.

\bibitem{Haas93}
Z. Hass,
``The ``staggering switch'': An electronically controlled optical packet switch,''
\emph{IEEE Journal of Lightwave Technology}\/,
vol.~11, pp.~925--936, May/June~1993.

\bibitem{CF94}
I. Chlamtac and A. Fumagalli,
``Quadro-star: A high performance optical WDM star network,''
\emph{IEEE Transactions on Communications}\/,
vol.~42, pp.~2582--2591, August~1994.

\bibitem{CFKMNPCCFHLMSSW96}
I. Chlamtac, A. Fumagalli, L. G. Kazovsky, P. Melman,
W. H. Nelson, P. Poggiolini, M. Cerisola, A. N. M. M. Choudhury, T. K. Fong,
R. T. Hofmeister, C.-L. Lu, A. Mekkittikul, D. J. M. Sabido IX, C.-J. Suh, and E. W. M. Wong,
``Cord: contention resolution by delay lines,''
\emph{IEEE Journal on Selected Areas in Communications}\/,
vol.~14, pp.~1014--1029, June~1996.

\bibitem{CT96}
R. L. Cruz and J.-T. Tsai,
``COD: alternative architectures for high speed packet switching,''
\emph{IEEE/ACM Transactions on Networking}\/,
vol.~4, pp.~11--21, February~1996.

\bibitem{HCA97}
D. K. Hunter, D. Cotter, R. B. Ahmad, D. Cornwell, T. H. Gilfedder, P. J. Legg, and I. Andonovic,
``$2 \times 2$ buffered switch fabrics for traffic routing, merging and shaping in photonic cell networks,''
\emph{IEEE Journal of Lightwave Technology}\/,
vol.~15, pp.~86--101, January~1997.

\bibitem{HCG98}
D. K. Hunter, W. D. Cornwell, T. H. Gilfedder, A. Franzen, and I. Andonovic,
``SLOB: a switch with large optical buffers for packet switching,''
\emph{IEEE Journal of Lightwave Technology}\/,
vol.~16, pp.~1725--1736, October~1998.

\bibitem{Varvarigos98} E. Varvarigos,
``The ``packing'' and the ``scheduling packet'' switch architectures for almost all-optical lossless networks,''
\emph{IEEE Journal of Lightwave Technology}\/,
vol.~16, pp.~1757--1767, October~1998.

\bibitem{CFS00}
I. Chlamtac, A. Fumagalli, and C.-J. Suh,
``Multibuffer delay line architectures for efficient contention resolution in optical switching nodes,''
\emph{IEEE Transactions on Communications}\/,
vol.~48, pp.~2089--2098, December~2000.

\bibitem{CCCL07a}
Y.-T. Chen, C.-S. Chang, J. Cheng, and D.-S. Lee,
``Feedforward SDL constructions of output-buffered multiplexers and switches with variable length bursts,''
in \emph{Proceedings IEEE International Conference on Computer Communications (INFOCOM'07)}\/,
Anchorage, AK, USA, May~6--12, 2007.

\bibitem{CFS97}
I. Chlamtac, A. Fumagalli, and C.-J. Suh,
``Optimal $2\times 1$ multi-stage optical packet multiplexer,''
in \emph{Proceedings IEEE Global Telecommunications Conference (GLOBECOM'97)}\/,
Phoenix, AZ, USA, November~3--8, 1997, pp.~566--570.



\bibitem{CCLC06}
C.-C. Chou, C.-S. Chang, D.-S. Lee and J. Cheng,
``A necessary and sufficient condition for the construction of 2-to-1 optical FIFO multiplexers
by a single crossbar switch and fiber delay lines,''
\emph{IEEE Transactions on Information Theory}\/,
vol.~52, pp.~4519--4531, October~2006.

\bibitem{Cheng07}
J. Cheng,
``Constructions of fault tolerant optical 2-to-1 FIFO multiplexers,''
\emph{IEEE Transactions on Information Theory}\/,
vol.~53, pp.~4092--4105, November~2007.

\bibitem{Cheng08}
J. Cheng,
``Constructions of optical 2-to-1 FIFO multiplexers with a limited number of recirculations,''
\emph{IEEE Transactions on Information Theory}\/,
vol.~54, pp.~4040--4052, September~2008.


\bibitem{WJH08e}
X. Wang, X. Jiang, and S. Horiguchi,
``Improved bounds on the feedfoward design of optical multiplexers,''
in \emph{Proceedings International Symposium on Parallel Architectures, Algorithms, and Networks (I-SPAN'08)}\/,
Sydney, Australia, May~7--9, 2008, pp.~178--183.

\bibitem{LT06} S.-Y. R. Li and X. J. Tan,
``Fiber memory,''
in \emph{Proceedings Annual Allerton Conference on Communication, Control, and Computing (Allerton'06)}\/,
Monticello, IL, USA, September~27--29, 2006.
Full version of this paper is submitted to \emph{IEEE Transactions on Information Theory}\/.

\bibitem{CCL06}
C.-S. Chang, Y.-T. Chen, and D.-S. Lee,
``Constructions of optical FIFO queues,''
\emph{IEEE Transactions on Information Theory}\/,
vol.~52, pp.~2838--2843, June~2006.


\bibitem{SSB07} B. A. Small, A. Shacham, and K. Bergman,
``A modular, scalable, extensible, and transparent optical packet buffer,''
\emph{Journal of Lightwave Technology}\/,
vol.~25, pp.~978--985, April~2007.

\bibitem{HCCL07}
P.-K. Huang, C.-S. Chang, J. Cheng, and D.-S. Lee,
``Recursive constructions of parallel FIFO and LIFO queues with switched delay lines,''
\emph{IEEE Transactions on Information Theory}\/,
vol.~53, pp.~1778--1798, May~2007.

\bibitem{BG07}
N. Beheshti and Y. Ganjali,
``Packet scheduling in optical FIFO buffers,''
in \emph{Proceedings IEEE High-Speed Networks Workshop (HSN'07)}\/,
Anchorage, AK, USA, May~11, 2007.

\bibitem{WJH08d}
X. Wang, X. Jiang, and S. Horiguchi,
``A construction of shared optical buffer queue with switched delay lines,''
in \emph{Proceedings IEEE International Conference on High Performance Switching and Routing (HPSR'08)}\/,
Shanghai, China, May~15--17, 2008, pp.~86--91.

\bibitem{SA06}
A. D. Sarwate and V. Anantharam,
``Exact emulation of a priority queue with a switch and delay lines,''
\emph{Queueing Systems: Theory and Applications}\/,
vol.~53, pp.~115--125, July~2006.

\bibitem{CCCL07b}
H.-C. Chiu, C.-S. Chang, J. Cheng, and D.-S. Lee,
``A simple proof for the constructions of optical priority queues,''
\emph{Queueing Systems: Theory and Applications}\/,
vol.~56, pp.~73--77, June~2007


\bibitem{KK07a}
H. Kogan and I. Keslassy,
``Optimal-complexity optical router,''
in \emph{Proceedings IEEE International Conference on Computer Communications (INFOCOM'07 Minisymposium)}\/,
Anchorage, AK, USA, May~6--12, 2007.

\bibitem{RGG09}
H. Rastegarfar, M. Ghobadi, and Y. Ganjali,
``Emulation of Optical PIFO Buffers,''
in \emph{Proceedings IEEE Global Communications Conference (GLOBECOM'09)}\/,
Honolulu, HI, USA, November~30--December~4, 2009.

\bibitem{JLLR94}
F. Jordan, D. Lee, K. Y. Lee, and S. V. Ramanan,
``Serial array time slot interchangers and optical implementations,''
\emph{IEEE Transactions on Computers}\/,
vol.~43, pp.~1309--1318, 1994.

\bibitem{CCL10}
Y.-T. Chen, J. Cheng, and D.-S. Lee,
``Constructions of linear compressors, non-overtaking delay lines, and flexible delay lines for optical packet switching,''
\emph{IEEE/ACM Transactions on Networking}\/,
vol.~17, pp.~2014--2027, December~2009.
Conference version appeared in \emph{IEEE INFOCOM} 2006.

\bibitem{CCCL09}
C.-S. Chang, J. Cheng, T.-H. Chao, and D.-S. Lee,
``Optimal constructions of fault tolerant optical linear compressors and linear decompressors,''
\emph{IEEE Transactions on Communications}\/,
vol.~57, pp.~1140--1150, April~2009.
Conference version appeared in \emph{IEEE INFOCOM} 2007.





\bibitem{KK07b}
H. Kogan and I. Keslassy,
``Fundamental complexity of optical systems,''
in \emph{Proceedings IEEE International Conference on Computer Communications (INFOCOM'07 Minisymposium)}\/,
Anchorage, AK, USA, May~6--12, 2007.


\bibitem{LLJH09}
J. Liu, T. T. Lee, X. Jiang, and S. Horiguchi,
``Blocking and delay analysis of single wavelength optical buffer with
general packet size distribution,''
\emph{IEEE Journal of Lightwave Technology}\/,
vol.~27, pp.~955--966, April 2009.


\bibitem{HCA98}
D. K. Hunter, M. C. Chia, and I. Andonovic,
``Buffering in optical packet switches,''
\emph{IEEE Journal of Lightwave Technology}\/,
vol.~16, pp.~2081--2094, December~1998.

\bibitem{YMD00}
S. Yao, B. Mukherjee, and S. Dixit,
``Advances in photonic packet switching: An overview,''
\emph{IEEE Communications Magazine}\/,
vol.~38, pp.~84--94, February~2000.

\bibitem{HA00}
D. K. Hunter and I. Andonovic,
``Approaches to optical Internet packet switching,''
\emph{IEEE Communications Magazine}\/,
vol.~38, pp.~116--122, September~2000.

\bibitem{YMYD03} S. Yao, B. Mukherjee, S. J. B. Yoo, and S. Dixit,
``A unified study of contention-resolution schemes in optical packet-switched networks,''
\emph{Journal of Lightwave Technology}\/,
vol.~21, pp.~672--683, March~2003.

\bibitem{Yoo06} S. J. B. Yoo,
``Optical packet and burst switching technologies for the future photonic internet,''
\emph{Journal of Lightwave Technology}\/,
vol.~24, pp.~4468--4492, December~2006.

\bibitem{Miklos08} K. Mikl\'{o}s,
``Congestion resolution and buffering in packet switched all-optical networks,''
Ph.D. Dissertation, Budapest University of Techonology and Economics,
Budapest, Hungary, 2008.

\bibitem{BB06}
E. F. Burmeister and J. E. Bowers,
``Integrated gate matrix switch for optical packet buffering,''
\emph{IEEE Photonics Technology Letters}\/,
vol.~18, pp.~103--105, January~2006.

\bibitem{LG97}
C. P. Larsen and M. Gustavsson,
``Linear crosstalk in $4\times 4$ semiconductor optical amplifier gate switch matrix,''
\emph{IEEE Journal of Lightwave Technology}\/,
vol.~15, pp.~1865--1870, October~1997.

\bibitem{GWCMKDLKY05}
R. Geldenhuys, Z. Wang, N. Chi, I. Tafur Monroy, A. M. J. Koonen, H. J. S. Dorren,
F. W. Leuschner, G. D. Khoe, and S. Yu,
``Multiple recirculations through Crosspoint switch fabric for recirculating optical buffering,''
\emph{Electronics Letters}\/,
vol.~41, pp.~1136--1137, September~2005.


\end{thebibliography}
\end{document}